\documentclass[aps,twocolumn]{revtex4}

\RequirePackage[T2A]{fontenc}

\usepackage{amssymb}
\usepackage{amsmath}
\usepackage{bm}
\usepackage[dvips]{epsfig}
\usepackage{graphicx}

\begin{document}

\title{Reconstructions of the electron dynamics in 
magnetic field and the geometry of complex Fermi surfaces.}

\author{A.Ya. Maltsev}

\affiliation{
\centerline{\it L.D. Landau Institute for Theoretical Physics
of RAS}
\centerline{\it 142432 Chernogolovka, pr. Ak. Semenova 1A,
maltsev@itp.ac.ru}}

\begin{abstract}
The paper considers the semiclassical dynamics of electrons 
on complex Fermi surfaces in the presence of strong magnetic 
fields. The reconstructions of the general topological structure 
of such dynamics are accompanied by the appearance of closed 
extremal trajectories of a special form, closely related to 
geometry and topology of the Fermi surface. The study of 
oscillation phenomena on such trajectories allows, in particular, 
to propose a relatively simple method for refining the parameters 
of the dispersion relation in metals with complex Fermi surfaces. 
\end{abstract}

\maketitle

\vspace{5mm}

\section{Introduction}

 In this paper, we want to consider some issues related 
to the topological structure of dynamical system 
\begin{equation}
\label{MFSyst}
{\dot {\bf p}} \,\,\,\, = \,\,\,\, {e \over c} \,\,
\left[ {\bf v}_{\rm gr} ({\bf p}) \, \times \, {\bf B} \right]
\,\,\,\, = \,\,\,\, {e \over c} \,\, \left[ \nabla \epsilon ({\bf p})
\, \times \, {\bf B} \right] 
\end{equation}
that determines the semiclassical dynamics of electrons on 
the Fermi surface in the presence of an external magnetic field. 
More precisely, we are interested here in the reconstructions of 
the topological structure of the system (\ref{MFSyst}) that occur 
when the direction of the magnetic field changes, as well as the 
relationship between the general picture of such reconstructions 
and the geometry of the Fermi surface. 

  Below we explain in more detail what we mean by the reconstruction 
of the topological structure of the system (\ref{MFSyst}), and also 
give a description of a typical picture of the reconstructions under
consideration for rather complex Fermi surfaces. Certainly, we 
will be most interested in the possibility of experimentally 
determining the picture of reconstructions of (\ref{MFSyst}) on the 
angular diagram, as well as the possibility of using it to refine 
the geometry of the Fermi surface (especially in the case of complex 
surfaces). 

  As we will see below, both possibilities can actually be associated 
with the appearance of special extremal (closed) trajectories on 
the Fermi surface near the boundaries of the reconstructions of 
(\ref{MFSyst}). Moreover, the boundaries of the reconstructions 
of the structure of system (\ref{MFSyst}) can be defined as the 
boundaries at which trajectories of this type appear and disappear, 
and the positions of such trajectories on the Fermi surface can be 
effectively used to refine its geometry. An important role in 
determining the picture of reconstructions and in restoring the 
geometry of the Fermi surface will be played by special features 
of oscillation phenomena, and especially the cyclotron resonance 
phenomenon on trajectories of this type. 

 One of the features of the trajectories that we will consider 
is the presence of (two or more) special points on them at which 
the electron velocity along the trajectory is very small. This 
property is due to the proximity of such points to the singular 
points of the dynamical system (\ref{MFSyst}), which, as we have 
already said, is due to the connection of the trajectories under 
consideration with a change in the structure of the dynamical 
system on the Fermi surface with a change in the direction of
$\, {\bf B}$. The second important feature of the trajectories 
of interest to us is that they represent ``extremal'' orbits, 
i.e. trajectories, the circulation period along which or the area 
of which have an extremal value in comparison with 
trajectories close to them. It is not difficult to see that such 
properties of special trajectories make the study of the associated 
oscillation phenomena a convenient tool for studying them. 

 Here, of course, we will not consider in detail the theory of 
oscillation phenomena in metals (see, for example, 
\cite{Kittel,etm,Abrikosov}). We only note here that the main 
thing for us in the corresponding picture of oscillations having 
a purely classical (classical cyclotron resonance) or quantum 
origin (de Haas - Van Alphen effect, Shubnikov - de Haas effect) 
will be that it is given by the sum of a finite number of oscillation 
terms (with different periods) corresponding to extremal 
trajectories on the Fermi surface. The changes in the topological 
structure of the system (\ref{MFSyst}) should correspond then to 
a sharp change in the general picture of oscillations, consisting 
in the disappearance of some oscillation terms in the sum and their 
replacement with new ones. Thus, we expect the observation of sharp 
changes of the described type on some net of one-dimensional curves
(on the angular diagram) corresponding to reconstructions of the 
topological structure of the system (\ref{MFSyst}). 

 As for the special position of the trajectories under consideration 
on the Fermi surface, it is in fact due to the proximity of such 
trajectories to the very special singular trajectories of the 
dynamical system (\ref{MFSyst}). Trajectories of the latter type 
can arise only for special directions of $\, {\bf B} \, $ (defined 
by the picture of reconstructions of (\ref{MFSyst}) at the angular 
diagram) and are associated with exact geometric parameters of 
the Fermi surface. We will examine the situation in more detail in 
just a few paragraphs below. 

 As is well known, the geometry of the trajectories of the system 
(\ref{MFSyst}) plays a crucial role in the theory of galvanomagnetic 
phenomena in metals in the presence of strong magnetic fields. 
In particular, as was first shown in \cite{lifazkag}, the behavior 
of conductivity in strong magnetic fields should differ significantly 
in cases where only closed trajectories of (\ref{MFSyst}) present 
on the Fermi surface and when open trajectories also appear on it.
Further studies of the behavior of the trajectories of the system 
(\ref{MFSyst}) on various Fermi surfaces and effects related to their 
geometry were carried out very actively, starting from the middle of 
the last century, in particular, a huge role in the formation and 
development of this area was played by studies conducted by 
I.M. Lifshits school during this period (see 
\cite{lifpes1,lifpes2,lifkag1,lifkag2,lifkag3,etm,ElProv,KaganovPeschansky}).
It should also be noted that studies related to the geometry of 
trajectories of (\ref{MFSyst}) have become the source of a number 
of approaches and methods for studying electronic spectra in metals, 
which have not lost their relevance at present. 
 
\begin{figure}[t]
\begin{center}
\vspace{0.5cm}
\includegraphics[width=0.9\linewidth]{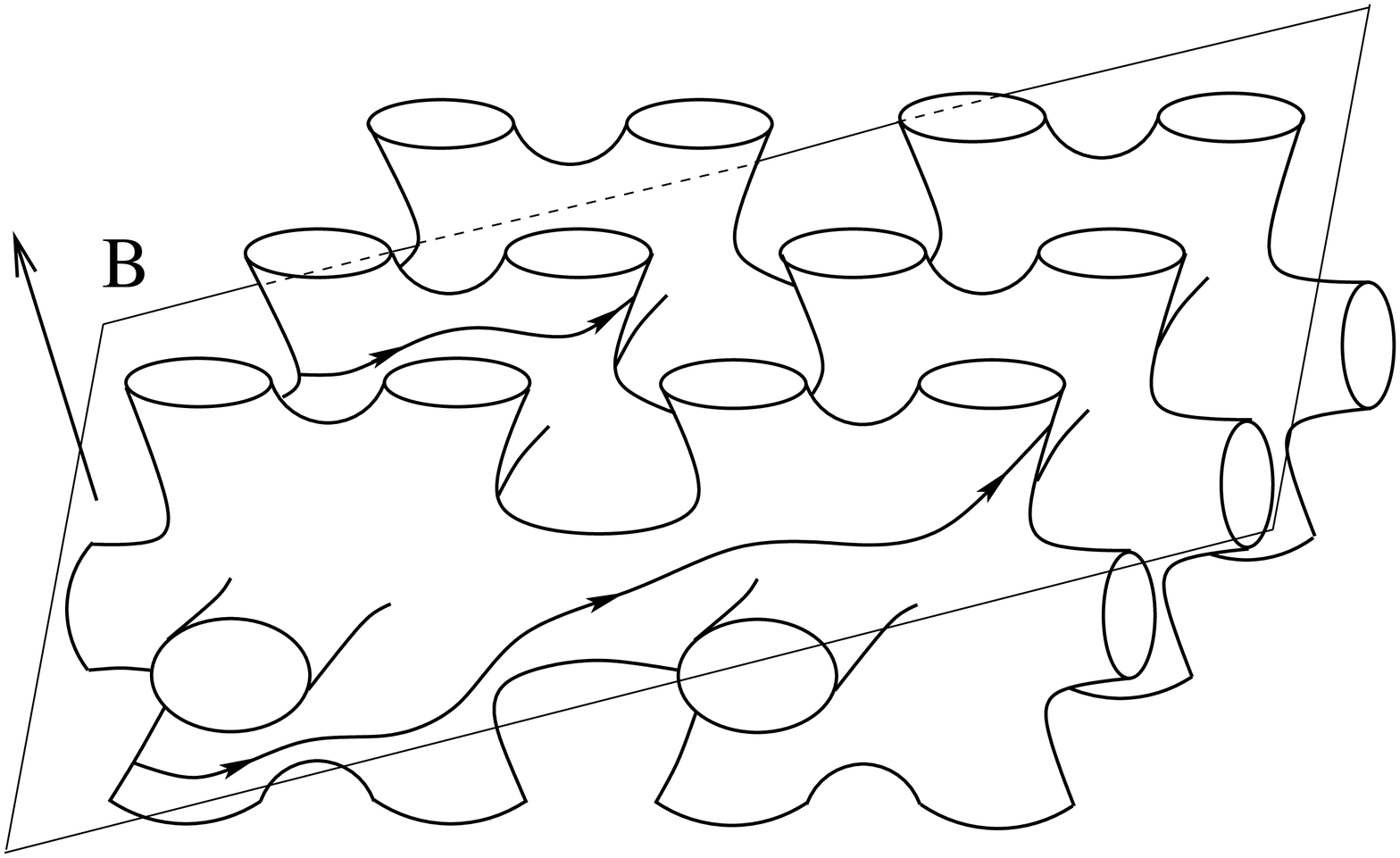}
\end{center}
\caption{The intersection of a rather complex periodic surface in 
the space of quasimomenta by a family of planes orthogonal to the 
magnetic field.}
\label{fig1}
\end{figure}

 The problem of a complete description of all types of trajectories 
of the system (\ref{MFSyst}) for an arbitrary periodic relation 
$\, \epsilon ({\bf p}) \, $ was first set by S.P. Novikov in 
\cite{MultValAnMorseTheory} and intensively studied in his topological 
school (S.P. Novikov, A.V. Zorich, S.P. Tsarev, I.A. Dynnikov 
\cite{MultValAnMorseTheory,zorich1,dynn1992,Tsarev,dynn1,zorich2,DynnBuDA,
dynn2,dynn3}). In studying the Novikov problem, in particular, new 
types of open trajectories (\cite{Tsarev, dynn2}) were found that 
were previously unknown for the systems (\ref{MFSyst}). Note that 
since the system (\ref{MFSyst}) conserves the electron energy 
$\, \epsilon ({\bf p}) \, $ and the projection of the quasimomentum 
$\, {\bf p} \, $ on the direction of the magnetic field, the Novikov 
problem can also be posed as the problem of describing the geometry of 
all possible sections of an arbitrary periodic surface by a family of 
parallel planes in $\, {\bf p} $ - space (Fig. \ref{fig1}). 
At present, a complete classification of various types of trajectories 
for arbitrary dispersion relations $\, \epsilon ({\bf p}) \, $ 
is obtained (see \cite{DynnBuDA, dynn2, dynn3}), among which the most 
important are stable open trajectories of (\ref{MFSyst}).

 For the description of the galvanomagnetic phenomena in strong 
magnetic fields, the angular diagrams for magnetoconductivity are 
usually used, which reflect the appearance of various types of 
trajectories (primarily open) on the Fermi surface for different 
directions of $\, {\bf B} \, $. Since the appearance of various 
types of trajectories in this case is directly related to changes 
in the structure of the system (\ref{MFSyst}), such diagrams are, 
in fact, closely related to angular diagrams reflecting changes 
in the topological structure of (\ref{MFSyst}), which we will 
consider below. In particular, the appearance of trajectories 
of the type of interest to us is also associated with the appearance 
of open trajectories, so that they always arise when the directions 
of the magnetic field are close to the directions of the appearance 
of open trajectories on the Fermi surface. The appearance of our 
trajectories in this case is connected with (infinite) sequences 
of reconstructions of trajectories of system (\ref{MFSyst}) when 
open trajectories appear. As a result of this, the richest structures
at the angular diagrams describing the reconstructions of the 
topological structure of the system (\ref{MFSyst}) are, in fact, 
associated with open trajectories of (\ref{MFSyst}), and such 
diagrams are in some sense ``attached'' to the angular diagrams 
for magnetoconductivity in the case of rather complex Fermi surfaces. 

 In the most general case, however, the appearance of trajectories 
of the type of interest to us may not be related to the appearance 
or disappearance of open trajectories, however, it always corresponds 
to a reconstruction of the ``topological structure'' generated by  
trajectories of (\ref{MFSyst}) on the Fermi surface. Moreover, to 
describe the reconstructions of the topological structure of 
(\ref{MFSyst}) it is enough, in fact, to monitor changes of closed 
trajectories, which explains the close connection of the trajectories 
of interest to us with such reconstructions. In a sense, one can say 
even more, namely, for a sufficiently comprehensive description of 
the topological structure of the system (\ref{MFSyst}) 
on a two-dimensional periodic surface, in fact, information on the 
structure of closed trajectories arising on it is sufficient. 
The formulated statement can be explained as follows. The indication 
of the complete set $\, {\cal M} \, $ of closed trajectories of the 
system (\ref{MFSyst}) also uniquely determines its complement on the 
Fermi surface, consisting of a finite number of (nonequivalent) 
connected components carrying open trajectories of the system 
(\ref{MFSyst}). As follows from rather deep topological studies
(see
\cite{zorich1,dynn1992,Tsarev,dynn1,zorich2,DynnBuDA,dynn2,dynn3,DeLeoPhysB}),
the main features of the geometry of open trajectories of (\ref{MFSyst}) 
are completely determined by the topology of the components that support 
them (more precisely, by the genus of the corresponding compact manifolds
obtained after factorization by the reciprocal lattice vectors). Thus, 
the indication of the complete set of nonequivalent closed trajectories 
of the system (\ref{MFSyst}) on the Fermi surface actually determines 
the complete structure of the system (\ref{MFSyst}) on it, that is, 
allows us to effectively describe all the trajectories arising on it.
Below we discuss the statements made in more detail. 

\begin{figure}[t]
\begin{center}
\vspace{0.5cm}
\includegraphics[width=0.95\linewidth]{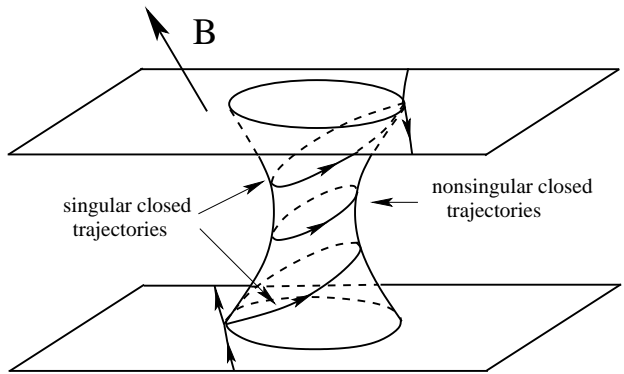}
\end{center}
\caption{The cylinder of closed trajectories of system (\ref{MFSyst}) 
on the Fermi surface, bounded by singular trajectories on its bases.}
\label{fig2}
\end{figure}

 All nonsingular closed trajectories of system (\ref{MFSyst}) on the 
Fermi surface are combined into a finite number of (nonequivalent) 
cylinders of such trajectories having singular points of system 
(\ref{MFSyst}) on their bases (Fig. \ref{fig2}). For 
structures in general position, we can assume that exactly one 
singular point presents on each of the bases of the cylinders of closed 
trajectories of (\ref{MFSyst}). A change in the structure of the 
trajectories of the system (\ref{MFSyst}) is always associated with 
a change in the set of cylinders of closed trajectories (for example, 
with rotations of the direction of $\, {\bf B} \, $ or a change in the 
Fermi level), i.e. the disappearance of part (or all) of such cylinders 
and the appearance of new ones (or the complete disappearance of closed
trajectories on the Fermi surface). It is not difficult to see that the
disappearance (or appearance) of a cylinder of closed trajectories must 
be associated with the appearance of special singular trajectories of 
the system (\ref{MFSyst}) (or ``cylinders of zero height'') connecting 
more than one singular point of (\ref{MFSyst}).

 As we have already said, we are primarily interested here in a change 
in the structure of the trajectories of the system (\ref{MFSyst}) 
under rotations of the magnetic field direction. According to the 
above, the entire space of directions of $\, {\bf B} \, $ (that is, 
the unit sphere $\, \mathbb{S}^{2} $) can be divided into regions 
(as well as limit sets of such regions), in each of which the 
topological structure of system (\ref{MFSyst}) on the Fermi surface 
remains unchanged. The boundaries of these regions are determined 
by the disappearance (or appearance) of at least one of the cylinders 
of closed trajectories, i.e. the presence of at least one 
``cylinder of zero height'' at the corresponding directions 
of $\, {\bf B} \, $. It is easy to see that near the boundaries 
of the reconstructions of the structure of (\ref{MFSyst}) we should 
expect the appearance of cylinders of closed trajectories of very 
small height, giving in the limit ``cylinders of zero height'' on the 
Fermi surface. A special feature of closed trajectories on cylinders 
of small height is their proximity to singular trajectories and, 
in particular, to the critical singular trajectory that appears 
at the moment of reconstruction of the structure of (\ref{MFSyst}).
As with any cylinder of closed trajectories, on cylinders of small 
height up to the moment of reconstruction, extremal closed trajectories 
with the minimal circulation period among all trajectories 
of the cylinder are preserved. It is precisely such trajectories that 
we will mainly consider in this paper.

 As we will see below, reconstructions of system (\ref{MFSyst}) 
can have different topology, and the corresponding special extremal 
trajectories can have different geometry. Here we will list the 
topological types of reconstruction of (\ref{MFSyst}) and present the 
corresponding geometric types of special trajectories. In this paper, 
we will examine in more detail one of the simplest types, which occurs, 
in reality, in the vast majority of cases for real Fermi surfaces. 
As an example of the use of special trajectories, we will give here one 
of the possible methods for partial restoration of the shape of the 
Fermi surface using trajectories of this type. In the general case, 
however, it can be noted that the above method is not the only possible 
one.

 In general, we would like to describe here typical pictures of 
reconstructions of the structure of system (\ref{MFSyst}) on the 
angular diagram for all directions of $\, {\bf B} $, discuss the 
topological types of such reconstructions and the geometry of the 
corresponding special extremal trajectories, and discuss special 
features of oscillation phenomena under structural reconstructions 
of (\ref{MFSyst}), as well as the possibility of using information 
about such reconstructions for the determination of the shape of 
the Fermi surface.

\vspace{1mm}

 As an example of describing the complete structure of the system 
(\ref{MFSyst}) in the presence of open trajectories and its 
relationship to closed trajectories, we give in Chapters 2 and 3 
a brief description of the topological structure of system 
(\ref{MFSyst}) on the Fermi surface in the presence of stable open 
trajectories of this system. In this situation, we will mainly be 
interested in the procedure of crossing the boundary of the region 
corresponding to the existence of stable open trajectories of the 
system (\ref{MFSyst}) on the Fermi surface, as well as the structure 
of closed trajectories near its borders. As we will see, in addition 
to the above-mentioned features (the existence of sections with a very 
low trajectory speed), the closed trajectories that appear here may 
have many additional interesting properties arising from their 
geometry. Note here that such regions, as is well known, can only 
occur for Fermi surfaces of sufficiently complex shape (open Fermi 
surfaces). In Chapter 3, we will also then consider the general 
structure of diagrams describing reconstructions of the structure 
of system (\ref{MFSyst}) and their relationship to the angular 
diagrams for magnetic conductivity, and discuss the complexity 
classes of such diagrams.

 In Chapter 4, we will look in more detail at the schemes of
``elementary'' reconstructions of the structure of system 
(\ref{MFSyst}) and discuss the probability of different types 
of reconstructions appearing on real Fermi surfaces.

 In Chapter 5, we will consider the relationship of the phenomena 
we are considering with the general geometry of the Fermi surface 
and, in particular, discuss the possibilities of using the observation 
of such phenomena to study the features of the electron spectrum.

In Chapter 6, we will discuss some features of oscillation phenomena 
associated with trajectories of the type we are interested in, where 
we will consider in particular the observation of the phenomenon of 
cyclotron resonance on such trajectories.

\section{Topological structure of the system (\ref{MFSyst}) in the 
presence of stable open trajectories on the Fermi surface}
\setcounter{equation}{0}

 For the purposes of this work, we will need, in particular, 
a description of topological structure of the system (\ref{MFSyst}) 
in the presence of stable open trajectories on the Fermi surface.
Trajectories of this type are stable with respect to small rotations 
of the direction of $\, {\bf B} \,$, and their presence means in fact 
that the corresponding direction of $\, {\bf B} \, $ belongs to some 
``Stability Zone'' in the space of directions of $\, {\bf B} \, $ 
(i.e. on the unit sphere $\, \mathbb{S}^{2} $).
 
 As we have already said, in description of the topological structure 
of dynamical system (\ref{MFSyst}), the structure of the set of closed
trajectories arising on the Fermi surface plays the most important role. 
Let us illustrate here the relationship of this structure to the 
complete structure of the system in the presence of stable open 
trajectories on the Fermi surface.
 
\begin{figure}[t]
\begin{center}
\vspace{0.5cm}
\includegraphics[width=0.9\linewidth]{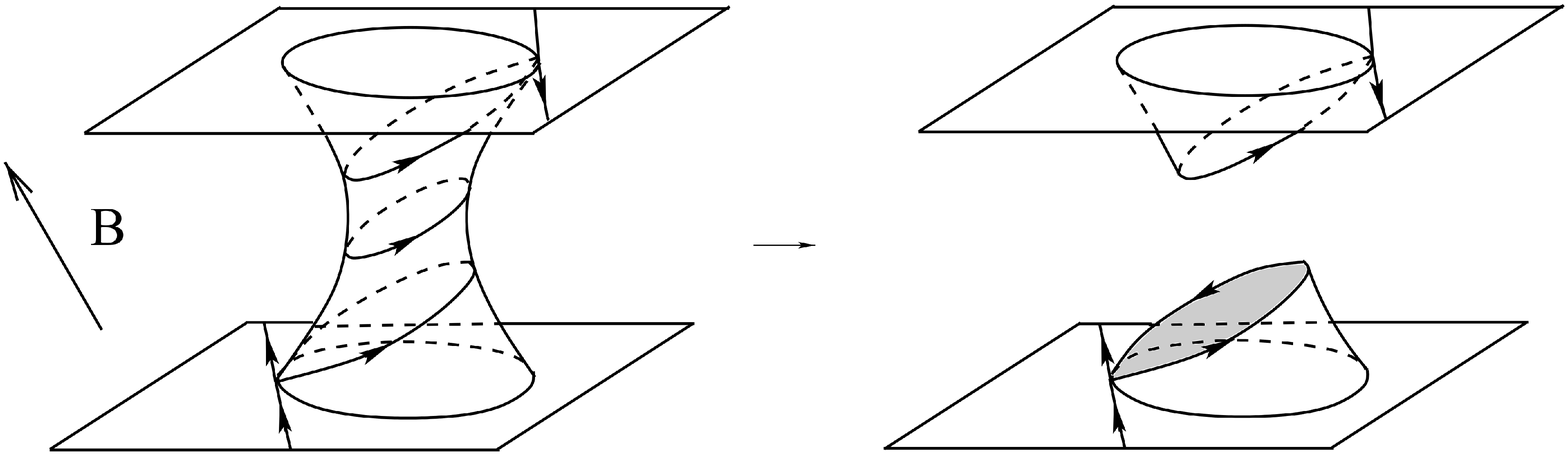}
\end{center}
\caption{Removing the cylinder of closed trajectories from the 
Fermi surface, followed by gluing the holes with disks orthogonal 
to the magnetic field.}
\label{fig3}
\end{figure}

 Using the information about the set $\, {\cal M}
 \, $ of cylinders 
of closed trajectories, we define the following procedure of
simplifying (reducing) the Fermi surface (see \cite{dynn3}) for 
a given direction of the magnetic field:

\vspace{1mm}

1) Remove all cylinders of closed trajectories from the Fermi 
surface (Fig. \ref{fig3}). (The rest of the Fermi surface 
contains only singular and open trajectories of the system 
(\ref{MFSyst}), so its connected components can be called ``carriers'' 
of open trajectories).

\vspace{1mm}

2) Glue the resulting holes with flat disks in the $\, {\bf p}$ - space, 
orthogonal to $\, {\bf B}$. The result of the reduction will again be 
a periodic two-dimensional surface in $\, {\bf p}$ - space, now carrying 
only open trajectories of system (\ref{MFSyst}).

\vspace{1mm}

\begin{figure}[t]
\begin{center}
\vspace{5mm}
\includegraphics[width=0.9\linewidth]{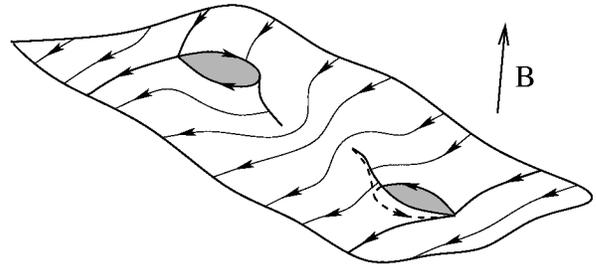}
\end{center}
\caption{The connected component of the reduced Fermi surface,
carrying stable open trajectories of the system (\ref{MFSyst}).}
\label{fig4}
\end{figure}

The most important property (see \cite{zorich1,dynn1}) of a reduced 
Fermi surface in the presence of stable open trajectories of the system
(\ref{MFSyst}) is that all its connected components in this case are
periodically deformed planes having the same integral direction 
in $\, {\bf p}$ - space (Fig. \ref{fig4}). Thus, all carriers 
of open trajectories form a set of parallel periodically deformed 
planes with a common (two-dimensional) direction generated by two 
arbitrary reciprocal lattice vectors. The number of nonequivalent 
such planes is an even number, and for Fermi surfaces of a not very 
large genus cannot actually exceed two.

\begin{figure}[t]
\begin{center}
\vspace{5mm}
\includegraphics[width=\linewidth]{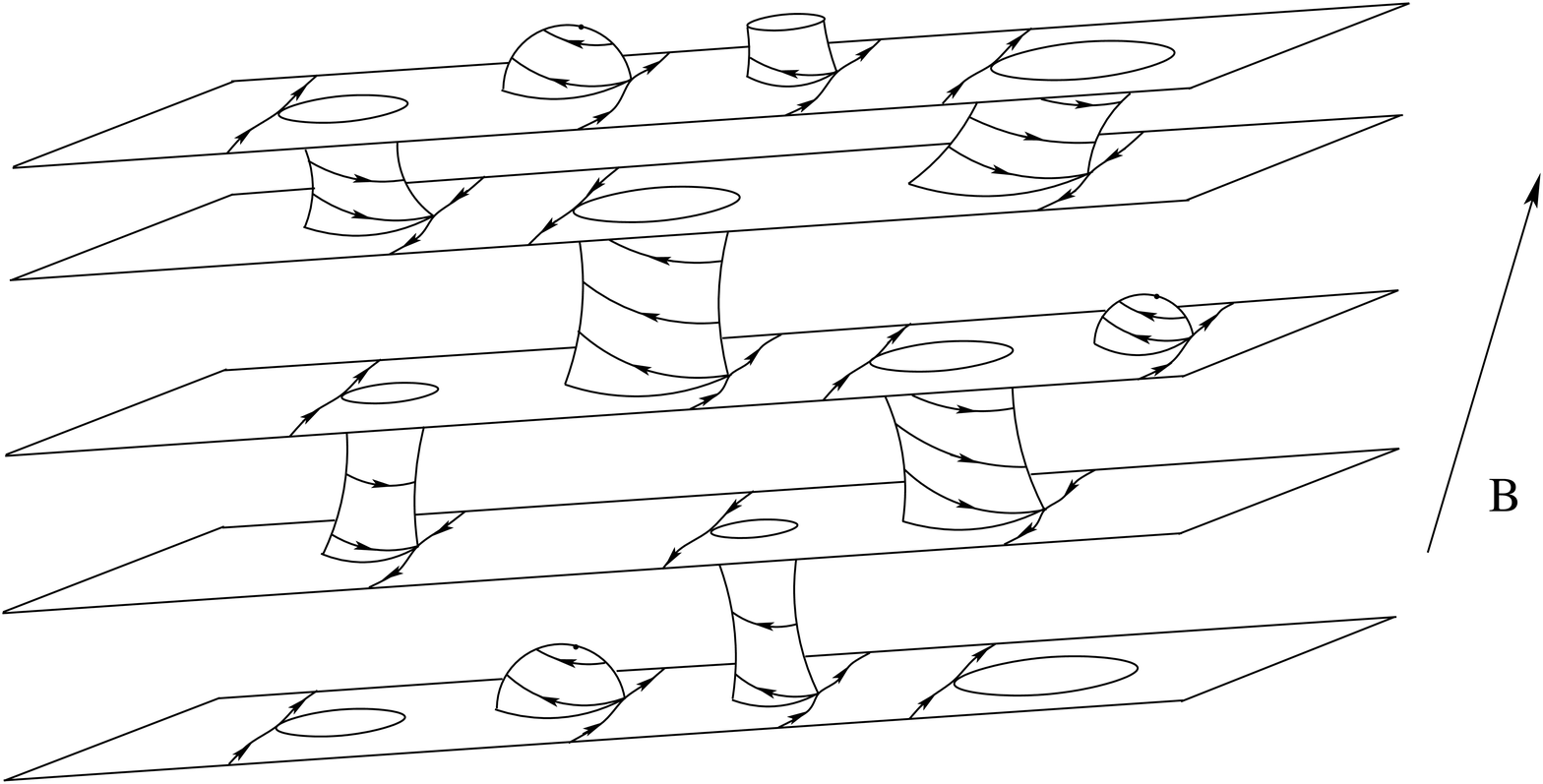}
\end{center}
\caption{A schematic representation of a complex Fermi surface 
carrying stable open trajectories of system (\ref{MFSyst}).}
\label{fig5}
\end{figure}

 In general, the structure of the original Fermi surface carrying 
stable open trajectories of system (\ref{MFSyst}) can be 
schematically represented by the picture shown at
Fig. \ref{fig5}. We can see that this structure 
remains the same for all rotations of the direction of 
$\, {\bf B} \, $ until at least one of the cylinders of closed 
trajectories connecting the carriers of open trajectories in 
$\, {\bf p}$ - space is destroyed. The invariance of the topological 
structure of the system (\ref{MFSyst}) on the Fermi surface means, 
in particular, the invariance of homological cycles defined by the 
circles that ``cut'' the Fermi surface into carriers of open 
trajectories, and the invariance of the topology of 
the carriers of open trajectories, as well as the topological 
class of embedding of carriers in the $\, {\bf p}$ - space
(more precisely, the homological class of embedding of 
two-dimensional tori $\, \mathbb{T}^{2} \,$, i.e. carriers 
factorized by the vectors of the reciprocal lattice, into the 
three-dimensional torus $\, \mathbb{T}^{3} \,$, i.e. the compactified 
Brillouin zone, defined by the integral two-dimensional direction 
of embedding of carriers in the $\, {\bf p}$ - space). 
Fig. \ref{fig5}, of course, is only a schematic 
representation of the topological structure of the system 
(\ref{MFSyst}) on the Fermi surface, and the real picture may 
look much more complex from a visual point of view. In particular, 
cylinders of closed trajectories can have a fairly small height 
and pass through a large number of Brillouin zones for structures 
with large homological classes of embedding 
$\, \mathbb{T}^{2} \, \rightarrow \, \mathbb{T}^{3} \, $.

\vspace{1mm}

\begin{figure}[t]
\begin{center}
\vspace{5mm}
\includegraphics[width=0.9\linewidth]{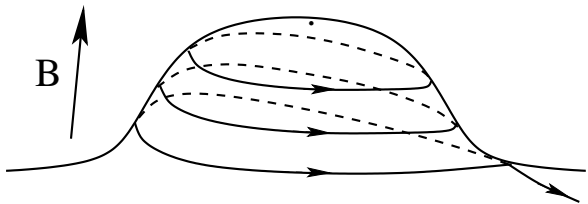}
\end{center}
\caption{A ``simple'' cylinder of closed trajectories that has 
a singular point of the system (\ref{MFSyst}) as one of its bases.}
\label{fig6}
\end{figure}

 As can also be seen from Fig. \ref{fig5}, among 
the cylinders of closed trajectories on the Fermi surface, ``simple'' 
cylinders of closed trajectories can occur, one of the bases of which 
is tightened into a single singular point (Fig. \ref{fig6}).

 The appearance and disappearance of cylinders of this type, 
generally speaking, changes the general topological structure 
of the system (\ref{MFSyst}) on the Fermi surface. Usually, 
however, the presence or absence of such cylinders does not play 
a big role in describing the complex geometric properties of 
trajectories of system (\ref{MFSyst}). Here, too, we are mainly 
interested only in reconstructions of the structure of 
(\ref{MFSyst}) with the disappearance or appearance of 
``non-trivial'' cylinders of closed trajectories 
(Fig. \ref{fig2}).

\vspace{1mm}

\begin{figure}[t]
\begin{center}
\includegraphics[width=\linewidth]{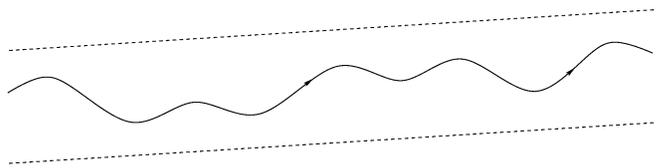}
\end{center}
\caption{The form of a stable open trajectory of system 
(\ref{MFSyst}) in a plane orthogonal to $\, {\bf B} $.}
\label{fig7}
\end{figure}

 Open trajectories of system (\ref{MFSyst}) are defined by 
intersections of the carriers of open trajectories with planes 
orthogonal to $\, {\bf B}\,$. The described structure of the 
Fermi surface allows us to specify the most important properties 
of stable open trajectories in $\, {\bf p}$ - space. Namely,

\vspace{1mm}

1) Each stable open trajectory of system (\ref{MFSyst}) in 
$\, {\bf p}$ - space lies in a straight strip of finite width 
in some plane orthogonal to $\, {\bf B}\, $ (Fig. \ref{fig7}), 
passing through it (see \cite{zorich1, dynn1992}).

\vspace{1mm}

2) All stable open trajectories for a given direction of 
$\, {\bf B} \, $ have the same mean direction, given by the 
intersection of the plane orthogonal to $\, {\bf B}\,$, and some 
(constant for a given Stability Zone $\, \Omega$) integral 
(i.e. generated by two reciprocal lattice vectors) plane 
$\, \Gamma\,$.

\vspace{1mm}

 The indicated properties of stable open trajectories play 
a crucial role in the behavior of the conductivity of normal 
metals in sufficiently strong magnetic fields. In particular, 
the directions of the integral planes $\, \Gamma_{\alpha} \, $, 
defined for different Stability Zones $\, \Omega_{\alpha} \,$, 
play the role of integer topological invariants observed in 
conductivity in strong magnetic fields (\cite{PismaZhETF,UFN}).
The corresponding directions can actually be given by irreducible 
triples of integers 
$\, (M^{1}_{\alpha}, M^{2}_{\alpha}, M^{3}_{\alpha})\,$, 
which can be called topological numbers observable in the 
conductivity of normal metals.

\vspace{1mm}

 The topological representation of the Fermi surface in the 
form shown at Fig. \ref{fig5} is not the only one, 
in particular, such representations are different for different 
Stability Zones $\, \Omega_ {\alpha} \,$. It is easy to see that 
the specific structure of representation of the Fermi surface in 
this form (as well as the numbers 
$\, (M^{1}_{\alpha}, M^{2}_{\alpha}, M^{3}_{\alpha})$) 
is determined by specifying cylinders of closed trajectories 
on the Fermi surface.

 Thus, we can see that determination of the set $\, {\cal M}\, $ 
of cylinders of closed trajectories completely determines the 
topology of the system (\ref{MFSyst}) on the Fermi surface in the 
described situation. Strictly speaking, the complete description 
of trajectories of (\ref{MFSyst}) also includes the rotation number 
on two-dimensional tori $\, \mathbb{T}^{2} \, $ (or the mean 
direction of open trajectories in $\, {\bf p}$ - space), which 
are not defined just by the topology of division of the Fermi surface  
into the described components, but are easily calculated in the 
described situation for any given direction of $\, {\bf B} \, $
(${\bf B}/B \, \in \, \Omega_{\alpha}$). As we will show in the 
next chapter, the appearance of stable open trajectories on the 
Fermi surface actually entails the existence of a large set of 
directions of $ \, {\bf B} \, $, for which special extremal closed 
trajectories of interest to us arise.

\vspace{1mm}

 We note at once that stable open trajectories are not the only 
type of open trajectories that can arise on rather complex Fermi 
surfaces (see \cite{Tsarev, DynnBuDA, dynn2}). It can be noted, 
however, that in this case too, a change in the structure of the 
trajectories of system (\ref{MFSyst}) is always associated 
with the appearance or disappearance of cylinders of closed 
trajectories and, in a sense, the presence of the set $\, {\cal M} \, $ 
(or its absence), as well as its structure, determines the complete 
topological structure of the system (\ref{MFSyst}) on the Fermi 
surface. Different questions related to the classification of open 
trajectories of the system (\ref{MFSyst}), including their physical
applications, as well as studies of the structure of unstable open 
trajectories of various types, can be found in 
\cite{BullBrazMathSoc,JournStatPhys,Zorich1996,ZorichAMS1997,zorich3,
DeLeo1,DeLeo2,ZorichLesHouches,DeLeoDynnikov1,Skripchenko1,Skripchenko2,
DynnSkrip1,DynnSkrip2,AvilaHubSkrip1,AvilaHubSkrip2,DeLeo2017,UMNObzor,
GutierrezRomoMatheus,ObzorJetp}. It is also natural to note that the 
set $ \, {\cal M} \, $ determines the structure of the system 
(\ref{MFSyst}) on the Fermi surface also if only closed trajectories 
appear on it, since in this case it contains information about all 
trajectories of the system (\ref{MFSyst}).
 In any case, any change 
in the topological structure of the system (\ref{MFSyst}) on the 
Fermi surface is always associated with the restructuring of the 
set $\, {\cal M} \, $ and, in particular, with the disappearance 
or appearance of individual cylinders of closed trajectories.

\section{Reconstructions of the structure of system (\ref{MFSyst}) 
and the appearance of special closed trajectories on the Fermi surface}
\setcounter{equation}{0}

\begin{figure}[t]
\begin{center}
\includegraphics[width=0.9\linewidth]{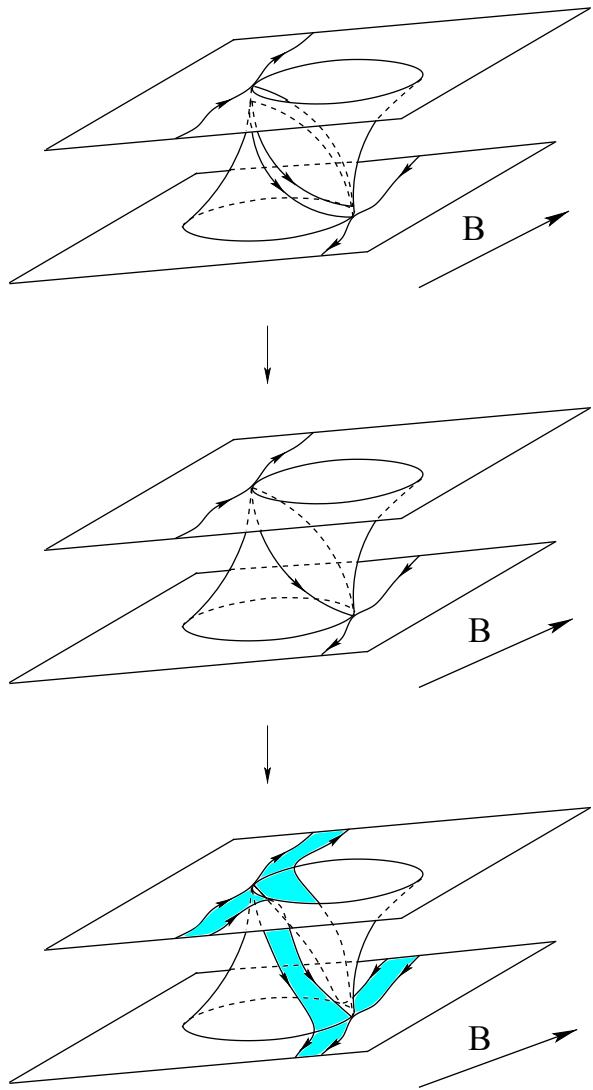}
\end{center}
\caption{The disappearance of a cylinder of closed trajectories 
on the Fermi surface during the rotation of the direction of 
$\, {\bf B} $.}
\label{fig8}
\end{figure}

 As we said above, a change in the structure of the system (\ref{MFSyst}) 
is always associated with the disappearance or appearance of cylinders 
of closed trajectories on the Fermi surface. A scheme of the disappearance 
(or appearance) of one of these cylinders is shown at Fig. \ref{fig8}.
It can be seen that with a correctly selected rotation of the direction 
of $\, {\bf B} \, $, the height of the cylinder of closed trajectories
decreases, then vanishes and then a narrow strip appears in which jumps 
of the trajectories between the former bases of the cylinder become 
possible.

\begin{figure}[t]
\begin{center}
\includegraphics[width=0.9\linewidth]{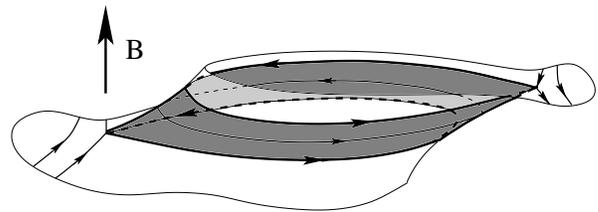}
\end{center}
\caption{A cylinder of closed trajectories of small height that 
accompanies a change in the topological structure of the system 
(\ref{MFSyst}) on the Fermi surface.}
\label{fig9}
\end{figure}

 Thus, it can be seen that a change in the topological structure of 
the system (\ref{MFSyst}) is always accompanied by the appearance 
of a cylinder of closed trajectories of very small height, bounded 
by singular trajectories on their bases (Fig. \ref{fig9}).  
The period of rotation along closed trajectories tends to infinity 
when approaching the bases of the cylinder (singular trajectories) 
and, thus, must have a minimum on some trajectory inside the cylinder.
The corresponding trajectory is ``extremal'', in particular, when 
considering the phenomenon of cyclotron resonance, since the main 
oscillating terms in the intensity of absorption of microwave 
radiation come precisely from such trajectories.  In this paper, 
we will be interested in precisely such trajectories. For convenience, 
we will call them special extremal trajectories. Most often 
(for example, for reconstructions near the boundaries of the regions 
where stable open trajectories appear on Fermi surfaces of not too 
large genus), such cylinders are centrally symmetric and then, if the 
extremal trajectory is unique (only this simplest case will be 
considered), it also has a central symmetry. In special cases, 
however, cylinders of closed trajectories can also appear in pairs 
and pass into each other under transformations of central symmetry. 
In this case, both cylinders of closed trajectories and extremal 
trajectories on them are not required to possess central symmetry. 
As we will see below, the presence of central symmetry of extremal 
trajectories provides a particularly convenient tool for studying 
the Fermi surface in the situation under consideration.

 The period $ \, T \, $ of rotation along an extremal trajectory 
also tends to infinity (according to the logarithmic law) as the 
height $\, h \, $ of the cylinder of closed trajectories tends to 
zero. The relatively slow growth of $ \, T \, $ in this case, 
however, allows us to maintain the condition 
$\, T / \tau \, \ll \, 1 \, $ and, thus, to observe oscillation 
phenomena on such trajectories up to very small deviations of 
$\, {\bf B} \, $ from the boundary of the reconstruction of the 
topological structure of the system (\ref{MFSyst}). It can be seen, 
therefore, that the study of oscillation phenomena can serve as a 
very good tool for determining the boundaries of the reconstructions 
of the topological structure of system (\ref{MFSyst}) on the Fermi 
surface. An exact study of the boundaries of the reconstructions 
of the system (\ref{MFSyst}) (at the angular diagram), in turn, 
can serve as a good tool for refining the shape of the Fermi 
surface and other dispersion relation parameters 
$\, \epsilon ({\bf p}) \, $ for metals with complex Fermi surfaces.

\begin{figure}[t]
\begin{center}
\includegraphics[width=0.9\linewidth]{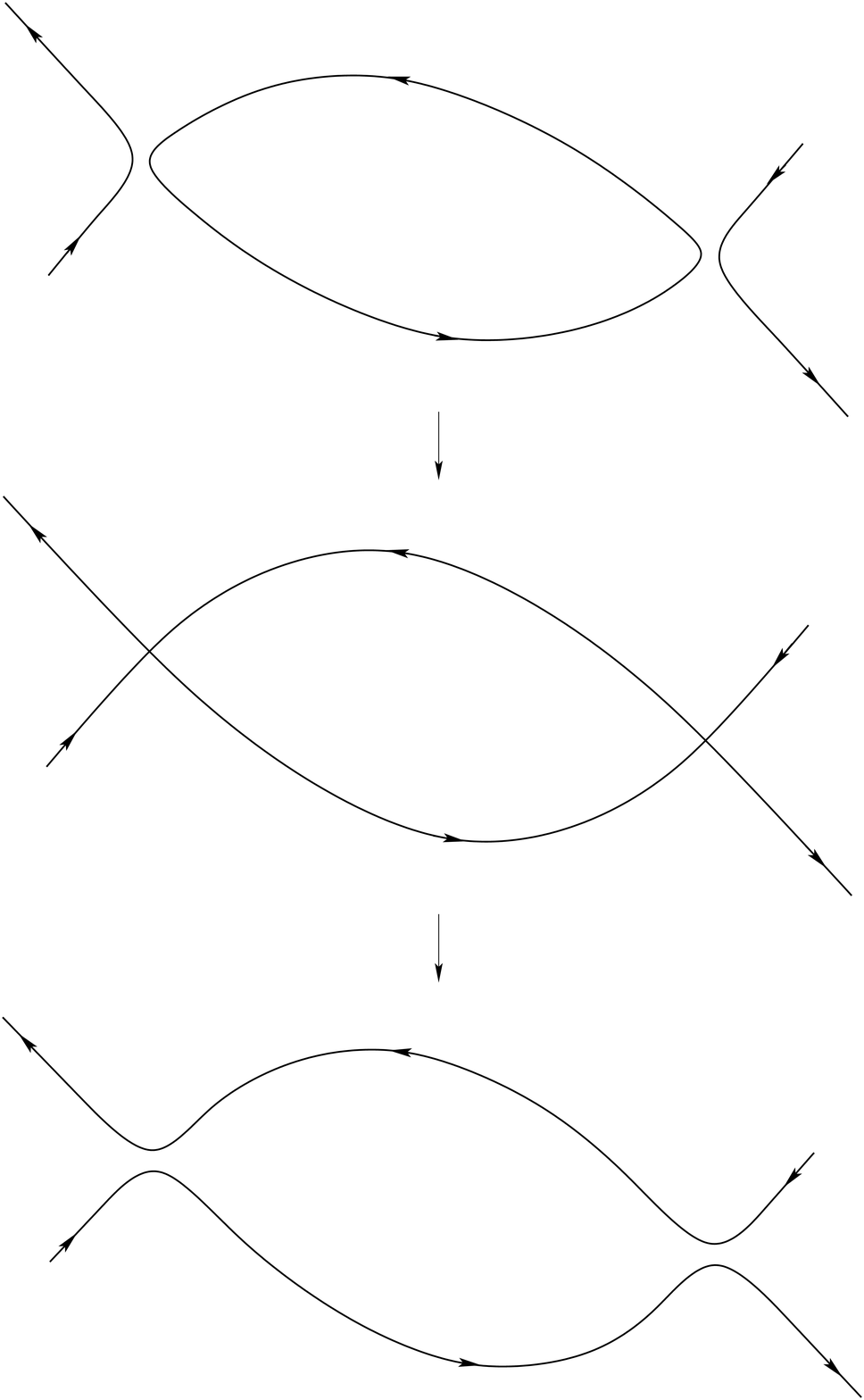}
\end{center}
\caption{Reconstruction of the extremal trajectory with a change 
in the topological structure of the system (\ref{MFSyst}) on the 
Fermi surface.}
\label{fig10}
\end{figure}

 The reason for the increase in the period $\, T \, $ when the 
height of $\, h \, $ tends to zero is the presence of points on 
an extremal trajectory very close to saddle singular points in 
$\, {\bf p}$ - space, associated with the reconstruction of this 
trajectory when changing the structure of the system (\ref{MFSyst}) 
(Fig. \ref{fig10}). It is near these points that the electron 
is delayed for a long time, running through the remaining sections 
of the trajectory rather quickly. When observing the phenomenon of 
cyclotron resonance, in particular, we will be interested in a 
situation where these points correspond to parts of the trajectory 
lying near the surface (inside the skin layer) of the sample under 
study, which, as we will see, leads to certain special features in the 
behavior of the oscillations of absorption of incident radiation.

 As we can see, the extremal trajectory exists only until the 
corresponding cylinder of closed trajectories disappears and 
breaks up into other trajectories after that. The 
resulting trajectories after reconstruction (Fig. \ref{fig10}) 
could apriori be both open and closed. As we will see later, however, 
whenever a change in the topological structure of the system 
(\ref{MFSyst}) is associated with the appearance of open 
trajectories, it is more complex and requires, as a rule, an 
infinite number of reconstructions presented above. 

\vspace{1mm}

 As we noted above, this circumstance makes it natural to draw 
a connection between the angular diagrams describing the 
reconstructions of the topological structure of (\ref{MFSyst}) 
with the angular diagrams describing the behavior of 
magnetoconductivity in metals in strong magnetic fields. 
Traditionally, angular diagrams for magnetoconductivity in 
conductors describe different conductivity behavior for different 
directions of $\, {\bf B} \, $, due to the appearance of various 
types of trajectories of the system (\ref{MFSyst}) on the Fermi 
surface. In a somewhat simplified formulation, we can say that 
the main goal of the angular diagram of magnetoconductivity is 
to separate the directions of $\, {\bf B} \, $, corresponding to 
the ``trivial'' behavior of the conductivity (the presence of 
only closed trajectories on the Fermi surface), and the directions 
corresponding to ``non-trivial'' behavior (the appearance of open 
trajectories). The main thing in the structure of the angular 
diagram for conductivity in strong magnetic fields is, therefore, 
the indication of areas (on the unit sphere) of the directions of 
$\, {\bf B} \, $, corresponding to the presence of only closed 
trajectories on the Fermi surface, as well as a description of 
the set of directions of $\, {\bf B} \, $, corresponding to the 
appearance of open trajectories of various types. Thus, the 
angular diagram for magnetoconductivity in the general case 
contains regions (Stability Zones) corresponding to the 
appearance of stable open trajectories of the system 
(\ref{MFSyst}) on the Fermi surface, as well as additional 
sets (one-dimensional curves or points) corresponding to the 
appearance of various types of unstable trajectories.

\begin{figure}[t]
\begin{center}
\includegraphics[width=\linewidth]{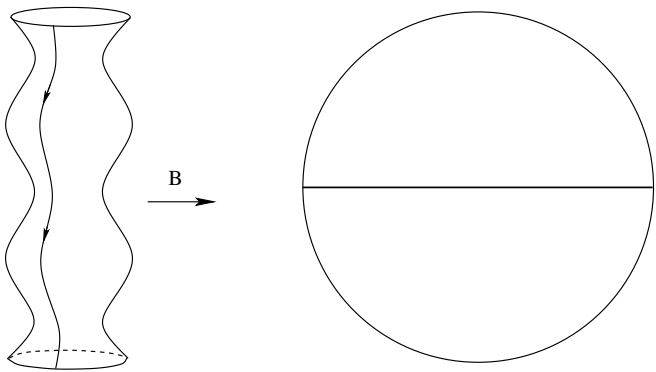}
\end{center}
\caption{Periodic trajectory and the conductivity diagram 
for a Fermi surface of the ``corrugated cylinder'' type.}
\label{fig11}
\end{figure}

\begin{figure}[t]
\begin{center}
\vspace{5mm}
\includegraphics[width=\linewidth]{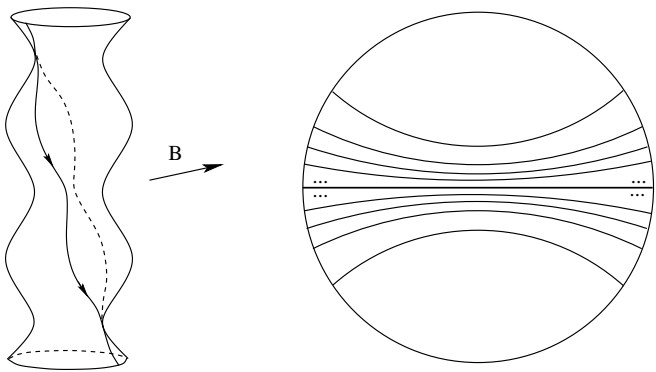}
\end{center}
\caption{An example of reconstruction of the topological 
structure of the system (\ref{MFSyst}) and the angular 
diagram describing the reconstructions of the structure of 
(\ref{MFSyst}) for a ``corrugated cylinder'' type Fermi surface.}
\label{fig12}
\end{figure}

  The angular diagram describing the reconstructions of the 
topological structure of the system (\ref{MFSyst}) on the 
Fermi surface, according to our definition, 
represents the structure (a net of one-dimensional directions)
of the set of directions of $\, {\bf B} \, $ corresponding to 
reconstructions of the system of cylinders of closed trajectories 
on the Fermi surface. It can be seen, however, that such a 
diagram should include the structure of diagram for 
magnetoconductivity as part of its overall structure. The 
reason for this, as we already noted above, is that when 
approaching the directions of $\, {\bf B} \, $ corresponding 
to the appearance of open trajectories, multiple reconstructions 
of closed trajectories must occur, giving in the limit the 
appearance of open trajectories of the system (\ref{MFSyst}). 
As an example, we can consider the simplest diagrams that arise 
for a Fermi surface of the type of ``corrugated cylinder'' 
(Fig. \ref{fig11}). It is easy to see that the angular 
diagram of conductivity represents here the unit sphere with 
only one distinguished large circle (the ``equator'') 
corresponding to the appearance of (unstable) periodic 
trajectories of the system (\ref{MFSyst}). The angular diagram 
describing the reconstructions of the topological structure of 
the system (\ref{MFSyst}) is more complicated and contains an 
infinite number of circles on $\, \mathbb{S}^{2} \, $, 
concentrating near the equator (Fig. \ref{fig12}).

 In general, the complexity of the angular diagrams describing 
the reconstructions of the topological structure of the system 
(\ref{MFSyst}) can be very different, and, as follows from the 
above arguments, their complexity is directly related to the 
complexity of the corresponding diagrams for magnetoconductivity. 

 It is easy to give examples where both the second and first 
diagrams are simply trivial (do not contain any elements), as, 
for example, for a Fermi surface close to a sphere. In general, 
angular diagrams for magnetoconductivity can be divided into 
several complexity classes (see, for example, \cite{UltraCompl}).
The simplest ones include naturally trivial diagrams corresponding 
to Fermi surfaces that do not allow the appearance of open 
trajectories at any direction of $\, {\bf B} $. The angular 
diagrams describing the reconstructions of the topological 
structure of the system (\ref{MFSyst}) are not required to be 
trivial here, thus, it is easy to construct examples of the 
Fermi surfaces on which the reconstructions of the structure 
of (\ref{MFSyst}) occur, but no open trajectories appear. 
The diagrams describing the reconstructions of the topological 
structure of the system (\ref{MFSyst}) contain in this case a 
finite net of one-dimensional directions of $\, {\bf B} \, $ 
(on $\, \mathbb{S}^{2}$) corresponding to the appearance of 
the ``cylinders of zero height'' described above.

 The second class of angle diagrams for magnetoconductivity 
is represented by angular diagrams containing a net of 
one-dimensional curves on $\, \mathbb{S}^{2} $, corresponding 
to the appearance of (unstable) periodic open trajectories on the 
Fermi surface. As can be seen from the example presented above, 
the angular diagrams describing the reconstructions of the 
topological structure of the system (\ref{MFSyst}) are in 
this case more complex and contain an infinite number of 
one-dimensional curves corresponding to the appearance 
of ``cylinders of zero height'' in the generic case.

 The most complex diagrams for magnetoconductivity include 
diagrams containing nontrivial regions of existence of open 
trajectories of the system (\ref{MFSyst}) (stable open 
trajectories). In fact (see \cite{UltraCompl}), such diagrams 
can also be divided into two classes differing in level of 
complexity. As we will see below, the angular diagrams 
describing the reconstructions of the topological structure 
of the system (\ref{MFSyst}) are in this case the most complex 
and have a very rich structure.

\vspace{1mm}

 In this paper, we would like, in particular, to provide a 
schematic description of the set of all directions of 
$\, {\bf B} \, $, corresponding to the reconstructions of 
the system (\ref{MFSyst}), on the angular diagrams of the most 
general, and, in particular, quite complex, type (containing at 
least one Stability Zone). We will need, among other things, to 
give a schematic description of the structure of this set near 
the boundary of one fixed Stability Zone $\, \Omega_{\alpha} \, $. 
 
\begin{figure}[t]
\begin{tabular}{lc}
\includegraphics[width=0.6\linewidth]{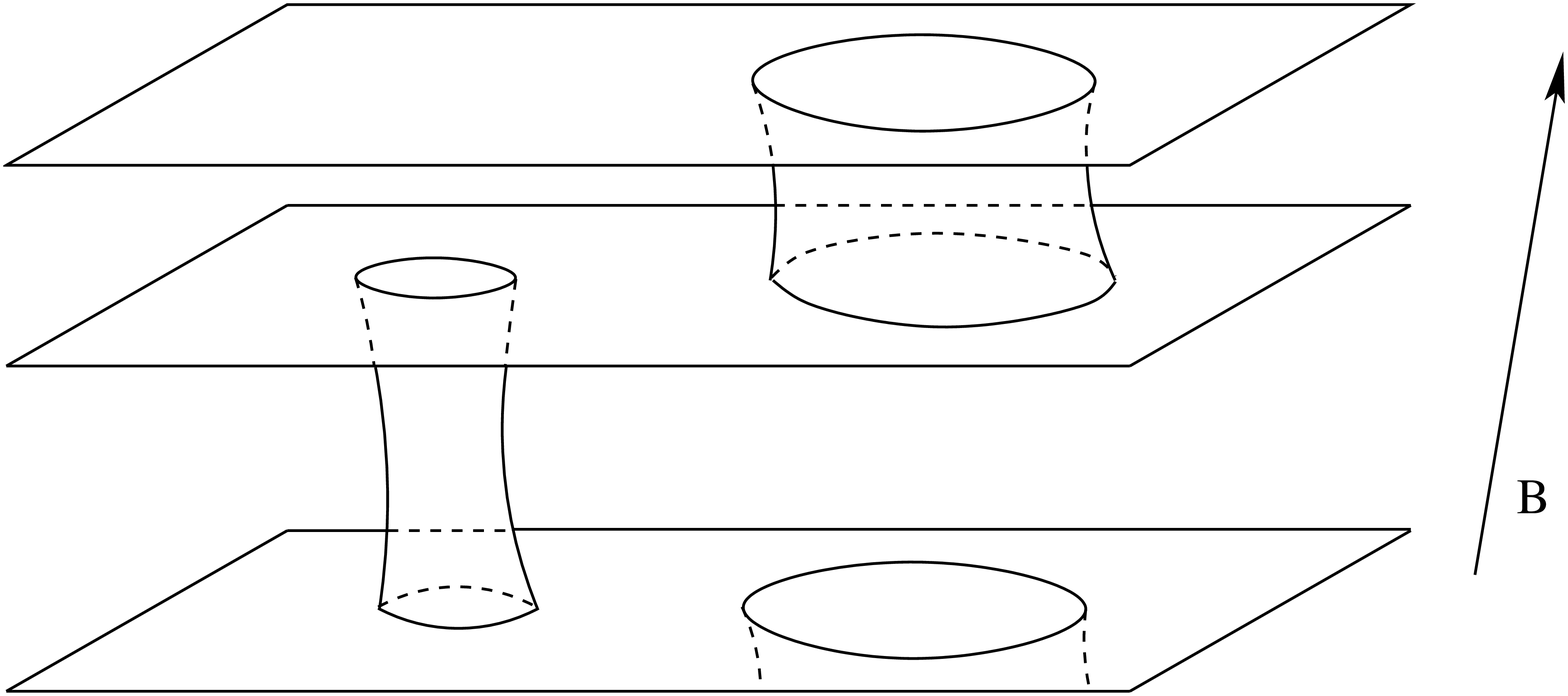}  &
\hspace{5mm}
\includegraphics[width=0.3\linewidth]{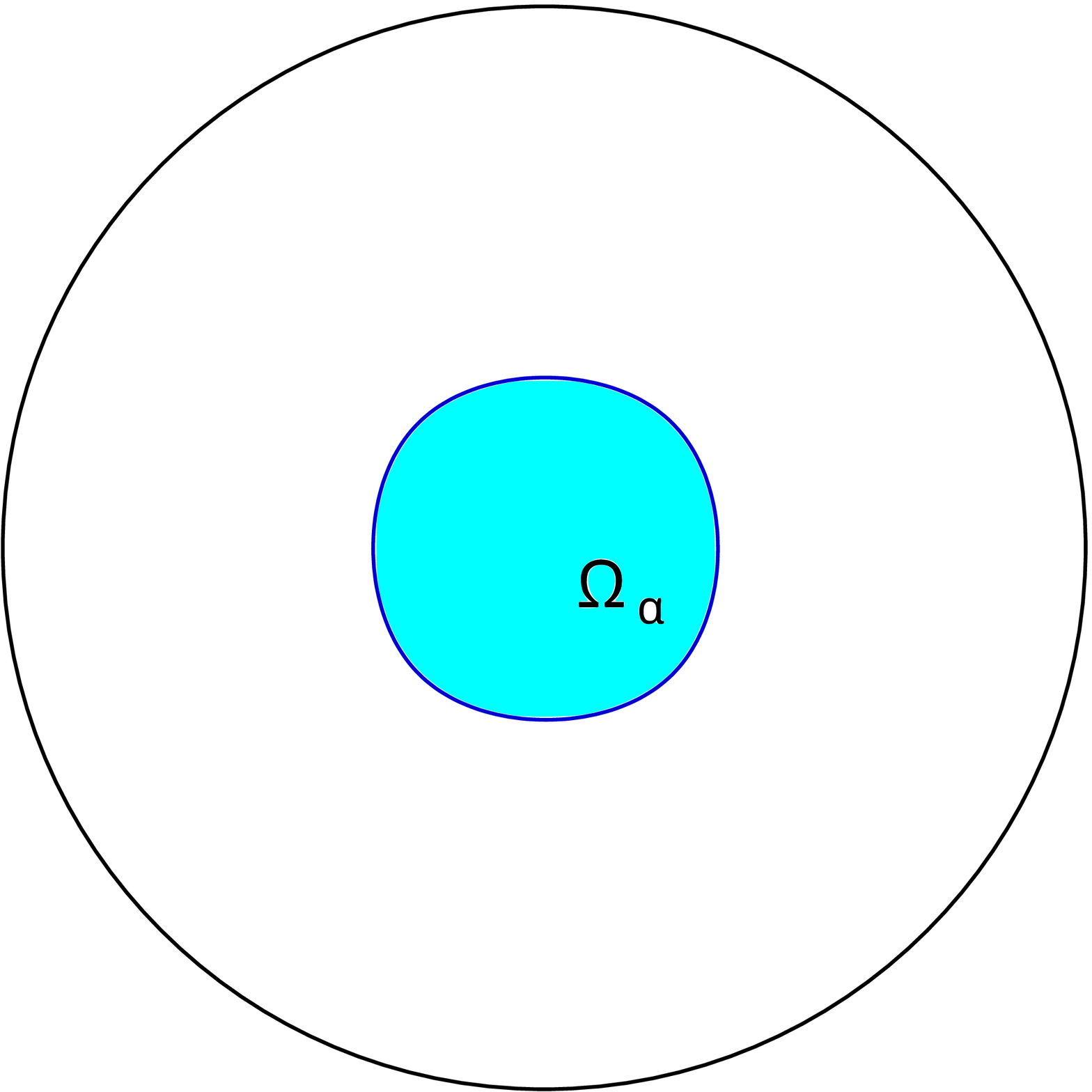}
\end{tabular}
\caption{A simplified representation of the structure of a Fermi 
surface, having a region of existence of stable open trajectories 
(a Stability Zone) in the full space of directions of $\, {\bf B} \, $ 
(on the sphere $\, \mathbb{S}^{2}$).}
\label{fig13}
\end{figure}

 For this purpose, let us consider a ``not too complicated'' case 
when the Fermi surface has genus 3 and extends in all three 
directions in the $ \, {\bf p} $ - space. To simplify as much as 
possible the visual representation of the structure of such a surface 
for the direction of $\, {\bf B} \, $ lying in one of the Stability 
Zones, one can imagine as an example the surface shown at 
Fig. \ref{fig13}. When the directions of $\, {\bf B} \, $ 
are close to the vertical, it is easy to visually distinguish 
a pair of non-equivalent carriers of open trajectories 
(integral planes), as well as two different cylinders of closed 
trajectories that separate these carriers (``thick'' and ``thin'' 
cylinders). Cylinders of closed trajectories have different types 
(``electron'' and ``hole'') and both have central symmetry in the 
described situation (the Stability Zone for a surface of genus 3, 
extending in three directions). The direction of $\, {\bf B} \, $ 
lies within the Stability Zone as long as both cylinders at 
Fig. \ref{fig13} contain non-singular closed trajectories 
of the system (\ref{MFSyst}). The boundary of the Stability Zone 
$\, \Omega_{\alpha} \, $ is determined by the disappearance of 
nonsingular closed trajectories on one of the cylinders 
(in this case, on the ``thick'' cylinder) shown at Fig. \ref{fig8} 
(that is, the height of the corresponding cylinder of closed 
trajectories vanishes).

 What can be said about the trajectories of the system 
(\ref{MFSyst}) for directions of $\, {\bf B} \, $ lying outside the 
Zone $\, \Omega_{\alpha} \, $ in the immediate vicinity of its 
boundary? One can see, in reality, that the described structure 
of the Fermi surface allows one to give a description of the 
trajectories of (\ref{MFSyst}) for such directions as well.
The reason for this is that, after crossing the boundary of 
$\, \Omega_{\alpha} \, $, non-singular closed trajectories remain 
on the second (``thin'') cylinder separating the carriers of open 
trajectories. As a result, the Fermi surface can now be represented 
as a union of pairwise ``merged'' former carriers of open trajectories, 
separated by cylinders of closed trajectories remaining on the 
``thin'' cylinders. In such a situation, the system still retains 
the memory of the integral plane $\, \Gamma_{\alpha} \, $, associated 
with the Zone $\, \Omega_{\alpha} \, $, which continues to play 
a significant role in the description of the trajectories of system 
(\ref{MFSyst}) on the Fermi surface.

\begin{figure}[t]
\begin{center}
\includegraphics[width=\linewidth]{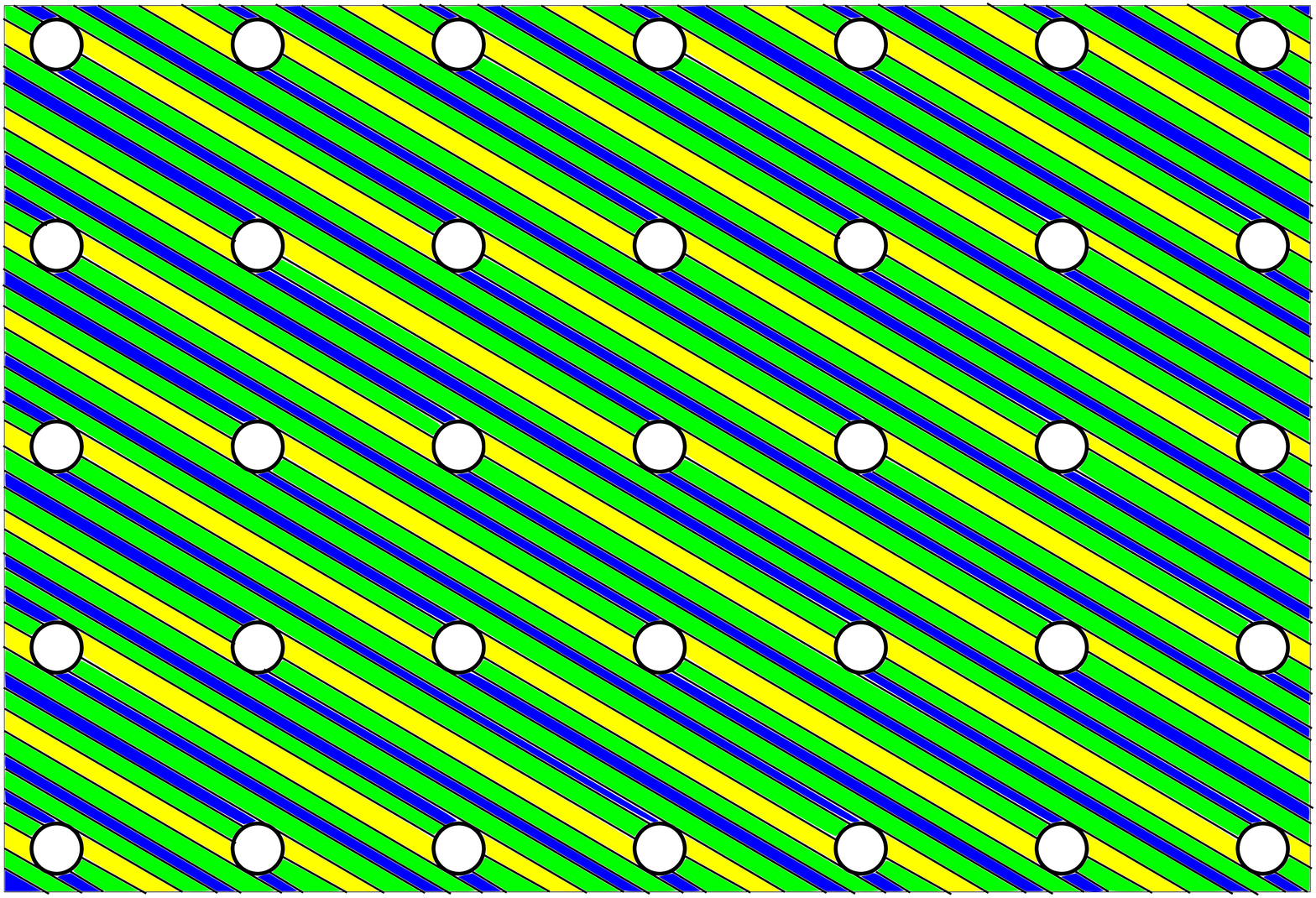}
\end{center}
\begin{center}
\includegraphics[width=\linewidth]{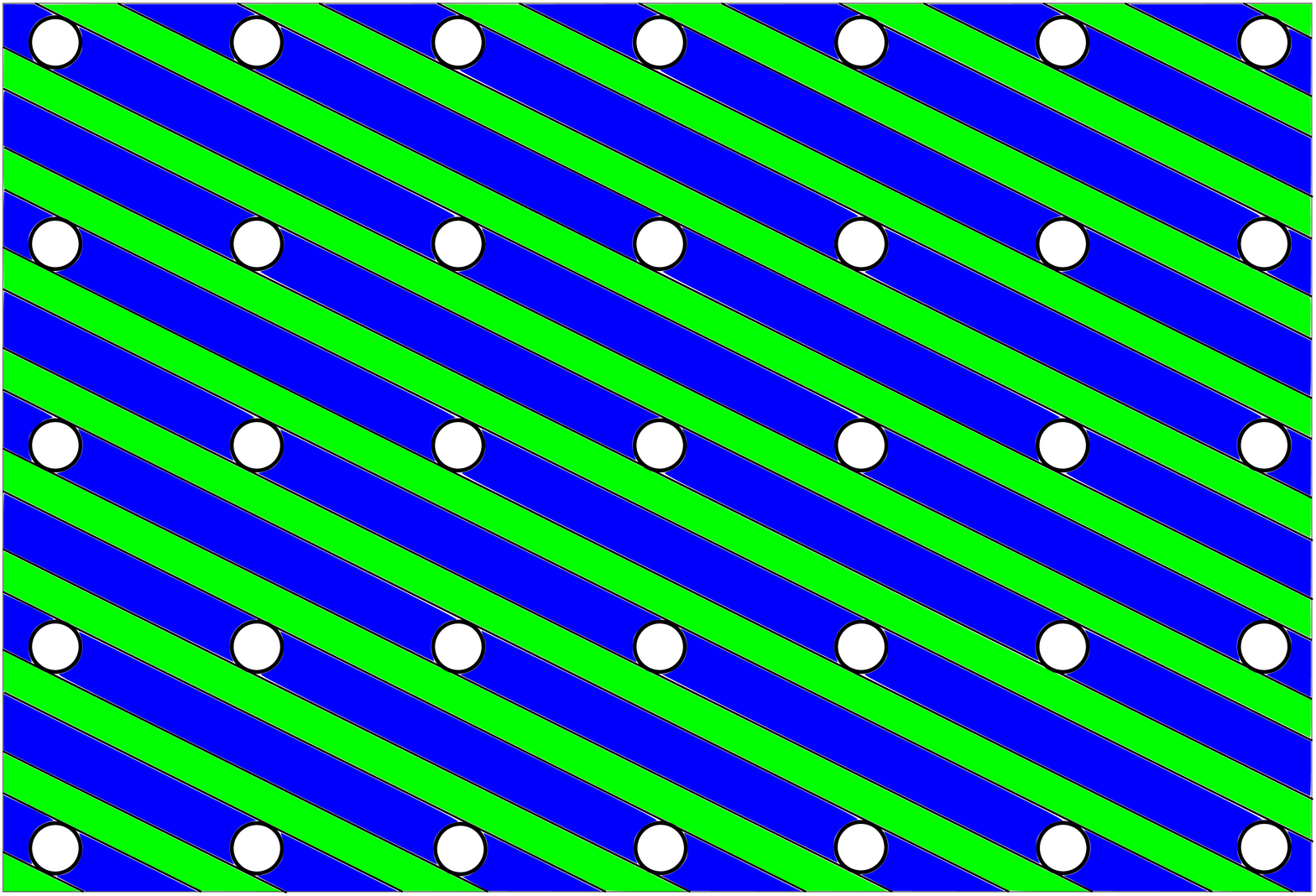}
\end{center}
\caption{A pair of former carriers of open trajectories 
(periodically deformed integral planes) carrying either 
closed trajectories or closed and periodic trajectories 
after crossing the boundary of $\, \Omega_{\alpha} \, $.
The circles indicate the places where the trajectory jumps 
from the ``upper'' carrier to the ``lower'' and vice versa.}
\label{fig14}
\end{figure}

 It is natural to divide the nonsingular trajectories arising 
for the directions of $\, {\bf B} \, $ near the boundary 
of the Zone $\, \Omega_{\alpha} \, $ into the trajectories 
lying on the ``thin'' cylinder and the trajectories, lying on 
a pair of ``merged'' carriers of open trajectories. It can be 
seen that the latter are either closed (if the plane orthogonal 
to $\, {\bf B} \, $ intersects $\, \Gamma_{\alpha} \, $ in an 
irrational direction), or can be periodic (if the plane orthogonal 
to $\, {\bf B} \, $ intersects $\, \Gamma_{\alpha} \, $ in an 
integer direction in $\, {\bf p}$ - space). Both situations are 
shown schematically at Fig. \ref{fig14}, which shows 
the division of a pair of former carriers of open trajectories 
into cylinders of the formed closed trajectories (upper picture), 
or into cylinders of closed trajectories and layers of periodic 
trajectories in $\, {\bf p} $ - space (bottom picture). As 
one can see, in the first case, exactly three nonequivalent 
cylinders of closed trajectories appear on a pair of former 
carriers, and in the second case - exactly one cylinder of 
closed trajectories (and two nonequivalent layers of periodic 
trajectories - one on the ``upper'' and one on the ``lower'' 
carrier). It can be seen, therefore, that in the situation 
under consideration, each of the cylinders of closed trajectories 
can only go into itself under the transformation 
$\, {\bf p} \, \rightarrow \, - {\bf p} \, \, $ and, therefore, 
possess central symmetry. As we have already said, extremal 
closed trajectories with such symmetry will be of particular 
interest to us in studying the Fermi surface.

\begin{figure}[t]
\begin{center}
\includegraphics[width=\linewidth]{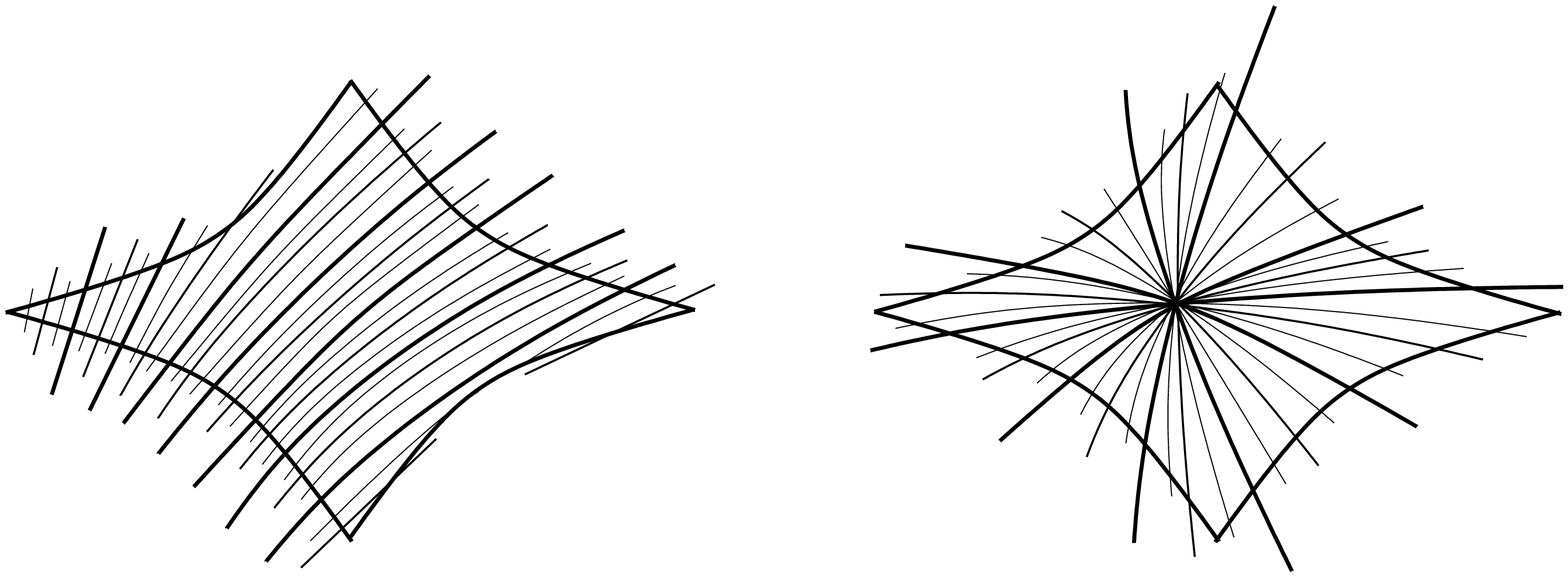}
\end{center}
\caption{Stability Zones and special segments corresponding to 
the appearance of periodic trajectories on the Fermi surface.}
\label{fig15}
\end{figure}

 It should be said right away that the appearance of periodic 
trajectories near the boundary of $\, \Omega_{\alpha} \, $ 
occurs only for non-generic directions of $\, {\bf B} \, $ 
(the intersection of the plane orthogonal to $\, {\bf B} \, $ 
and $\, \Gamma_{\alpha} \, $ represents a reciprocal lattice 
vector). It is easy to show that the corresponding directions 
of $\, {\bf B} \, $ form an infinite set of segments adjacent 
to the boundary of $\, \Omega_{\alpha} \, $, the length of 
which decreases rapidly with increasing module of the corresponding 
integer vector in $\, {\bf p} $ - space. For integer vectors of 
a small module, the corresponding segments can reach another 
Stability Zone and have no endpoints. Such segments are, 
in fact, extensions of segments passing through the entire 
Stability Zone and corresponding to the appearance of periodic 
trajectories on the Fermi surface (Fig. \ref{fig15}).  
(We note here that sometimes in the literature one can find 
the statement that open trajectories on the Fermi surface 
become periodic if the plane orthogonal to $\, {\bf B} \, $
contains an integer (rational) direction. Such a statement, 
generally speaking, is false. To formulate a more correct 
statement, we must first note that if we consider a change 
in stable open trajectories, then the direction of 
$\, {\bf B} \, $ belongs to some Stability Zone 
$\, \Omega_{\alpha} \, $ corresponding to a certain integral 
plane $\, \Gamma_{\alpha}
 \, $. In this case, stable open 
trajectories become periodic if and only if the intersection 
of the plane orthogonal to $\, {\bf B} \, $ and the plane 
$\, \Gamma_{\alpha} \, $ has a rational (integer) direction, 
i.e. if the plane orthogonal to $\, {\bf B} \, $ contains an 
integer direction that also belongs to the corresponding plane 
$\, \Gamma_{\alpha} \, $. It can be seen, therefore, that the 
appearance of periodic open trajectories among stable open 
trajectories is directly related to the structure of the 
system (\ref{MFSyst}) described above and, in fact, is closely
connected with the topological numbers 
$\, (M^{\alpha}_{1}, M^{\alpha}_{2}, M^{\alpha}_{3}) $).

\begin{figure}[t]
\begin{center}
\includegraphics[width=\linewidth]{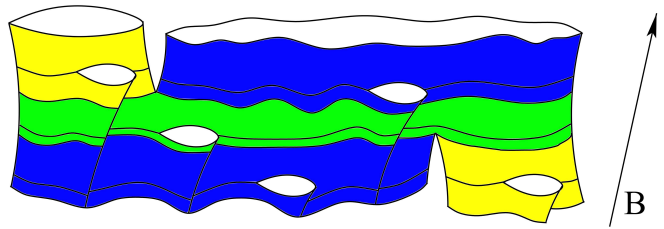}
\end{center}
\caption{Cylinders of closed trajectories arising on a pair 
of former carriers of open trajectories after crossing the 
boundary of a Stability Zone at an angular diagram.}
\label{fig16}
\end{figure}

\begin{figure}[t]
\begin{center}
\vspace{5mm}
\includegraphics[width=\linewidth]{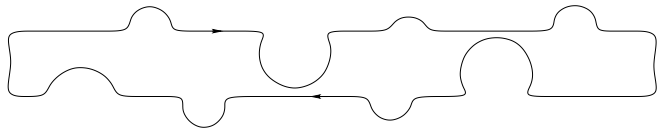}
\end{center}
\caption{Long closed trajectories arising on a pair of 
former carriers of open trajectories in the immediate 
vicinity of the boundary of a Stability Zone.}
\label{fig17}
\end{figure}

 The length of the arising closed trajectories is inversely 
proportional to the probability of ``hopping'' of the trajectory 
from one carrier to another and the smaller the farther the 
direction of $\, {\bf B} \, $ from the boundary of the Zone 
$\, \Omega_{\alpha} \, $. In the immediate vicinity of the 
boundary of $\, \Omega_{\alpha} \, $ for generic directions 
of $\, {\bf B} \, $ the length of the corresponding cylinders 
of closed trajectories is very long (see Fig. \ref{fig16}) 
and goes to infinity on the border itself. Closed trajectories 
arising on such cylinders also have a rather large length and 
a somewhat specific shape (Fig. \ref{fig17}) when approaching 
the boundary $\, \Omega_{\alpha} \, $. In particular, in the 
immediate vicinity of the boundary of $\, \Omega_{\alpha} \, $, 
the condition $\, \, T / \tau \, \ll \, 1 \, $ must be violated 
and the corresponding long closed trajectories become 
indistinguishable from open from an experimental point of view. 
As one of the consequences of this, the behavior of conductivity 
in strong magnetic fields for the corresponding directions of 
$\, {\bf B} \, $ also becomes quite complicated and may not be 
described by simple asymptotic formulas even for fairly large 
values of $\, B \, $ (see e.g. \cite{AnProp}). As indicated in 
\cite{CyclRes}, to determine the exact mathematical boundaries 
of the Stability Zones, it might be better to actually use 
observations of oscillation phenomena in strong magnetic fields 
that track the reconstructions of the topological structure of 
the system (\ref{MFSyst}) at the boundaries of 
$\, \Omega_{\alpha} \, $ (the disappearance of a cylinder 
of ``short'' closed trajectories), which we consider here. 
It can be noted that during the experimental observation of 
oscillation phenomena in this situation, it is specific that 
one of the oscillating terms disappears when crossing the 
boundary of the Stability Zone from the inside without 
other oscillating terms appearing immediately due to the 
large length of new closed trajectories near the boundary of 
$\, \Omega_{\alpha} \, $. When moving away from the boundary 
of $\, \Omega_{\alpha} \, $, the lengths of the corresponding 
trajectories, however, decrease rather quickly, and the 
trajectories themselves take on a ``normal'' shape. As a 
consequence of this, in the oscillation picture, it also 
becomes possible to observe the oscillation terms 
corresponding to the new cylinders of closed trajectories.

 A somewhat different situation is observed
if we approach (or move away) the boundary of a Zone 
$\, \Omega_{\alpha} \, $ along one of the adjacent segments 
corresponding to the appearance of periodic trajectories.   
In this case, as can be seen (Fig. \ref{fig14}, 
lower picture), the length of the cylinder of closed trajectories 
remains unchanged when moving along a given segment. However, its 
height increases rapidly (and the width of the layers of periodic 
trajectories decreases) with distance from the boundary of 
$\, \Omega_{\alpha} \, $, which is associated with an increase 
in the probability of hopping between former carriers of open 
trajectories (diameter of circles at Fig. \ref{fig14}). 
It can also be noted here that in this case the cylinders of 
closed trajectories are not generic, since 
they contain two singular points on each of their bases. The 
corresponding topological structure of the system (\ref{MFSyst}) 
is also, therefore, not a generic structure and arises only for 
a set of measure zero at the angular diagram. At the end point 
of the segment, the width of the layers of periodic trajectories 
vanishes and the former carriers of open trajectories completely 
disintegrate into cylinders of closed trajectories 
(Fig. \ref{fig14}, upper picture). The corresponding 
reconstruction of the topological structure of (\ref{MFSyst}) 
should also be attributed to non-generic reconstructions observed 
only at isolated points (and not along one-dimensional curves) 
at the angular diagram.

\vspace{1mm}

 Here, of course, it should be noted that the boundaries of 
the Stability Zones $\, \Omega_{\alpha} \, $ (as well as the 
special segments adjoining them) are completely special sets 
from the point of view of our definition of a change of the 
structure of system (\ref{MFSyst}) on the Fermi surface.  
On the one hand, as we have already seen, the boundary of 
a Zone $\, \Omega_{\alpha} \, $ itself is determined by the 
described reconstruction with the disappearance of a cylinder 
of closed trajectories. When approaching the boundary of 
$\, \Omega_{\alpha} \,$ from inside the topological structure 
of the system (\ref{MFSyst}) does not change. When approaching 
the boundary of $\, \Omega_{\alpha} \, $ from the outside, 
however, there is an infinite number of reconstructions of 
the system (\ref{MFSyst}) on the Fermi surface, so we can not 
say which structure has (\ref{MFSyst}) near the boundary 
of $\, \Omega_{\alpha} \, $. A similar pattern is also observed 
when approaching special segments (outside the Zone 
$\, \Omega_{\alpha} \,$) corresponding to the appearance of 
periodic trajectories on the Fermi surface.

\begin{figure}[t]
\begin{center}
\includegraphics[width=\linewidth]{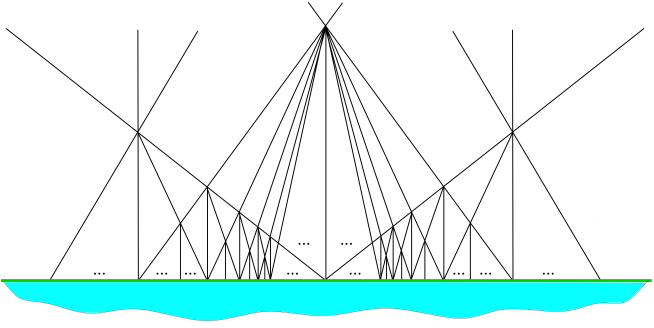}
\end{center}
\caption{The net of directions of $\, {\bf B} \, $, 
corresponding to the reconstructions of the structure of 
the system (\ref{MFSyst}) near the boundary of a Stability Zone 
$\, \Omega_{\alpha} \, $ (very schematically).}
\label{fig18}
\end{figure}

 The reconstructions of cylinders of closed trajectories on the 
former carriers of open trajectories represent changes in the 
topological structure of the system (\ref{MFSyst}) on the 
Fermi surface and are always accompanied by the disappearance 
(and the appearance of new ones) of such cylinders. Such 
reconstructions occur on certain lines at the angular diagram 
(see \cite{SecBound}), whose density tends to infinity when 
approaching the boundary of a Zone $\, \Omega_{\alpha} \, $, 
as well as the special segments corresponding to the appearance 
of periodic trajectories. When moving away from the boundary of 
the Zone $\, \Omega_{\alpha} \, $, as well as the special 
segments described above, the density of the ``net'' of 
directions of $\, {\bf B} \, $ corresponding to the 
reconstructions of the structure of (\ref{MFSyst}) is decreasing. 
Very schematically, the net of directions of $\, {\bf B} \, $, 
corresponding to the reconstructions of the structure of 
(\ref{MFSyst}) near the boundary of $\, \Omega_{\alpha} \, $, 
can be represented by Fig. \ref{fig18}.

\begin{figure}[t]
\begin{center}
\includegraphics[width=\linewidth]{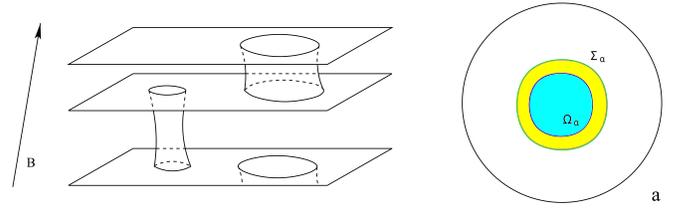}
\end{center}
\caption{The first and second boundaries of a Stability Zone 
on the angular diagram.}
\label{fig19}
\end{figure}

\begin{figure}[t]
\begin{center}
\vspace{5mm}
\includegraphics[width=\linewidth]{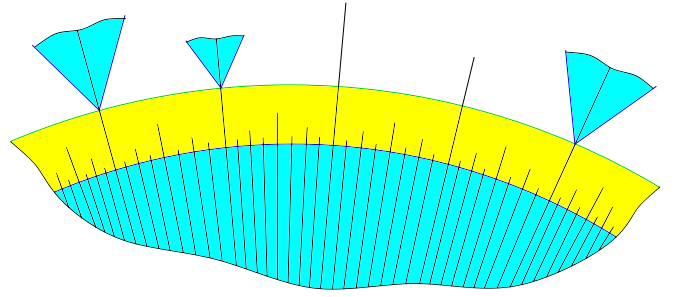}
\end{center}
\caption{Typical structure of the section of the second boundary 
of a Stability Zone $\, \Omega_{\alpha} \, $ on the angular diagram
(schematically).}
\label{fig20}
\end{figure}

 The given situation (the ability to describe the trajectories 
of the system (\ref{MFSyst})) persists until closed trajectories 
disappear on the second (``thin'') cylinder connecting pairs of 
former carriers of open trajectories (Fig. \ref{fig13}).  
It can be seen, therefore, that along with the ``first'' boundary 
of a Stability Zone $\, \Omega_{\alpha} \, $ it is to some extent 
natural to introduce its second boundary (see \cite{SecBound}), 
which defines the ``area of influence'' $\, \Sigma_{\alpha} \, $ 
of the Zone $\, \Omega_{\alpha} \, $ at the angular diagram 
(Fig. \ref{fig19}). The domain
$\, \Omega^{\prime}_{\alpha} \, = \, \Sigma_{\alpha} \backslash 
\Omega_{\alpha} \, $ can naturally be called the derivative of 
the Zone $\, \Omega_{\alpha} \, $, since the structure of the 
system (\ref{MFSyst}) for the corresponding directions of 
$\, {\bf B} \, $ is directly related to the structure of this 
system in the Zone $\, \Omega_{\alpha} \, $. Unlike the Stability 
Zones $\, \Omega_{\alpha} \, $ themselves, the derivatives of two 
different Stability Zones $\, \Omega^{\prime}_{\alpha} \, $, 
$\, \Omega^{\prime}_{\beta} \, $ may intersect with each other. 
In addition, unlike the first boundaries of the Stability Zones, 
the second boundaries are not such a complex set from the point of 
view of reconstructions of the structure of system (\ref{MFSyst}) 
and everywhere, with the exception of only a finite number of points, 
are described by a simple (elementary) reconstruction of this 
structure. The only exceptions are certain points of the second 
boundary (Fig. \ref{fig20}), where it can be crossed by the 
segments that correspond to the appearance of periodic trajectories, 
as well as other Stability Zones (see \cite{UltraCompl}).

\vspace{1mm}

\begin{figure}[t]
\begin{center}
\includegraphics[width=\linewidth]{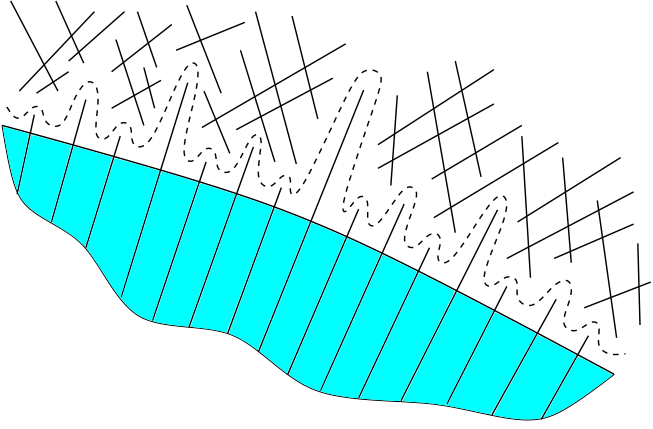}
\end{center}
\caption{The set (net) of experimentally observed directions 
of $\, {\bf B} \, $ near the boundary of a Stability Zone 
corresponding to reconstructions of the structure of 
system (\ref{MFSyst}) on the Fermi surface (schematically). }
\label{fig21}
\end{figure}

 One can see, therefore, that the set of directions of 
$\, {\bf B} \, $, corresponding to reconstructions of the 
topological structure of the system (\ref{MFSyst}), is rather rich 
for fairly complex Fermi surfaces. So, the presence of only one 
Stability Zone on the angular diagram entails in fact the existence 
of a rather complex net of one-dimensional curves on 
$\, \mathbb{S}^{2} \, $, corresponding to changes in the set 
$\, {\cal M} \, $ with the height of one of the cylinders of closed 
trajectories turning to zero. First of all, it is interesting to study 
here the oscillation phenomena at the first and second boundaries of 
the Zone $\, \Omega_{\alpha} \,$, which actually help to determine 
their exact location. No less interesting for studying the Fermi 
surface, however, is also the net of directions of $\, {\bf B} \, $ 
between the first and second boundaries of the Stability Zone, 
which is associated with the reconstructions of the structure 
of system (\ref{MFSyst}) on the former carriers of open trajectories.
As mentioned above, to study the features of oscillation phenomena 
that are interesting to us, the corresponding directions of 
$\, {\bf B} \, $ should not approach too close to the first boundary 
of the Zone $\, \Omega_{\alpha} \, $ or the adjacent segments 
corresponding to the appearance of periodic trajectories on the 
Fermi surface. From the point of view of modern experimental 
capabilities (for producing strong magnetic fields and materials 
with big electron mean free path), however, these restrictions 
are not really too strong and leave a significant part of this 
structure available for study (Fig. \ref{fig21}).

\begin{figure}[t]
\begin{center}
\includegraphics[width=\linewidth]{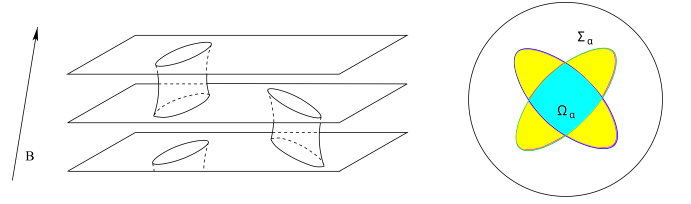}
\end{center}
\caption{An example of a model Fermi surface having a Stability 
Zone with composite first and second boundaries.}
\label{fig22}
\end{figure}

 It can also be noted here that the described structure of the 
regions $\, \Omega_{\alpha} \, $ and $\, \Sigma_{\alpha} \, $ is, 
in a sense, the simplest. For example, there may be cases when the 
first and second boundaries of a Zone $\, \Omega_{\alpha} \, $ 
are ``composite'', and the corresponding region 
$\, \Omega^{\prime}_{\alpha} \, $ is not connected 
(see e.g. Fig. \ref{fig22}). In addition, surfaces of a large 
genus can have a larger number of cylinders separating carriers of 
open trajectories (and even a larger number of pairs of carriers of 
open trajectories). Nevertheless, all the above remarks related 
to the topological structure of the trajectories near the Stability 
Zones are retained for more complex cases 
(see, for example, \cite{SecBound}) and, in particular, an increase 
in the number of Stability Zones at the angular diagram leads in 
general case to rapid increase in the set of directions of 
$\, {\bf B} \, $, corresponding to reconstructions of the 
topological structure of system (\ref{MFSyst}) on the Fermi surface.

\vspace{1mm}

 Thus, for conductivity diagrams, the complexity of which is 
limited by the presence of Stability Zones (and segments 
corresponding to the appearance of periodic trajectories), the set 
of directions of $\, {\bf B} \, $ corresponding to reconstructions 
of the topological structure of (\ref{MFSyst}) always includes the 
boundaries of the Stability Zones, as well as a rather rich ``net'' 
of one-dimensional directions of $\, {\bf B} \, $ in the space 
between the Zones. The indicated net of one-dimensional 
directions of $\, {\bf B} \, $ on $\, \mathbb{S}^{2} \, $ 
unlimitedly condenses near the boundaries of the Stability Zones, 
as well as the segments adjacent to them, corresponding to the 
appearance of periodic trajectories on Fermi surface. Observation 
of oscillatory phenomena corresponding to the reconstructions of 
the structure of (\ref{MFSyst}) is possible on this net of 
directions with not too close approach to the boundaries of the 
Stability Zones or adjacent segments, as well as at the boundaries 
of the Zones $\, \Omega_{\alpha} \, $ when approaching them from 
the inside.

\vspace{1mm}

 The sets of the directions of $\, {\bf B} \, $ described above 
are directly connected with the Stability Zones and, in particular, 
are associated with reconstructions of the structure of 
(\ref{MFSyst}) on the former carriers of open trajectories.  
At the same time, additional reconstructions of the structure of 
system (\ref{MFSyst}) that are not related to carriers of open 
trajectories (occurring on other parts of the Fermi surface) and, 
therefore, not related to these reconstructions, can occur on Fermi 
surfaces of a sufficiently large genus. The corresponding structure
is not connected with the above structure and the corresponding 
nets of directions of $ \, {\bf B} \, $ in this case ``overlap'' 
the sets of directions described above, in particular, they can 
lie inside the Zones $\, \Omega_{\alpha} \, $ and intersect with 
their boundaries.

 As follows from the consideration of the general complexity 
classes of angular diagrams of magnetoconductivity in metals 
(see \cite{UltraCompl}), angular diagrams corresponding to the 
appearance of only stable and periodic trajectories on the Fermi 
surface (complex diagrams of type A) represent a separate class 
of complex angular diagrams. As follows from general 
considerations, the appearance of complex diagrams of type A is 
apparently much more likely than the appearance of complex 
diagrams of the other class (complex diagrams of type B) even 
for conductors with very complex Fermi surfaces. Diagrams of 
type A contain in generic case a finite number of Stability 
Zones and the above picture of the set of directions of 
$\, {\bf B} \, $ corresponding to the reconstructions of the 
structure of system (\ref{MFSyst}) on the Fermi surface is 
common to diagrams of this type.

\begin{figure}[t]
\begin{tabular}{lc}
\includegraphics[width=0.45\linewidth]{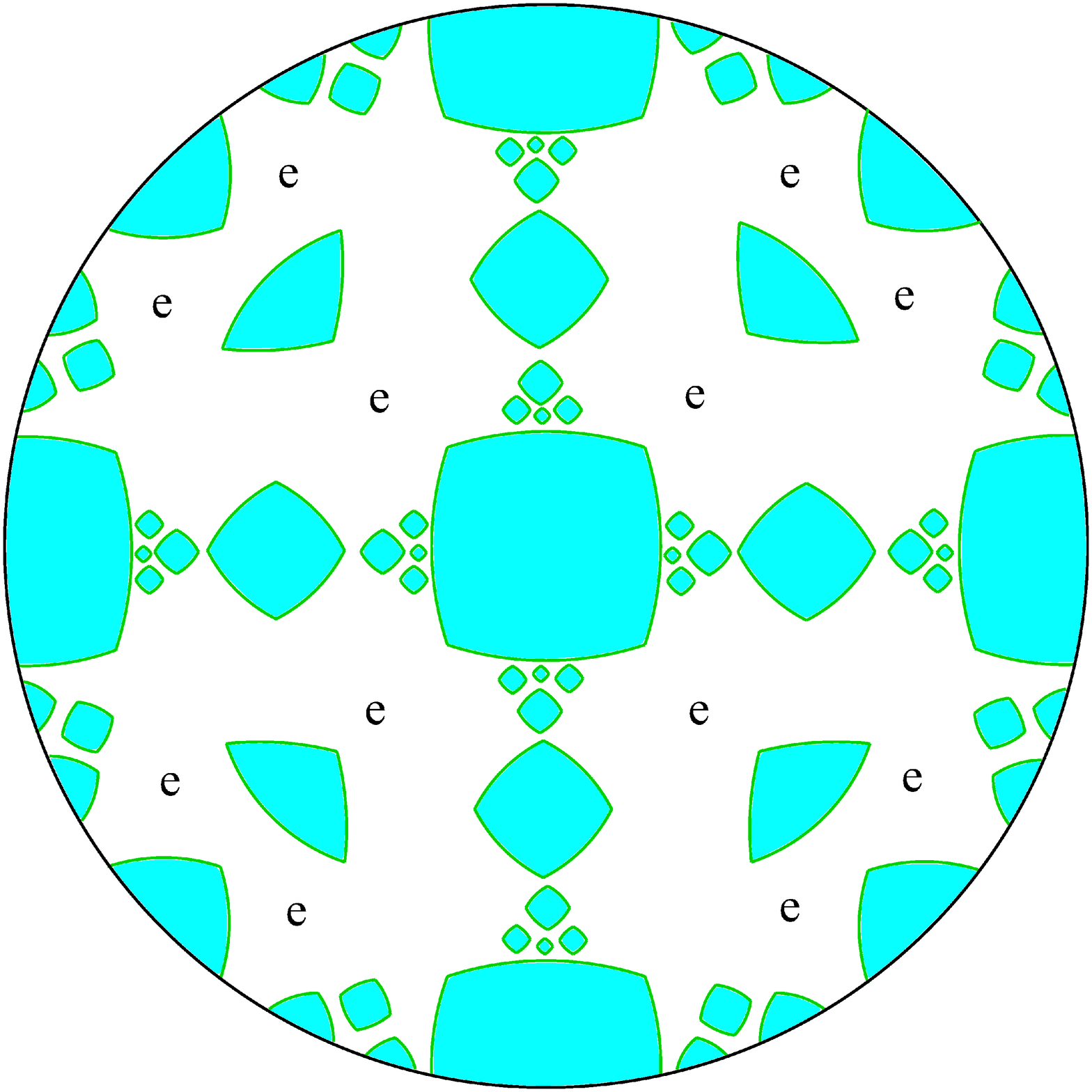} &
\hspace{5mm} \includegraphics[width=0.45\linewidth]{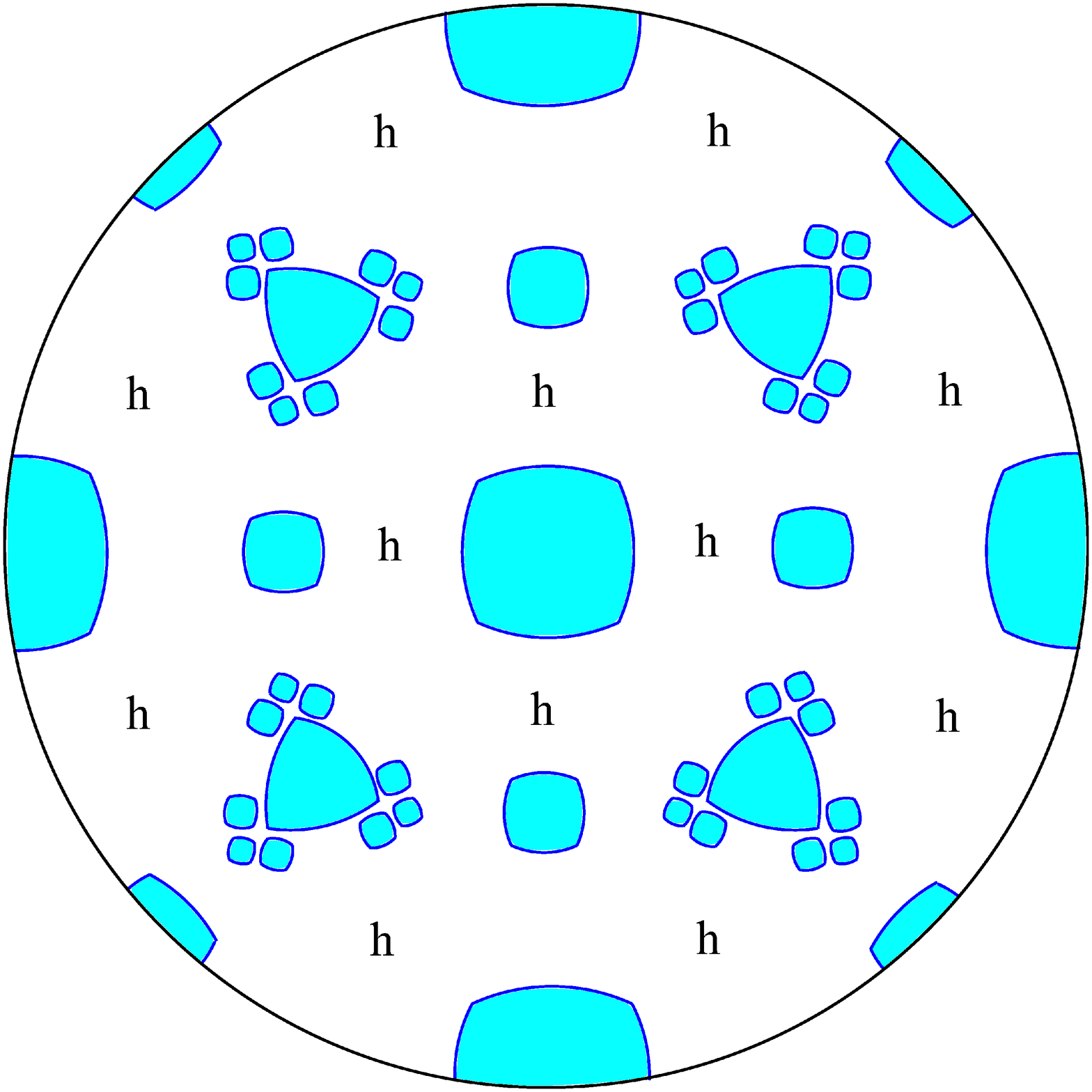}
\end{tabular}
\begin{tabular}{lc}
\includegraphics[width=0.45\linewidth]{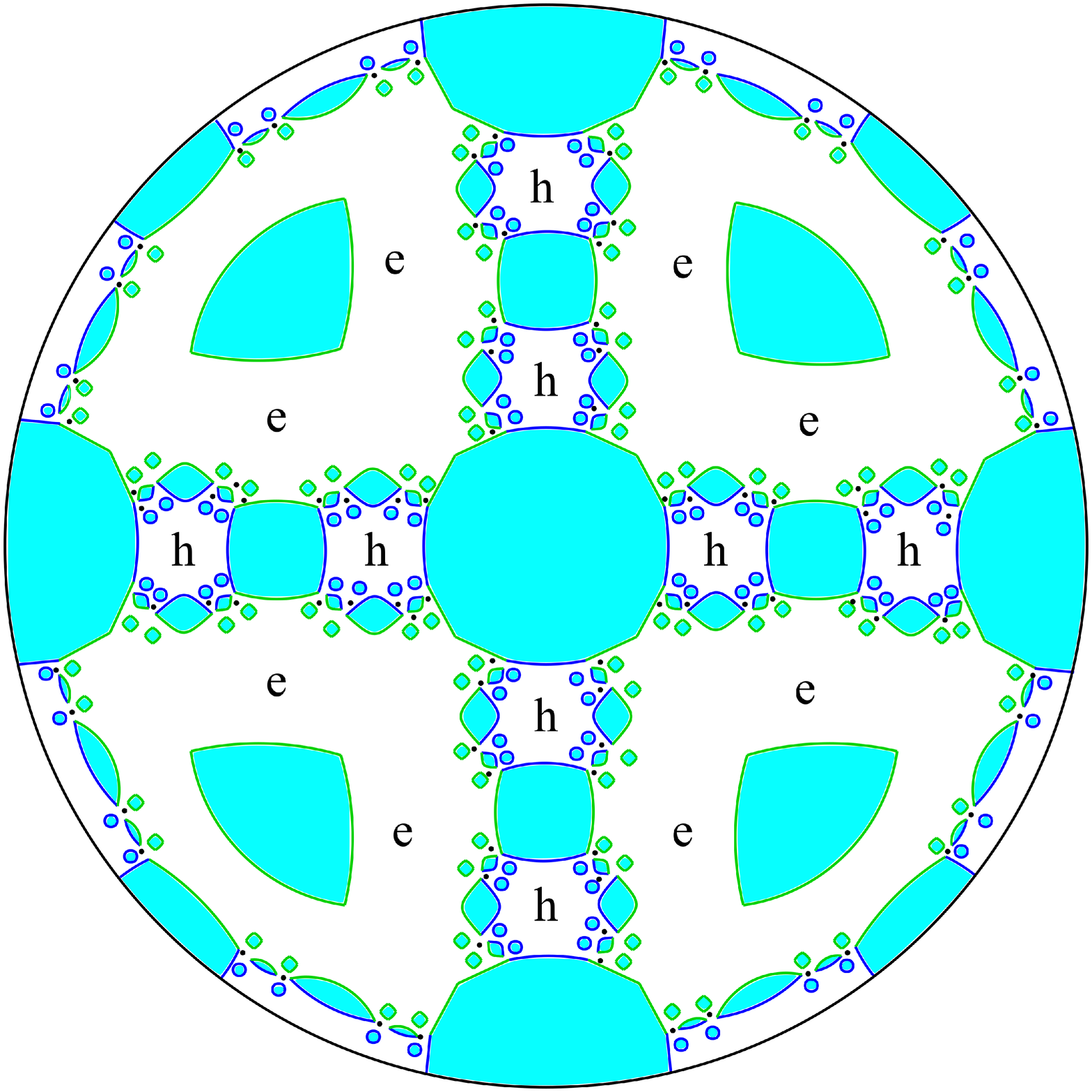} &
\hspace{5mm} \includegraphics[width=0.45\linewidth]{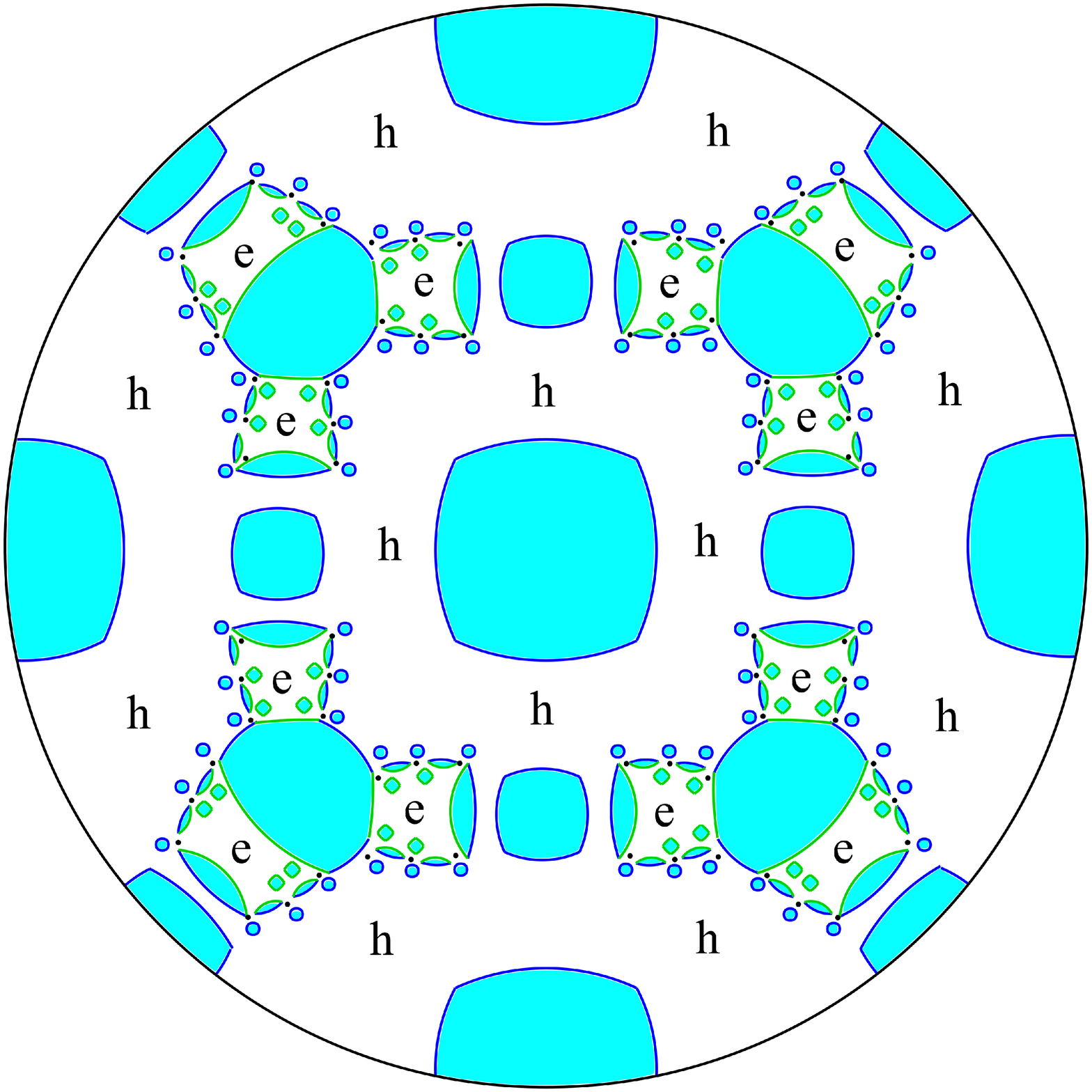}
\end{tabular}
\caption{Diagrams of type A (above) and B (below). The letters e 
and h denote the regions of electron and hole Hall conductivity 
in the absence of open trajectories on the Fermi surface. (Very 
schematically, only the Stability Zones and directions of 
$ \, {\bf B} \, $, corresponding to the appearance of 
``chaotic'' trajectories on the Fermi surface are shown).}
\label{fig23}
\end{figure}

 Complex diagrams of the second type (type B) can be defined 
as diagrams containing directions of $\, {\bf B}\, $ 
corresponding to the appearance of more complex (chaotic) 
unstable trajectories on the Fermi surface. In fact, as follows 
from the general consideration (see \cite{UltraCompl, SecBound}), 
diagrams of this type contain, in addition, an infinite number 
of Stability Zones in the generic case, and also differ from 
diagrams of type A by special behavior of Hall conductivity 
(alternating areas of electron and hole conductivity) in areas 
corresponding to the presence of only closed trajectories on the 
Fermi surface (Fig. \ref{fig23}).

\vspace{1mm}

 Thus, for the most complex Fermi surfaces (namely, surfaces 
on which trajectories that are more complex than stable and 
periodic open trajectories can appear), the structure of the 
set of directions of $\, {\bf B}\, $ that correspond to 
reconstructions of the topological structure of the system 
(\ref{MFSyst}) will be especially rich. In particular, the 
conductivity angular diagrams for such Fermi surfaces contain, 
in the generic case, an infinite number of Stability Zones 
that have condensation points corresponding to the special 
directions of $ \, {\bf B} \, $, connected with the 
appearance of especially complex (chaotic) trajectories 
on the Fermi surface. Near such directions of 
$\, {\bf B} \, $, the structure of the indicated set 
(as well as the structure of the system (\ref{MFSyst}) 
on the Fermi surface) becomes especially complicated. 
As in the case of approaching the boundaries of the 
Stability Zones $\, \Omega_{\alpha} \, $, approaching 
the special directions of $\, {\bf B} \, $, corresponding 
to the appearance of chaotic trajectories, requires an 
infinite number of ``elementary acts'' of reconstruction 
of the structure of system (\ref{MFSyst}). For the rest,
as regards the general features of the set of directions 
of $\, {\bf B} \, $, corresponding to changes in the 
topological structure of (\ref{MFSyst}), their description 
here contains the same basic details as for diagrams of 
type A.

\section{Elementary reconstructions of the structure of
system (\ref{MFSyst}) and the appearance of special closed 
trajectories on the Fermi surface.}
\setcounter{equation}{0}

 Based on the above picture, we can see that, in addition 
to the boundaries of the Stability Zones $\, \Omega_{\alpha}\, $, 
available to experimental study reconstructions of the structure of 
the system (\ref{MFSyst}) occur either in areas that correspond 
to the presence of only closed trajectories on the Fermi surface, 
or also inside the Stability Zones, without being associated with 
the appearance or disappearance of open trajectories. Such 
reconstructions can be called ``elementary'' and correspond to 
the disappearance (appearance) of cylinders of closed trajectories 
on both sides of the curve (on $\, \mathbb{S}^{2}$) that separates 
two different structures of the system (\ref{MFSyst}). Oscillation 
phenomena that indicate the corresponding reconstruction of the 
structure of (\ref{MFSyst}) must be observed on both sides of such 
a curve, so we must record the change of some oscillation terms by 
others at its intersection. In this chapter, we will look in more 
detail at the types of elementary reconstructions of the structure 
of system (\ref{MFSyst}) and discuss the probability of their 
occurrence on real Fermi surfaces.

\begin{figure}[t]
\begin{center}
\includegraphics[width=\linewidth]{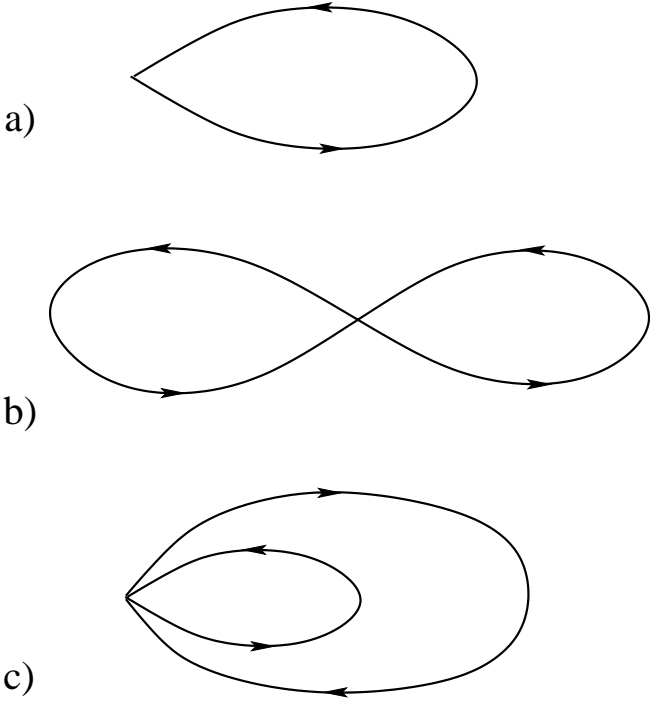}
\end{center}
\caption{Possible forms of bases of ``nontrivial'' 
cylinders of closed trajectories (in a plane orthogonal 
to $\, {\bf B}$).}
\label{fig24}
\end{figure}

 As we mentioned above, the bases of the cylinders of closed 
trajectories contain a saddle singular point of the 
system (\ref{MFSyst}) and can take one of the forms shown at 
Fig. \ref{fig24}. At the time of the reconstruction of the 
topological structure of system (\ref{MFSyst}), we must observe 
the appearance of a ``cylinder of zero height'', which represents 
actually two singular points connected by a set of singular 
trajectories.

\begin{figure}[t]
\begin{center}
\vspace{5mm}
\includegraphics[width=\linewidth]{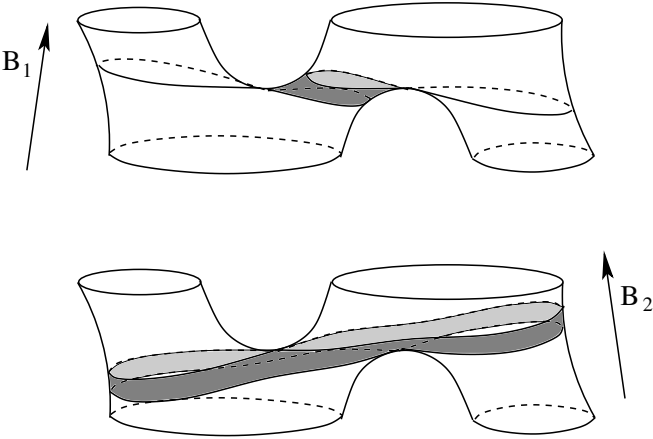}
\end{center}
\caption{The main type of ``elementary'' reconstruction of the 
structure of (\ref{MFSyst}) (only the part of the Fermi surface 
on which the reconstruction takes place is shown).}
\label{fig25}
\end{figure}

\begin{figure}[t]
\begin{center}
\includegraphics[width=\linewidth]{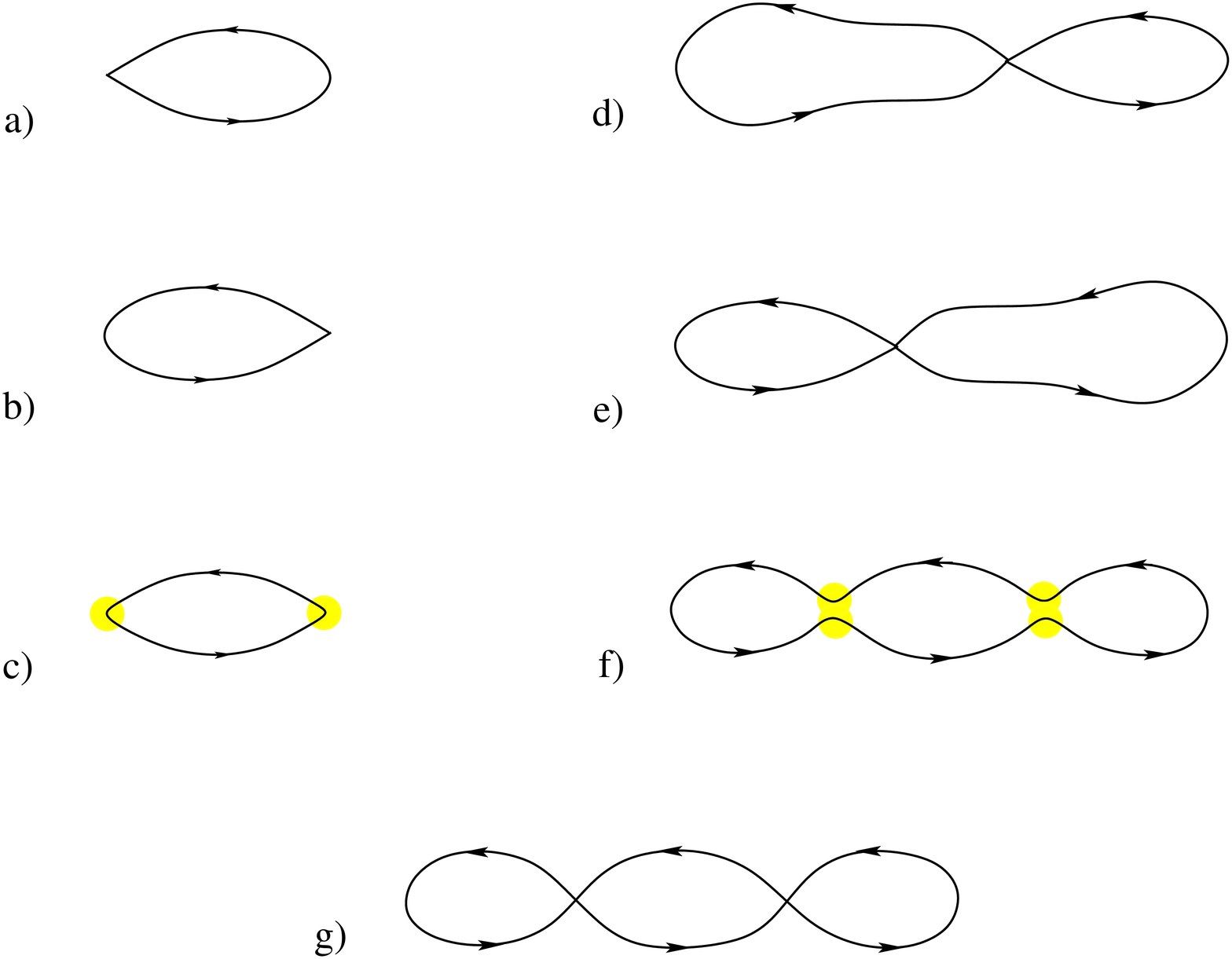}
\end{center}
\caption{The main elements involved in the reconstruction of the 
structure of system (\ref{MFSyst}), shown at Fig. \ref{fig25}. 
The lower and upper bases of the disappearing cylinder of closed 
trajectories and the extremal trajectory arising on it (a-c). 
The lower and upper bases of the appearing cylinder of closed 
trajectories and the extremal trajectory arising on it (d-f).
The ``cylinder of zero height'' that appears at the moment of 
reconstruction (g).}
\label{fig26}
\end{figure}

 Apparently, the most frequent type of the reconstruction 
of the structure of (\ref{MFSyst}) is the reconstruction shown 
at Fig. \ref{fig25}. Fig. \ref{fig26} represents 
the main components of the disappearing and appearing cylinders 
of closed trajectories (namely, their upper and lower bases, 
as well as the extremal trajectories arising on them), and the 
structure of the ``cylinder of zero height'' arising at the time 
of reconstruction. It is easy to see that the reconstruction 
shown at Fig. \ref{fig25}, leads to the disappearance of 
exactly one cylinder of closed trajectories and the appearance 
of exactly one new cylinder.

 In the general case, for generic reconstructions, the 
corresponding ``cylinder of zero height'' contains exactly two 
singular points of the system (\ref{MFSyst}) connected by 
singular trajectories (we will not consider here special
``degenerate'' cases with a larger number of singular points). 
We will consider both singular points on a 
``cylinder of zero height'' to be non-degenerate (which 
corresponds to the non-zero value of the Gaussian curvature 
at the corresponding points on the Fermi surface), in particular, 
this means that exactly two singular trajectories of the system 
(\ref{MFSyst}) enter and exit each such point. For elementary 
reconstructions of the structure of (\ref{MFSyst}), all singular 
trajectories that exit from a pair of such points must also end 
at these points.

 For Fermi surfaces of ``moderate complexity'' all the elements 
involved in the reconstruction shown at Fig. \ref{fig25} 
have central symmetry. In particular, it is present in extremal 
trajectories arising on the disappearing and appearing cylinders 
of closed trajectories. As a relatively simple consideration of 
Fermi surfaces of the most various types shows, the type of 
``elementary'' reconstruction shown at 
Fig. \ref{fig25} - \ref{fig26}, takes place, 
in fact, in the vast majority of cases for real Fermi surfaces.

\vspace{1mm}

\begin{figure}[t]
\begin{center}
\vspace{5mm}
\includegraphics[width=\linewidth]{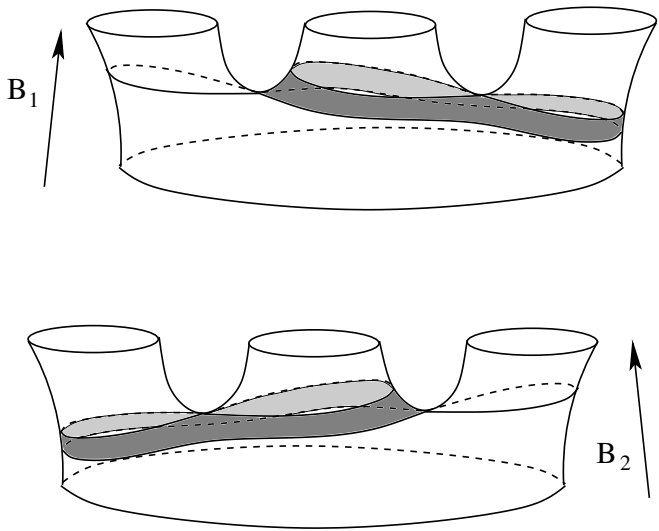}
\end{center}
\caption{An example of the reconstruction of the structure of 
system (\ref{MFSyst}) on a part of the Fermi surface. }
\label{fig27}
\end{figure}

\begin{figure}[t]
\begin{center}
\vspace{5mm}
\includegraphics[width=\linewidth]{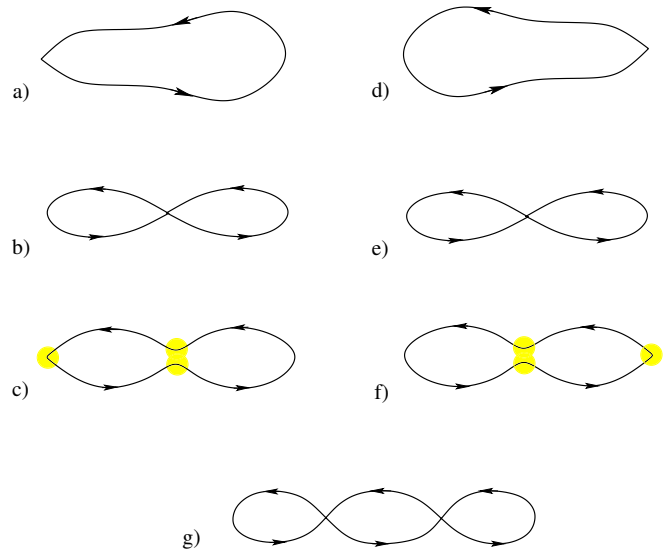}
\end{center}
\caption{The main elements involved in the reconstruction of the 
structure of system (\ref{MFSyst}), shown at Fig. \ref{fig27}. 
The lower and upper bases of the disappearing cylinder of closed 
trajectories and the extremal trajectory arising on it (a-c). 
The lower and upper bases of the appearing cylinder of closed 
trajectories and the extremal trajectory arising on it (d-f).
The ``cylinder of zero height'' that appears at the moment of 
reconstruction (g).}
\label{fig28}
\end{figure}

 The topological type of reconstruction shown at 
Fig. \ref{fig25} - \ref{fig26}, however, is not 
the only possible one. Fig. \ref{fig27} and 
\ref{fig28} show another type of reconstruction, 
different from the one shown at Fig. \ref{fig25} and 
\ref{fig26}.

 Here one can immediately note one important difference 
between the types of reconstruction shown at Fig. 
\ref{fig25} - \ref{fig26} and
\ref{fig27} - \ref{fig28}.
Namely, the reconstruction shown at
Fig. \ref{fig25} - \ref{fig26}, 
can possess central symmetry and thus occur just on one part 
of the Fermi surface. As for the reconstruction shown at 
Fig. \ref{fig27} - \ref{fig28}, it obviously 
cannot have a central symmetry and can thus only occur together 
with a similar reconstruction on another part of the Fermi 
surface that passes into this one under the central symmetry 
transformation. It can be seen, therefore, that for the 
reconstructions shown at 
Fig. \ref{fig27} - \ref{fig28}, the 
corresponding Fermi surface must have rather significant 
complexity (have a sufficiently large genus) even among 
traditionally complex examples of Fermi surfaces.

\begin{figure}[t]
\begin{center}
\includegraphics[width=0.9\linewidth]{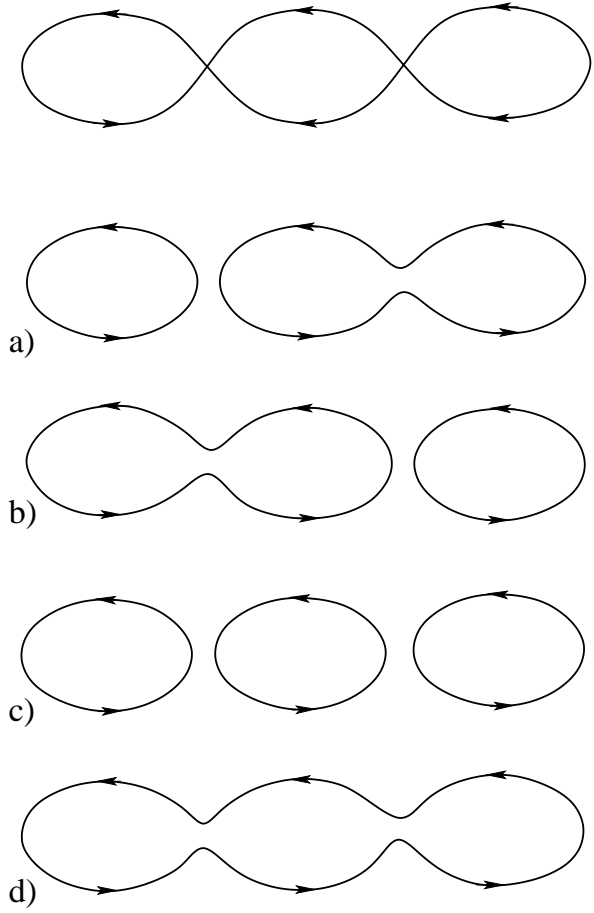}
\end{center}
\caption{``Cylinder of zero height'', corresponding to the 
examples shown at Fig. \ref{fig25} - \ref{fig26} 
and \ref{fig27} - \ref{fig28}, and possible 
options for splitting singular points with a parallel shift of 
the corresponding plane, orthogonal to $\, {\bf B} \, $, in the 
direction of the magnetic field. }
\label{fig29}
\end{figure}

 In general, to enumerate the various topological types 
of ``elementary'' reconstructions of the structure of 
system (\ref{MFSyst}) on the Fermi surface, it is actually 
enough to fix the topological structure of the 
``cylinder of zero height'' in the plane orthogonal to 
$\, {\bf B} \, $ and also indicate ways for splitting both 
saddle singular points with parallel displacement of this 
plane, for example, ``up'' (along the direction of the magnetic 
field). Fig. \ref{fig29} shows possible schemes 
of such a description for the ``cylinder of zero height'', 
corresponding to the examples shown at 
Fig. \ref{fig25} - \ref{fig26} and 
\ref{fig27} - \ref{fig28}.

 We note here that the direction of the arrows at 
Fig. \ref{fig29} is somewhat arbitrary and should 
be reversed when replacing 
$\, {\bf B} \, \rightarrow \, - {\bf B} \, $. 
If we are only interested in changing the topology of the 
carriers of the trajectories on the indicated part of the 
Fermi surface, we can, in reality, not indicate these 
directions. Each of the schemes (a-d) shown at 
Fig. \ref{fig29} fixes the structure of gluing 
to the ``cylinder of zero height'' other cylinders of closed 
trajectories ``from above''. With a parallel shift of the 
plane orthogonal to $\, {\bf B} \, $ ``down'' (against the 
direction of $\, {\bf B} $), the way of splitting each of the 
singular points changes to the opposite 
($(a) \leftrightarrow (b) \, $, $\, \, (c) \leftrightarrow (d) $), 
which also fixes the structure of gluing to the 
``cylinder of zero height'' other cylinders of closed 
trajectories ``from below''.

 In addition to fixing the structure of gluing to the 
``cylinder of zero height'' of other cylinders of closed 
trajectories at the moment of reconstruction, the specified 
scheme of splitting singular points also defines the structure 
of the system (\ref{MFSyst}) before and after it. Indeed, the 
reconstruction of the structure of (\ref{MFSyst}) occurs at 
small rotations of the magnetic field direction, which entail 
corresponding rotations of the plane orthogonal to 
$\, {\bf B} \, $. Those rotations of $\, {\bf B} \, $, at 
which the singular points remain in this plane, correspond to 
the motion along the curve of the reconstruction of the 
structure of (\ref{MFSyst}), and the rotations at which the 
singular points cannot simultaneously remain in the same plane, 
orthogonal to $\, {\bf B} \, $, correspond to the crossing 
this curve at the angular diagram. To monitor the reconstruction 
of the structure of (\ref{MFSyst}), one can observe the rotation 
of the plane orthogonal to $\, {\bf B} \, $ and containing some 
fixed point on the interval between two singular points. 
It is easy to see that during rotations corresponding to a 
reconstruction of the structure of (\ref{MFSyst}), such a plane 
goes upward near one of the singular points and downward near 
the other, and the rules for splitting the singular points in 
the rotating plane must comply with the rules specified initially. 
Then displacing the rotated plane parallel to itself and using 
the same fixed rules for reconstruction of trajectories as it 
passes through singular points, it is easy to restore the shape 
of the appearing (or disappearing) cylinders of closed trajectories 
of small height, as well as the structure of gluing to them other 
cylinders of closed trajectories before or after the reconstruction.

 It can also be noted here that the terms ``disappearing'' and 
``appearing'' structures are also arbitrary to some extent and 
depend on the direction of rotation of the magnetic
 field. 
Actually, it is more correct to speak of two different structures 
of the system (\ref{MFSyst}) adjacent to each other along a line 
of reconstruction at the angular diagram. From this point of view, 
in particular, it can be seen, as above, that replacing the 
splitting rules at once for both singular points in the 
above-described scheme does not essentially change the 
topological type of reconstruction taking place on a fixed curve. 
One can therefore also say that to describe the topological types 
of reconstructions of the structure of system (\ref{MFSyst}), 
it suffices to list the topological structures of the 
corresponding ``cylinders of zero height'' in the plane orthogonal 
to $\, {\bf B} \, $ and indicate for each cylinder, whether the 
directions of the group velocity at singular points are 
co-directional or opposite to each other. 
  
\begin{figure}[t]
\vspace{5mm}
\begin{tabular}{lc}
\includegraphics[width=0.5\linewidth]{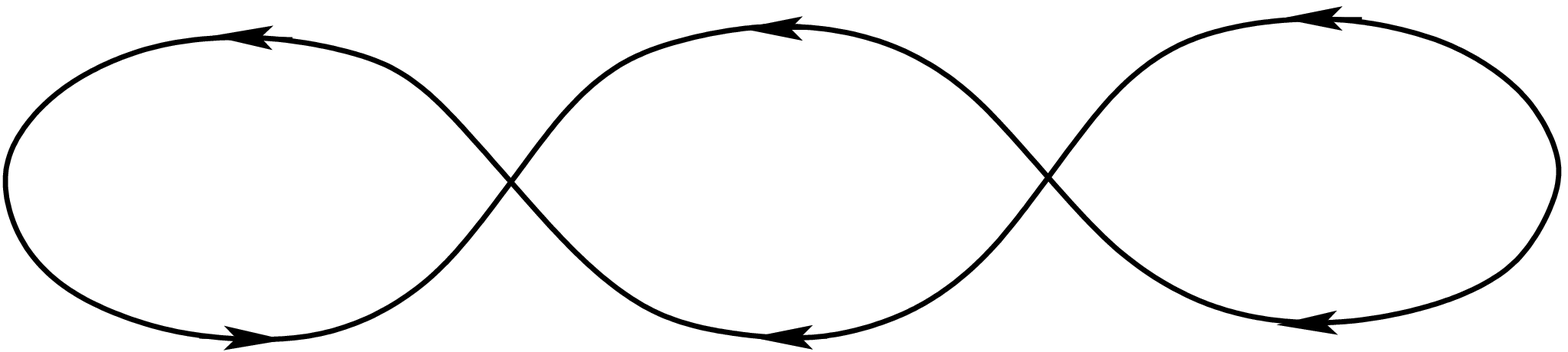}  &
\hspace{5mm}
\includegraphics[width=0.4\linewidth]{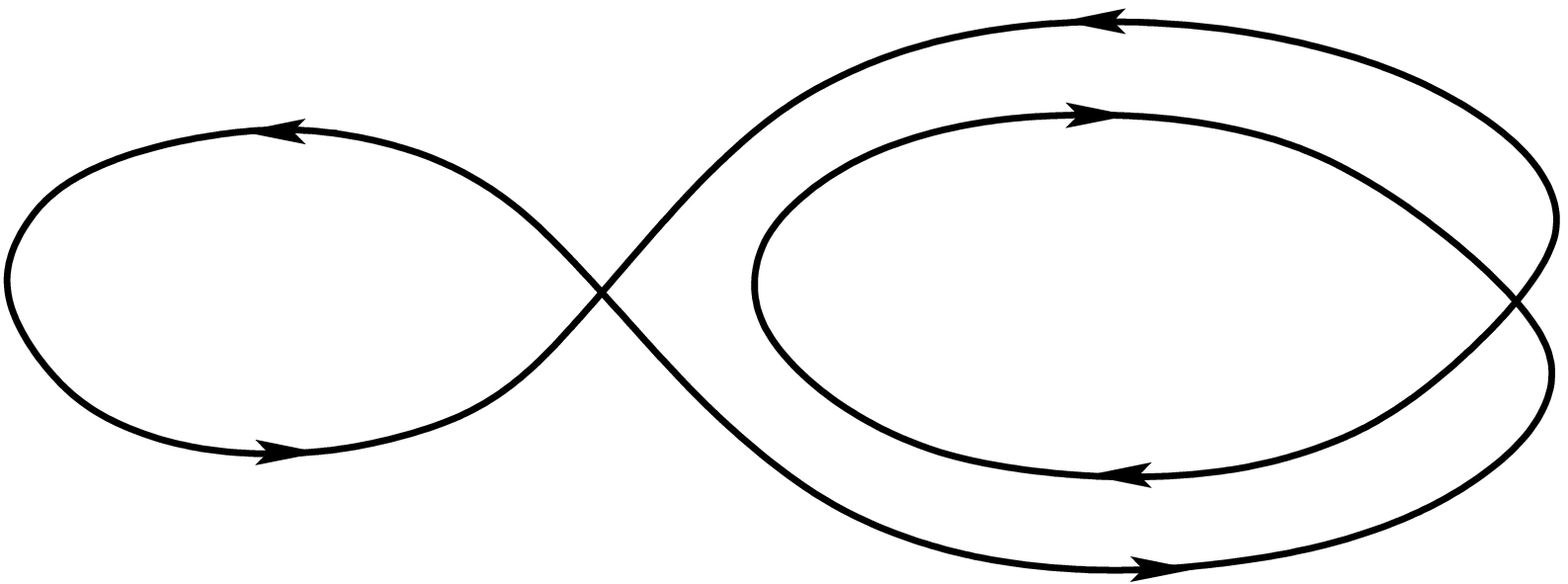}
\end{tabular}
\vspace{5mm}
\begin{tabular}{lc}
\includegraphics[width=0.4\linewidth]{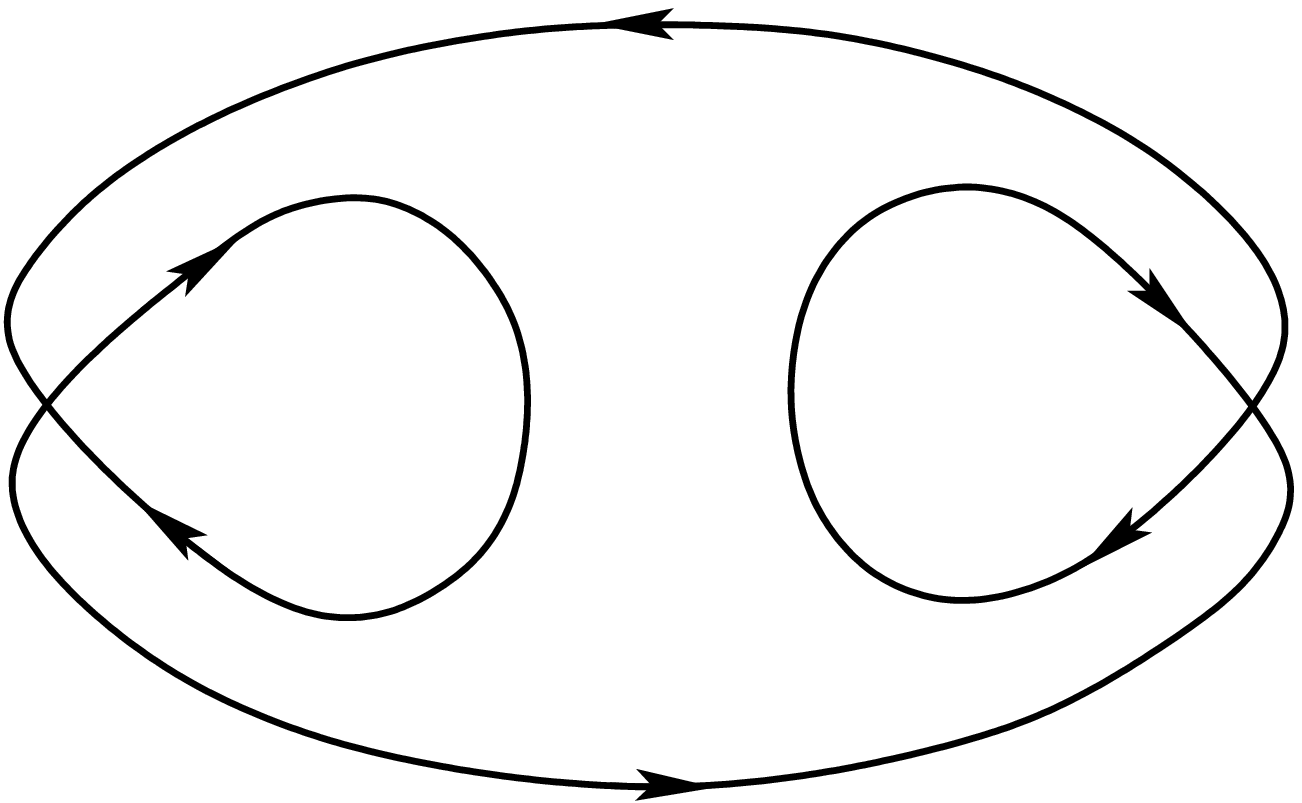}  &
\hspace{5mm}
\includegraphics[width=0.5\linewidth]{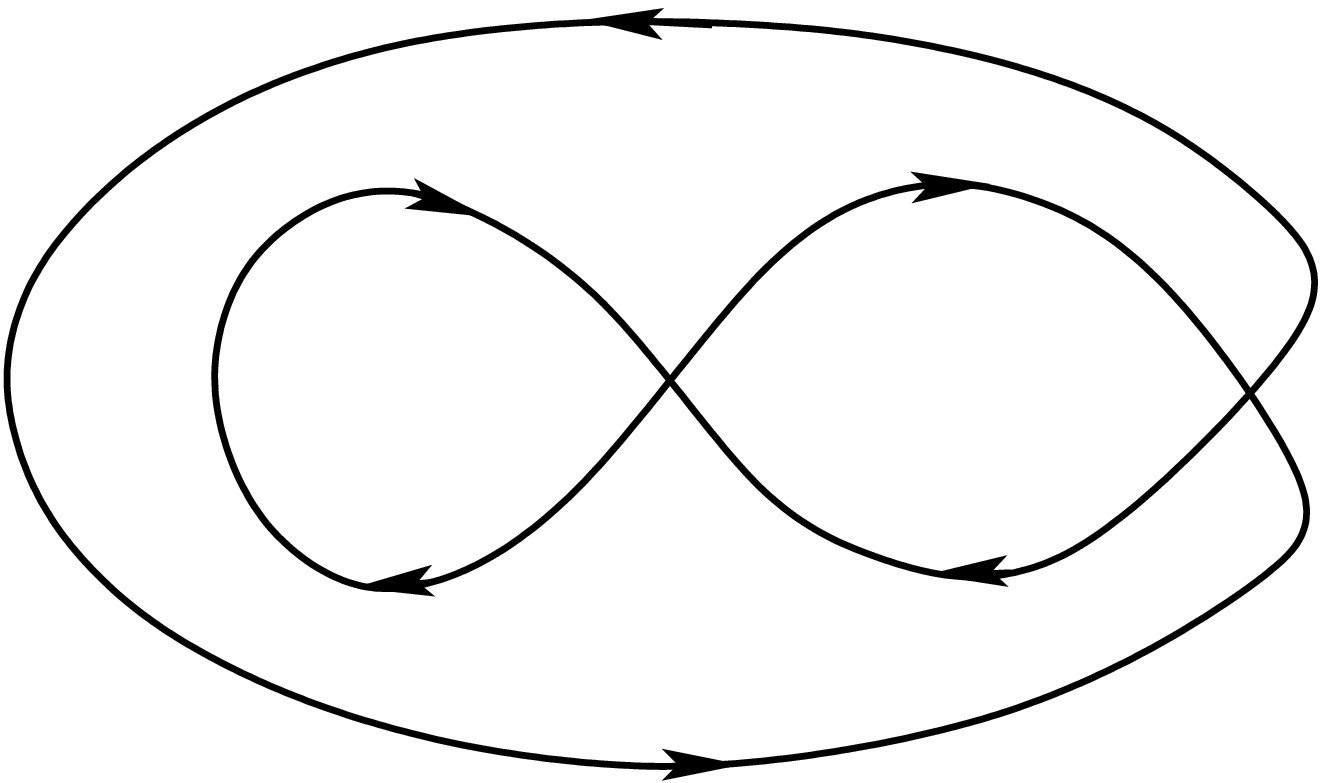}
\end{tabular}
\begin{tabular}{lc}
\includegraphics[width=0.5\linewidth]{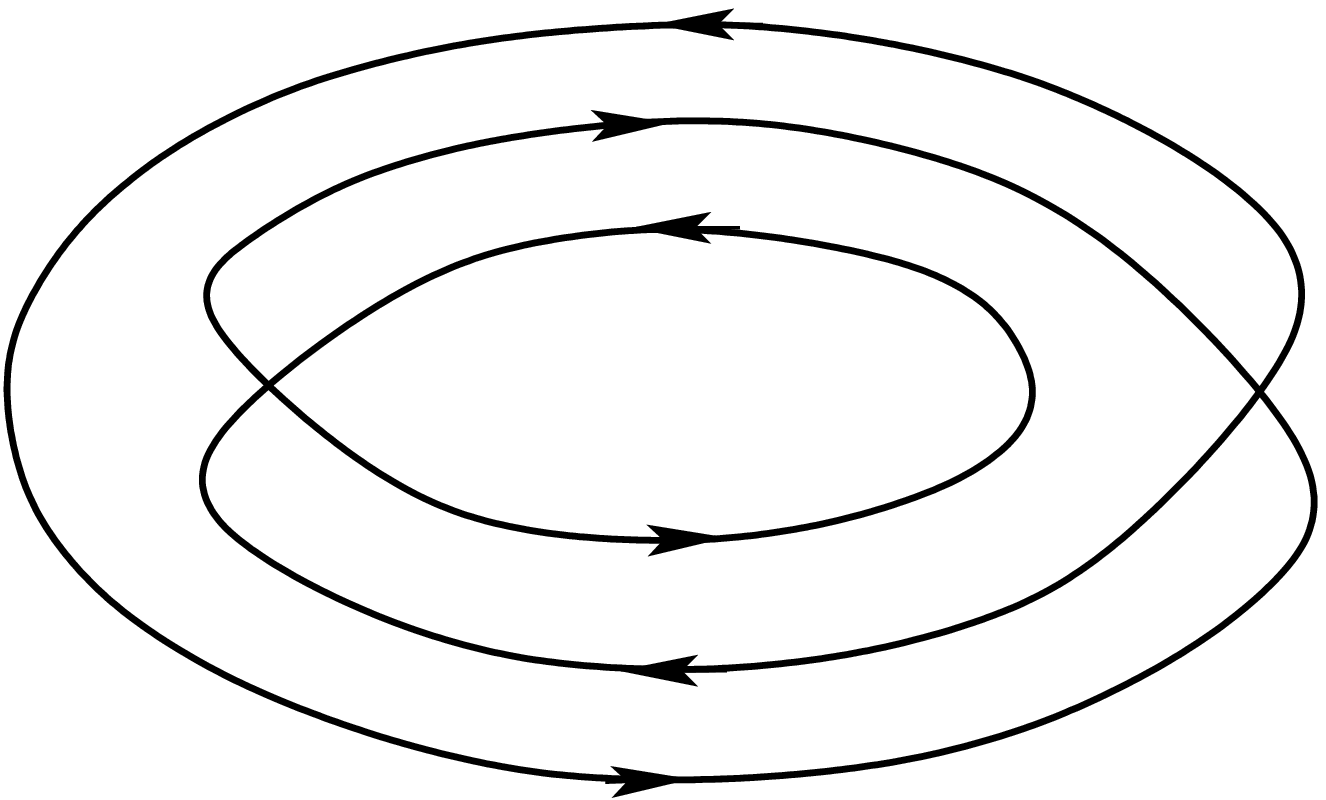}  &
\hspace{5mm}
\includegraphics[width=0.4\linewidth]{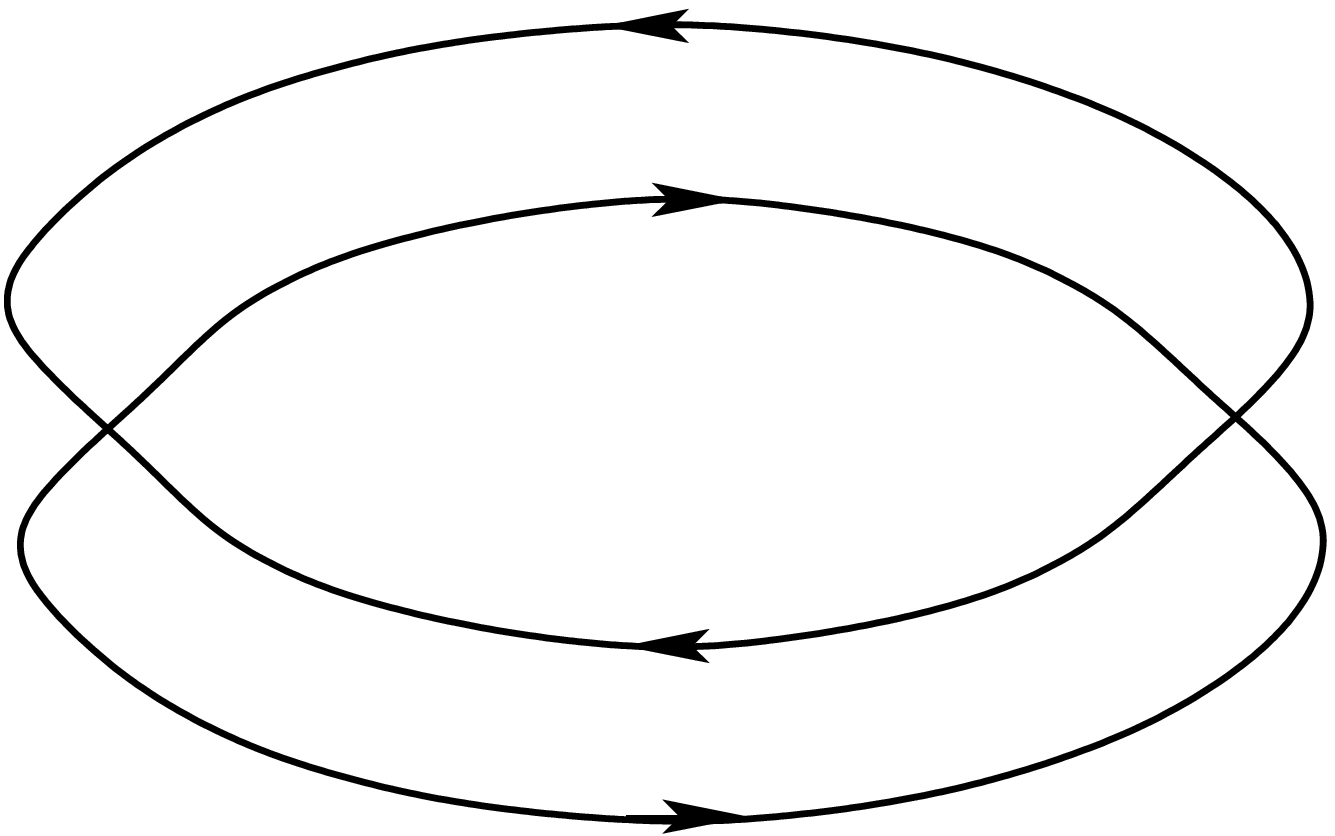}
\end{tabular}
\caption{Different types of ``cylinders of zero height'' 
arising at the moment of elementary reconstruction of the 
topological structure of the system (\ref{MFSyst}) on the 
Fermi surface.}
\label{fig30}
\end{figure}

\begin{figure}[t]
\begin{center}
\includegraphics[width=\linewidth]{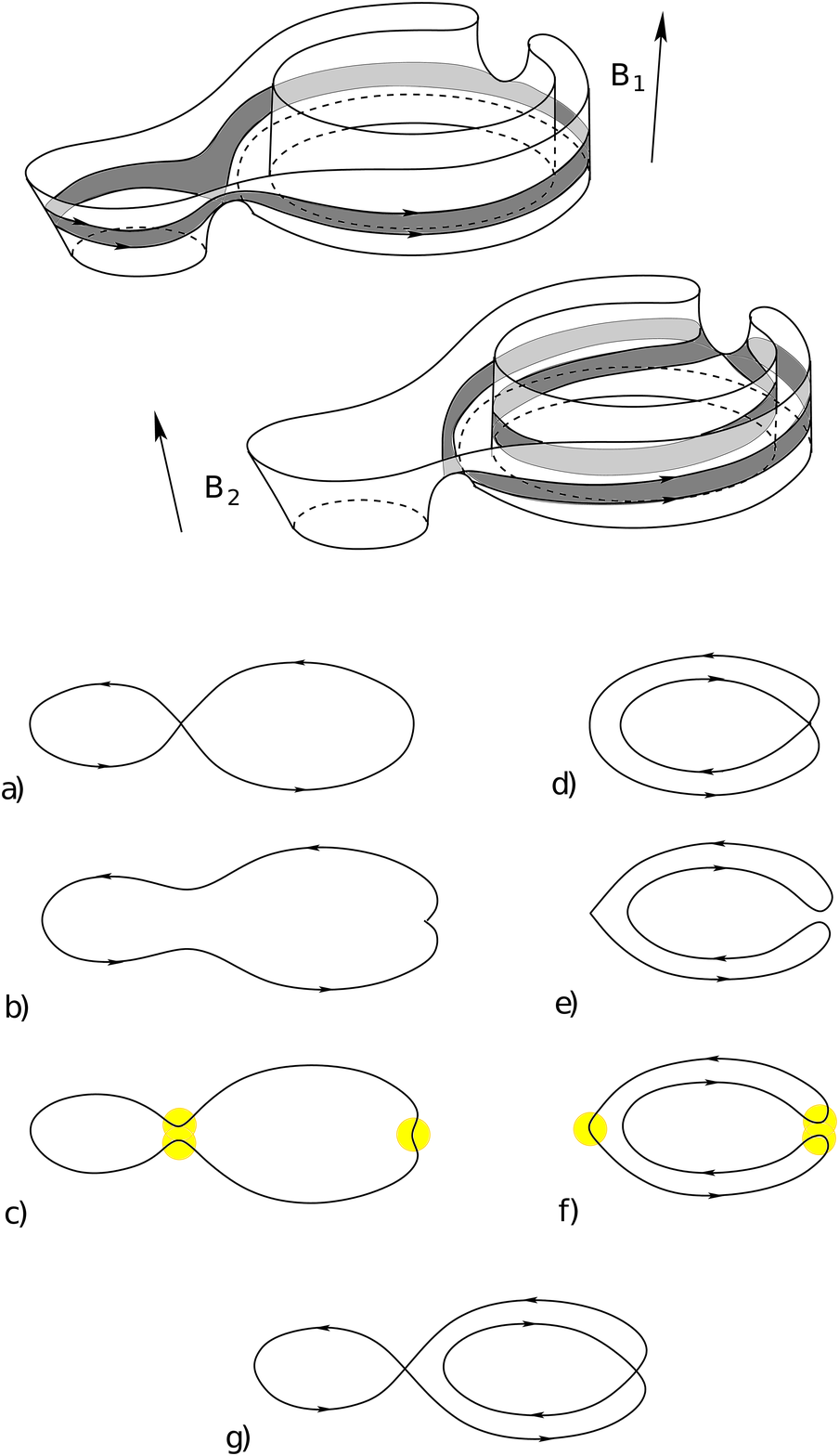}
\end{center}
\caption{A reconstruction of the structure of system 
(\ref{MFSyst}) on a part of the Fermi surface.  
The lower and upper bases of the disappearing cylinder of closed 
trajectories and the extremal trajectory arising on it (a-c). 
The lower and upper bases of the appearing cylinder of closed 
trajectories and the extremal trajectory arising on it (d-f).
The ``cylinder of zero height'' that appears at the moment of 
reconstruction (g).}
\label{fig31}
\end{figure}

\begin{figure}[t]
\begin{center}
\includegraphics[width=\linewidth]{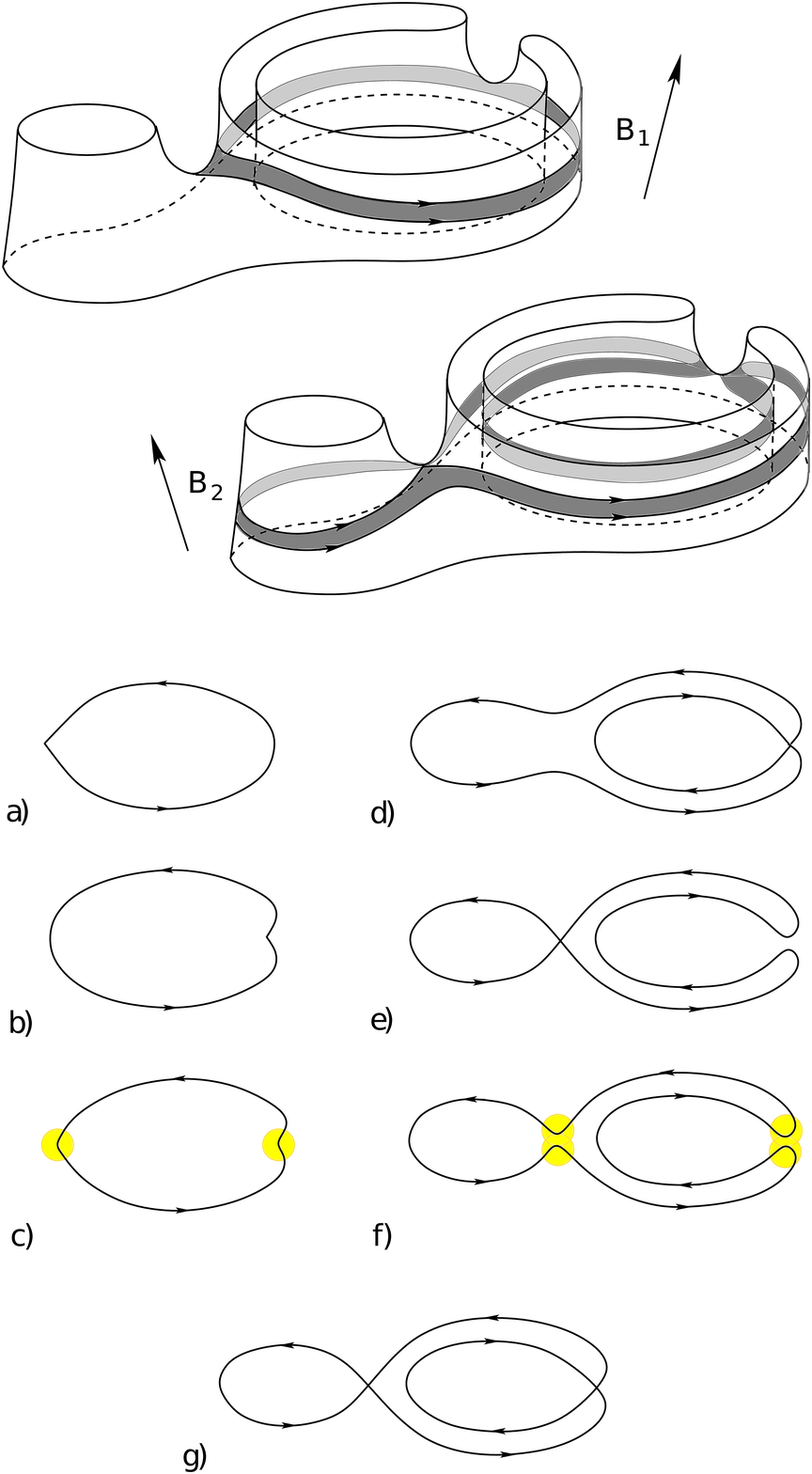}
\end{center}
\caption{A reconstruction of the structure of system 
(\ref{MFSyst}) on a part of the Fermi surface.  
The lower and upper bases of the disappearing cylinder of closed 
trajectories and the extremal trajectory arising on it (a-c). 
The lower and upper bases of the appearing cylinder of closed 
trajectories and the extremal trajectory arising on it (d-f).
The ``cylinder of zero height'' that appears at the moment of 
reconstruction (g).}
\label{fig32}
\end{figure}

\begin{figure}[t]
\begin{center}
\includegraphics[width=\linewidth]{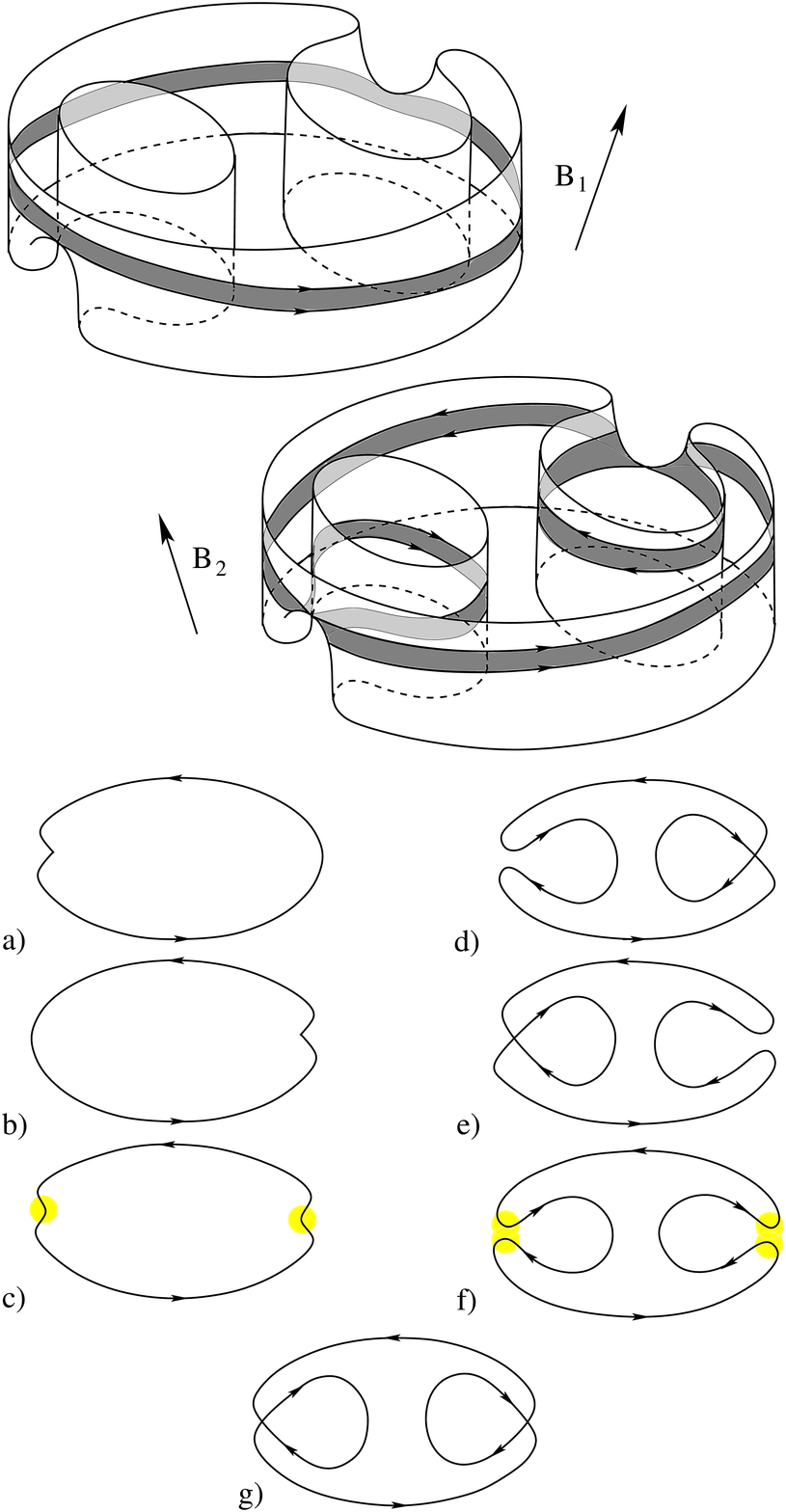}
\end{center}
\caption{A reconstruction of the structure of system 
(\ref{MFSyst}) on a part of the Fermi surface.  
The lower and upper bases of the disappearing cylinder of closed 
trajectories and the extremal trajectory arising on it (a-c). 
The lower and upper bases of the appearing cylinder of closed 
trajectories and the extremal trajectory arising on it (d-f).
The ``cylinder of zero height'' that appears at the moment of 
reconstruction (g).}
\label{fig33}
\end{figure}

\begin{figure}[t]
\begin{center}
\includegraphics[width=\linewidth]{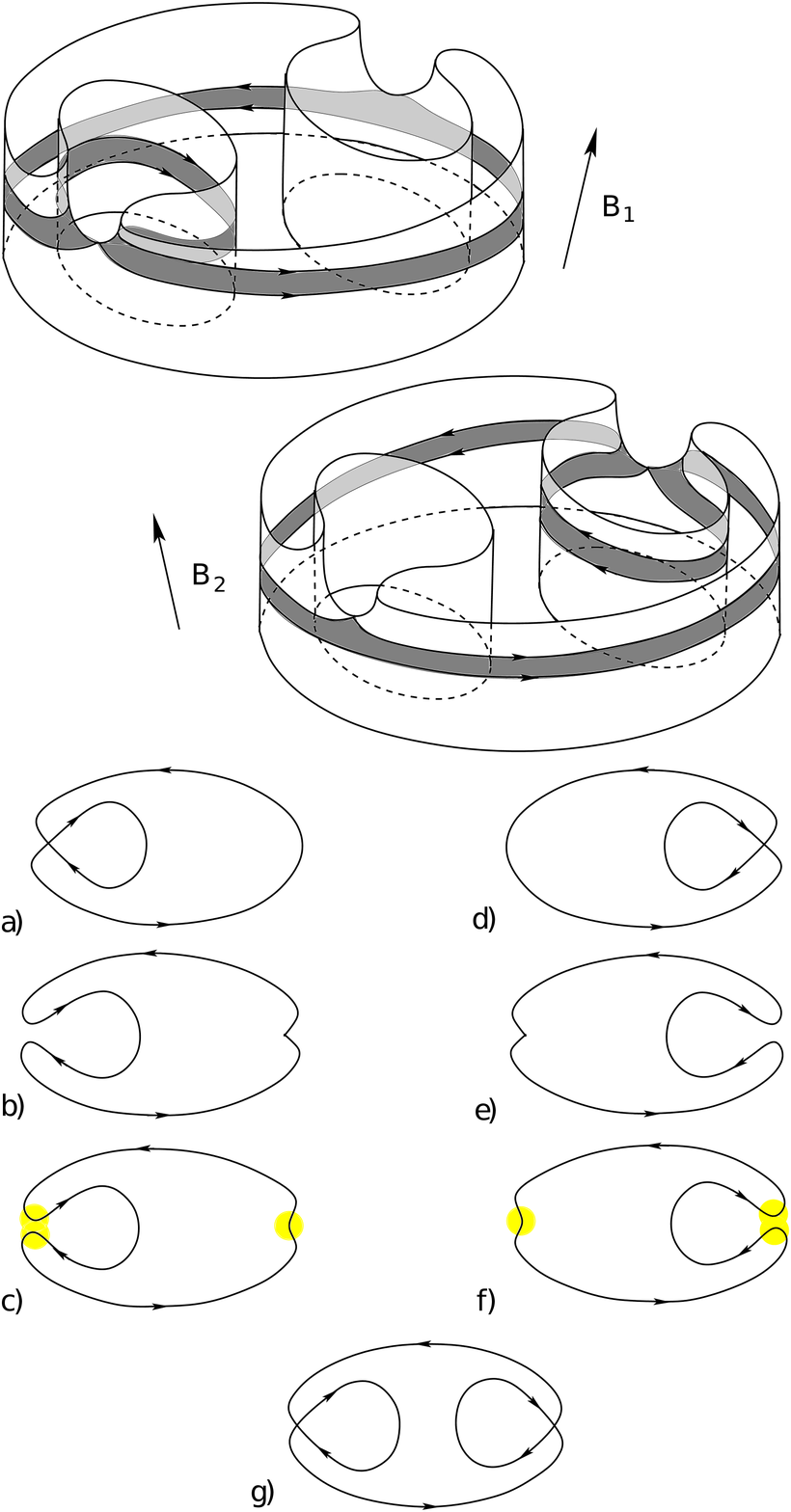}
\end{center}
\caption{A reconstruction of the structure of system 
(\ref{MFSyst}) on a part of the Fermi surface.  
The lower and upper bases of the disappearing cylinder of closed 
trajectories and the extremal trajectory arising on it (a-c). 
The lower and upper bases of the appearing cylinder of closed 
trajectories and the extremal trajectory arising on it (d-f).
The ``cylinder of zero height'' that appears at the moment of 
reconstruction (g).}
\label{fig34}
\end{figure}

\begin{figure}[t]
\begin{center}
\includegraphics[width=\linewidth]{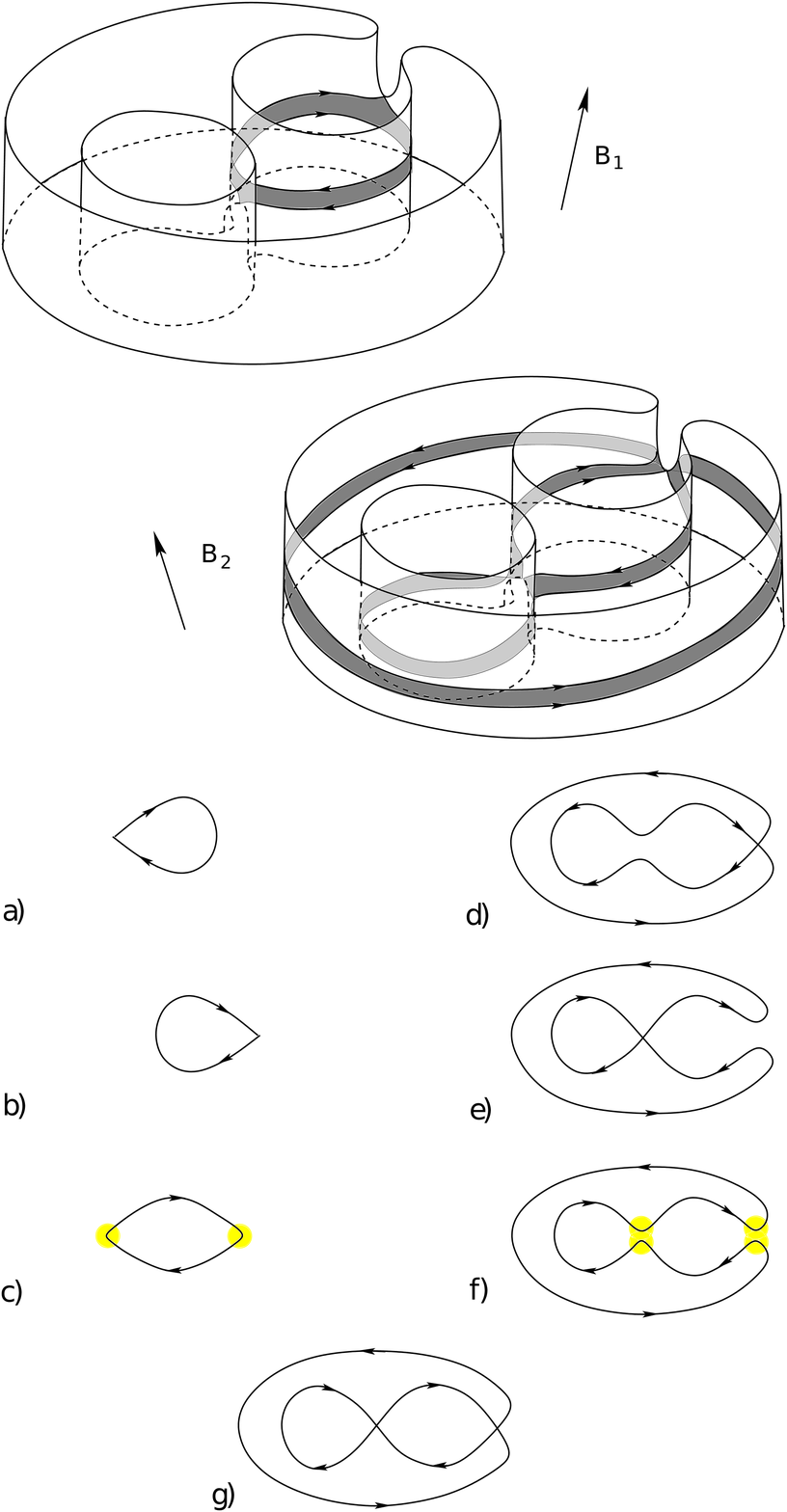}
\end{center}
\caption{A reconstruction of the structure of system 
(\ref{MFSyst}) on a part of the Fermi surface.  
The lower and upper bases of the disappearing cylinder of closed 
trajectories and the extremal trajectory arising on it (a-c). 
The lower and upper bases of the appearing cylinder of closed 
trajectories and the extremal trajectory arising on it (d-f).
The ``cylinder of zero height'' that appears at the moment of 
reconstruction (g).}
\label{fig35}
\end{figure}

\begin{figure}[t]
\begin{center}
\includegraphics[width=\linewidth]{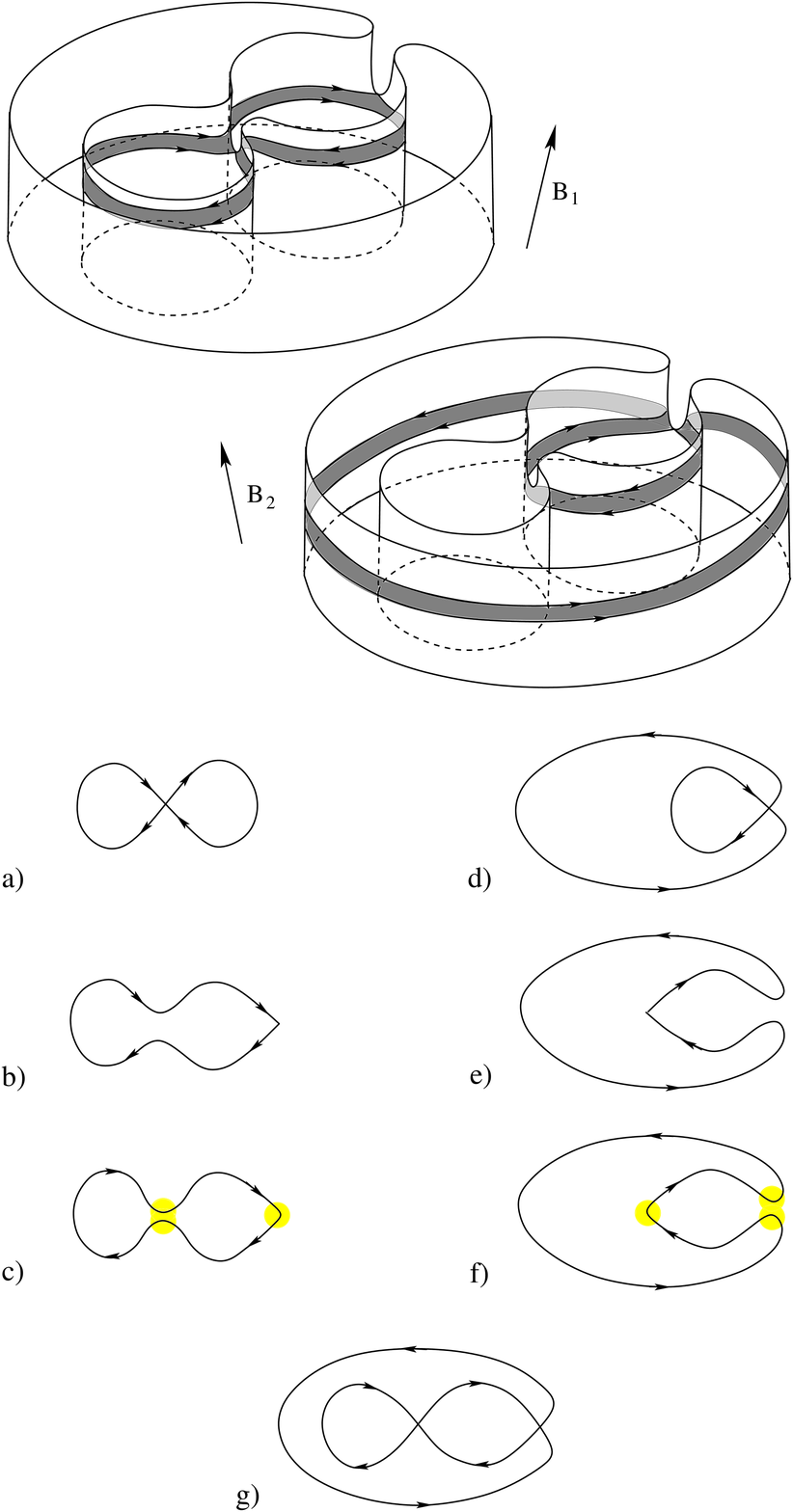}
\end{center}
\caption{A reconstruction of the structure of system 
(\ref{MFSyst}) on a part of the Fermi surface.  
The lower and upper bases of the disappearing cylinder of closed 
trajectories and the extremal trajectory arising on it (a-c). 
The lower and upper bases of the appearing cylinder of closed 
trajectories and the extremal trajectory arising on it (d-f).
The ``cylinder of zero height'' that appears at the moment of 
reconstruction (g).}
\label{fig36}
\end{figure}

\begin{figure}[t]
\begin{center}
\includegraphics[width=\linewidth]{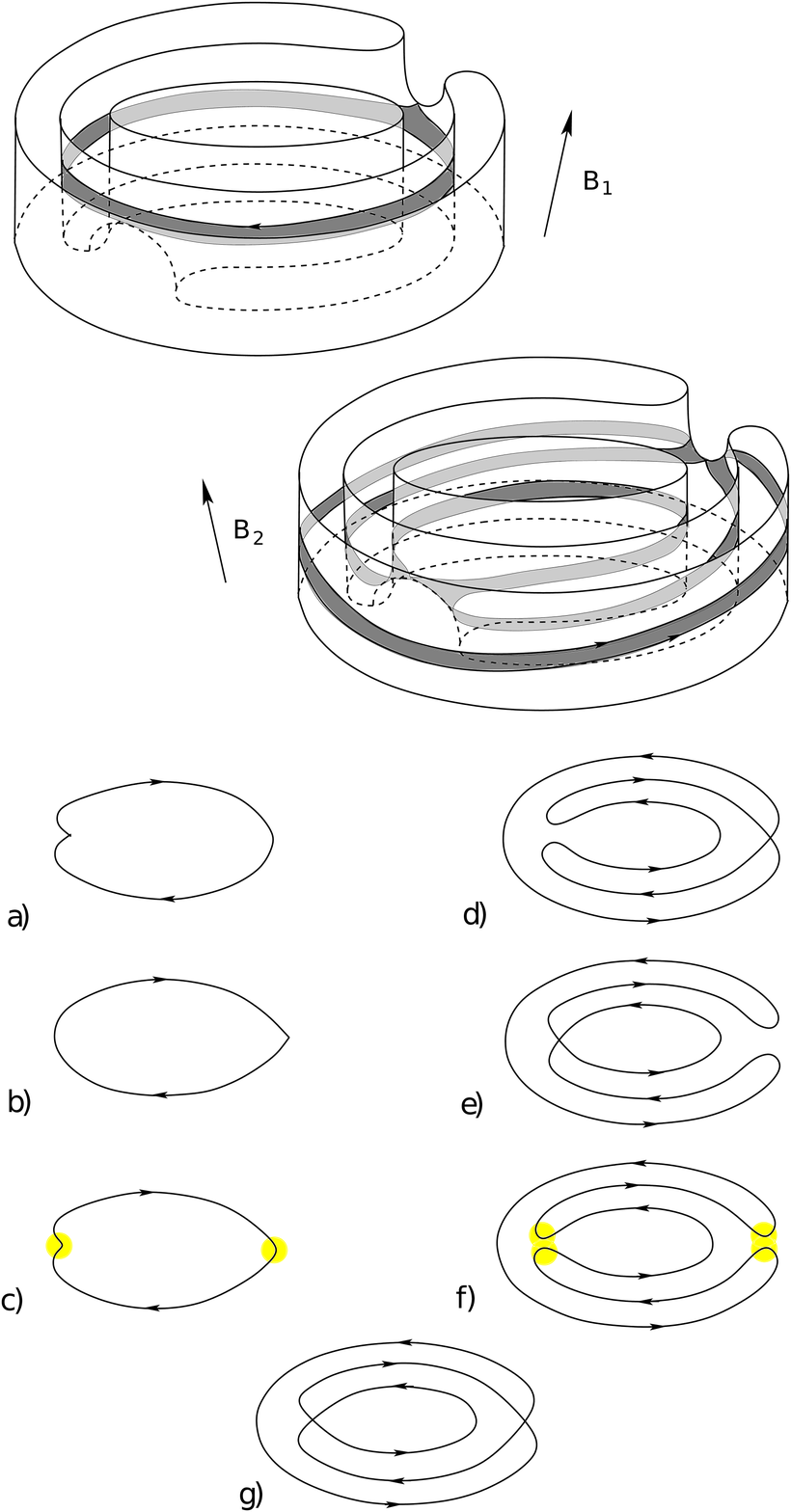}
\end{center}
\caption{A reconstruction of the structure of system 
(\ref{MFSyst}) on a part of the Fermi surface.  
The lower and upper bases of the disappearing cylinder of closed 
trajectories and the extremal trajectory arising on it (a-c). 
The lower and upper bases of the appearing cylinder of closed 
trajectories and the extremal trajectory arising on it (d-f).
The ``cylinder of zero height'' that appears at the moment of 
reconstruction (g).}
\label{fig37}
\end{figure}

\begin{figure}[t]
\begin{center}
\includegraphics[width=\linewidth]{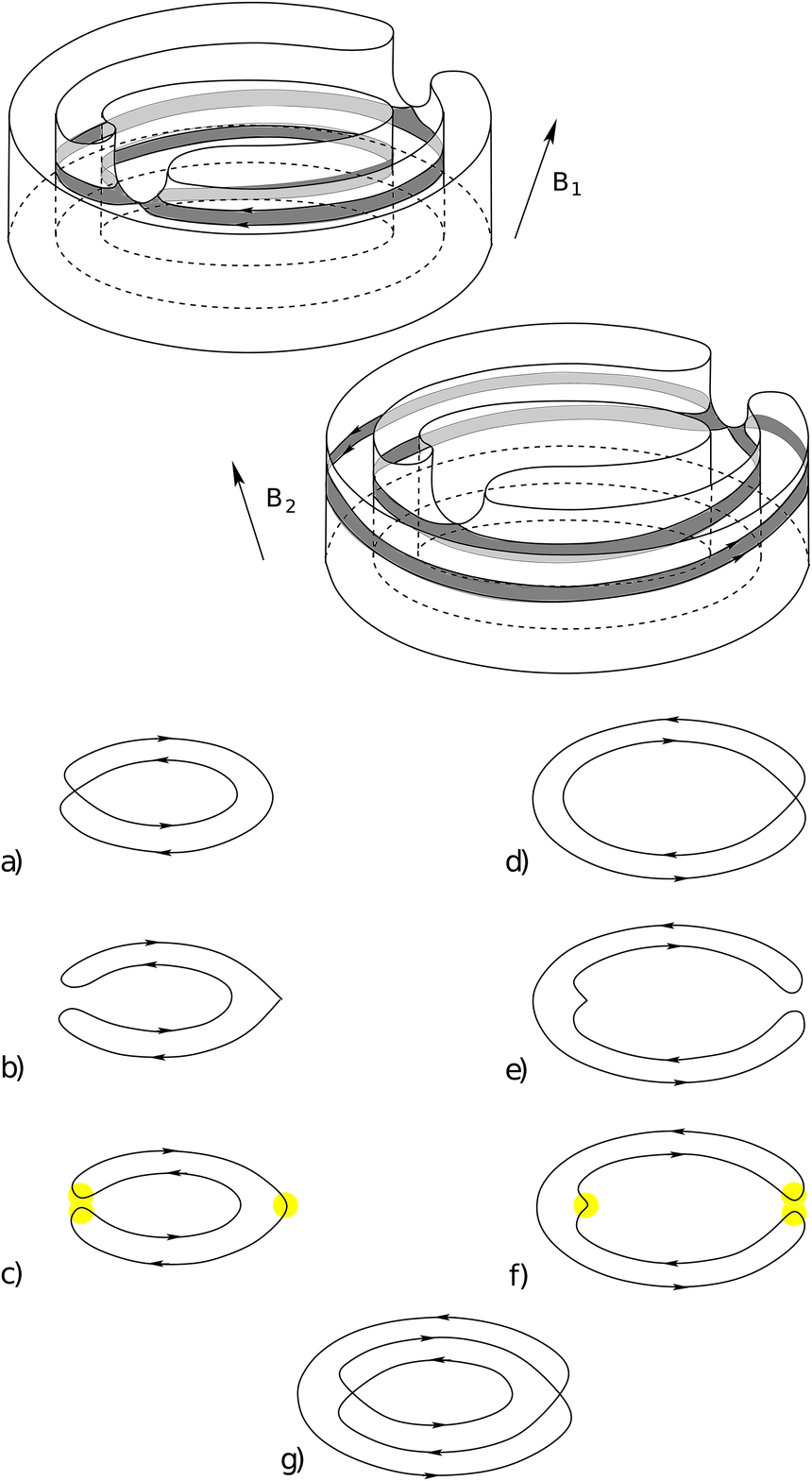}
\end{center}
\caption{A reconstruction of the structure of system 
(\ref{MFSyst}) on a part of the Fermi surface.  
The lower and upper bases of the disappearing cylinder of closed 
trajectories and the extremal trajectory arising on it (a-c). 
The lower and upper bases of the appearing cylinder of closed 
trajectories and the extremal trajectory arising on it (d-f).
The ``cylinder of zero height'' that appears at the moment of 
reconstruction (g).}
\label{fig38}
\end{figure}

\begin{figure}[t]
\begin{center}
\includegraphics[width=\linewidth]{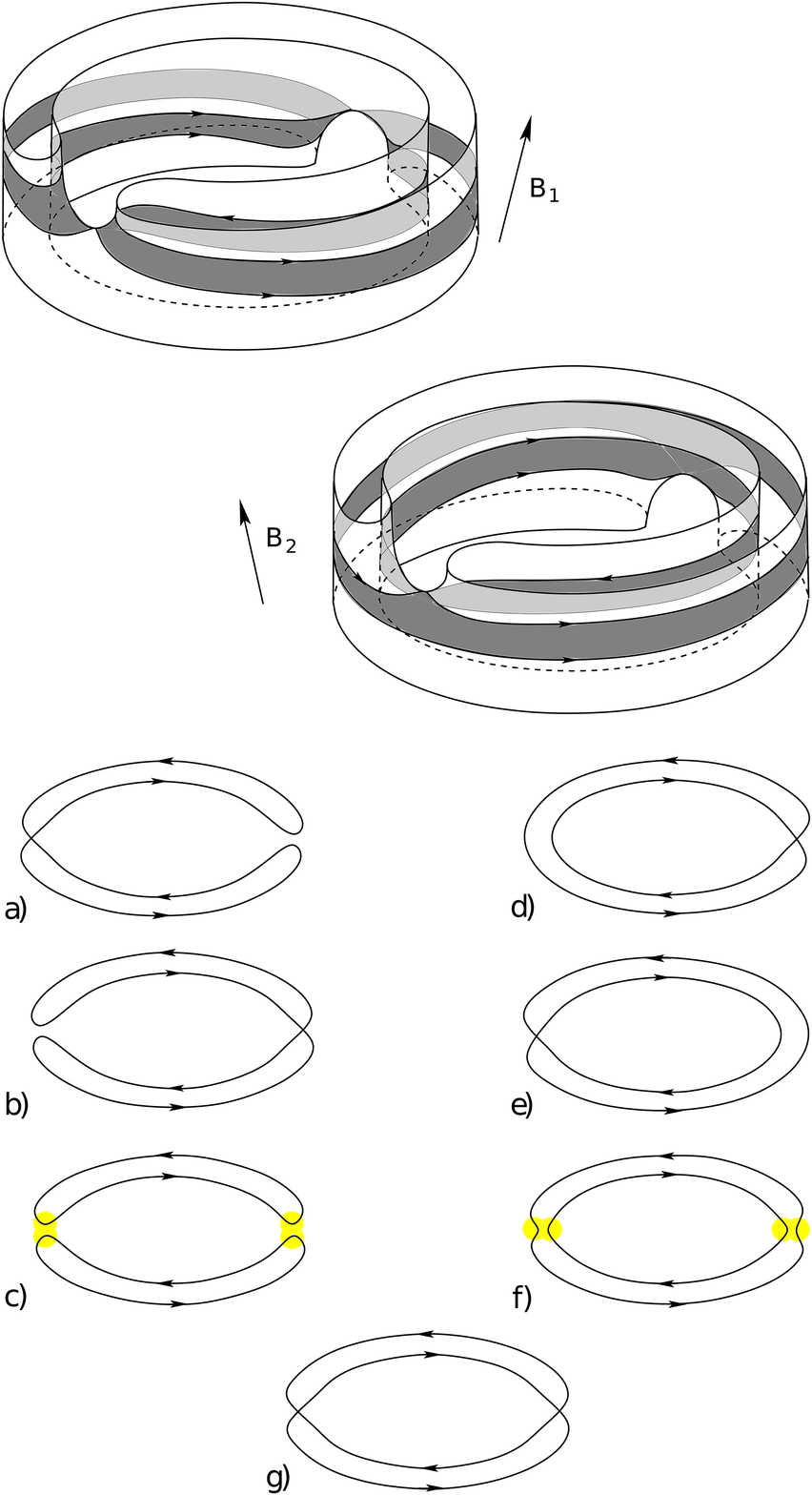}
\end{center}
\caption{A reconstruction of the structure of system 
(\ref{MFSyst}) on a part of the Fermi surface.  
The lower and upper bases of a pair of disappearing cylinders 
of closed trajectories and a pair of extremal trajectories 
arising on them (a-c). The lower and upper bases of a pair of 
appearing cylinders of closed trajectories and a pair of 
extremal trajectories arising on them (d-f).
The ``cylinder of zero height'' that appears at the moment of 
reconstruction (g).}
\label{fig39}
\end{figure}

\begin{figure}[t]
\begin{center}
\includegraphics[width=\linewidth]{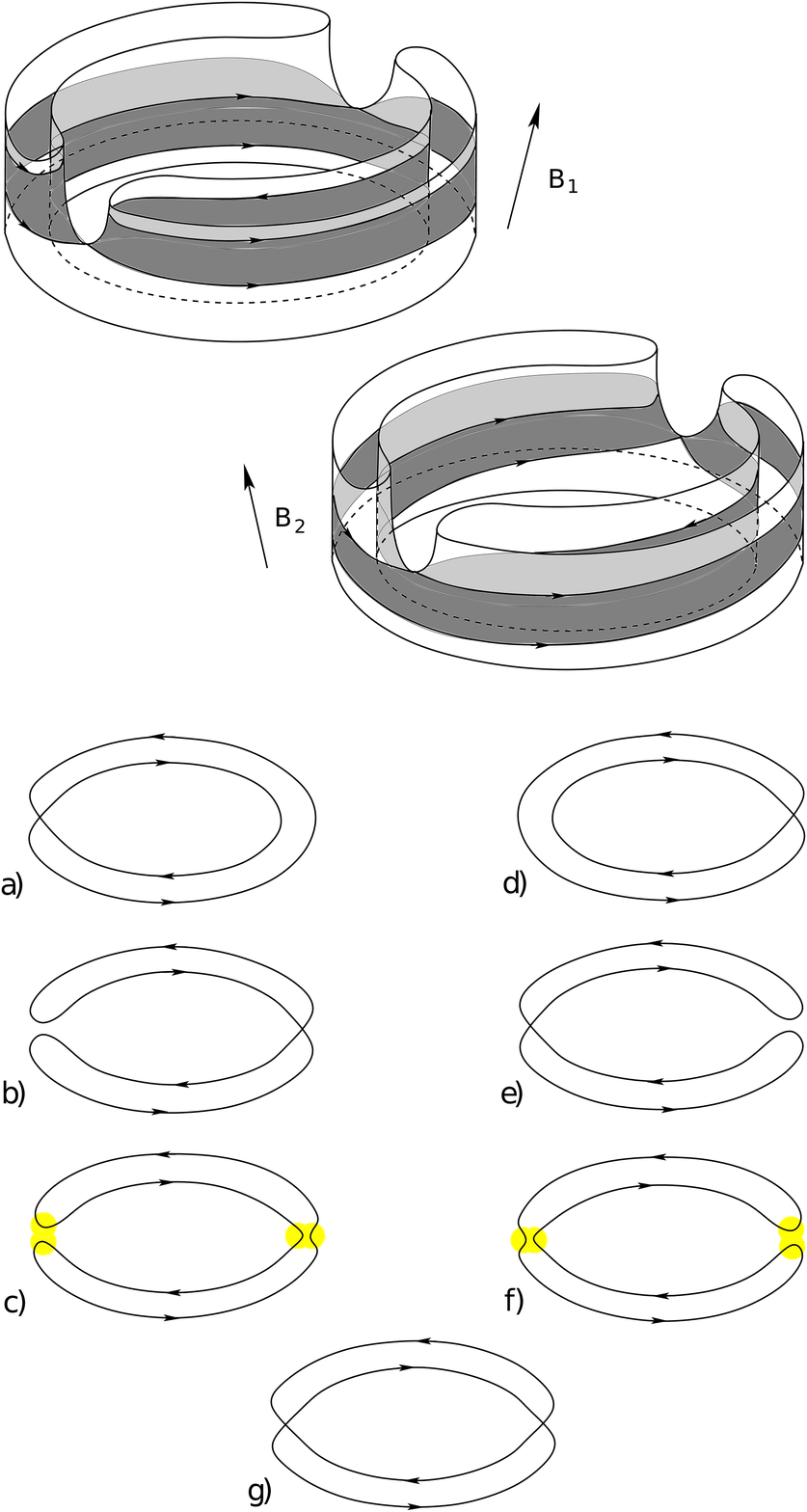}
\end{center}
\caption{A reconstruction of the structure of system 
(\ref{MFSyst}) on a part of the Fermi surface.  
The lower and upper bases of the disappearing cylinder of closed 
trajectories and the extremal trajectory arising on it (a-c). 
The lower and upper bases of the appearing cylinder of closed 
trajectories and the extremal trajectory arising on it (d-f).
The ``cylinder of zero height'' that appears at the moment of 
reconstruction (g).}
\label{fig40}
\end{figure}

 So, in particular, from this point of view, diagrams (a) 
and (b) at Fig. \ref{fig29} correspond to the same 
topological type of reconstruction shown at 
Fig. \ref{fig25} - \ref{fig26}, and diagrams (c) 
and (d) are of the topological type shown at 
Fig. \ref{fig27} - \ref{fig28}. 
Fig. \ref{fig30} presents all different types of
``cylinders of zero height'', which can arise in elementary 
reconstructions of the structure of the system (\ref{MFSyst}) 
on the Fermi surface. At Fig. \ref{fig31} - \ref{fig40} 
one can see the corresponding changes in the topological structure 
of the system (\ref{MFSyst}) for different orientations of the 
group velocity at saddle singular points for cases other than 
those presented at Fig. \ref{fig25} - \ref{fig26} 
and \ref{fig27} - \ref{fig28}.

 Among the presented topological types of reconstructions of the 
structure of system (\ref{MFSyst}) we can immediately distinguish 
cases of possible central symmetry of the corresponding section 
of the Fermi surface. It is easy to see that the presented sections 
of the Fermi surface cannot possess central symmetry if it is not 
possessed by the corresponding ``cylinders of zero height''.
In addition, central symmetry cannot be observed also in those 
cases when the group velocities at the singular points on the 
``cylinder of zero height'' are co-directed with each other. 
All the corresponding types of reconstructions of the system 
(\ref{MFSyst}) can, therefore, arise only in pairs on the Fermi 
surface, which, in turn, requires its sufficient complexity.

 It can be stated, therefore, that of the types of reconstruction
presented at Fig. \ref{fig31} - \ref{fig40}, only the 
reconstructions shown at Fig. \ref{fig33} and \ref{fig39} 
can have central symmetry. It is easy to see at the same time that 
the reconstruction shown at Fig. \ref{fig39} leads to the 
disappearance of a pair of cylinders of closed trajectories and 
the appearance of a pair of new cylinders, which, in turn, also 
requires sufficient complexity of the Fermi surface.

\vspace{1mm}

\begin{figure}[t]
\begin{center}
\includegraphics[width=0.9\linewidth]{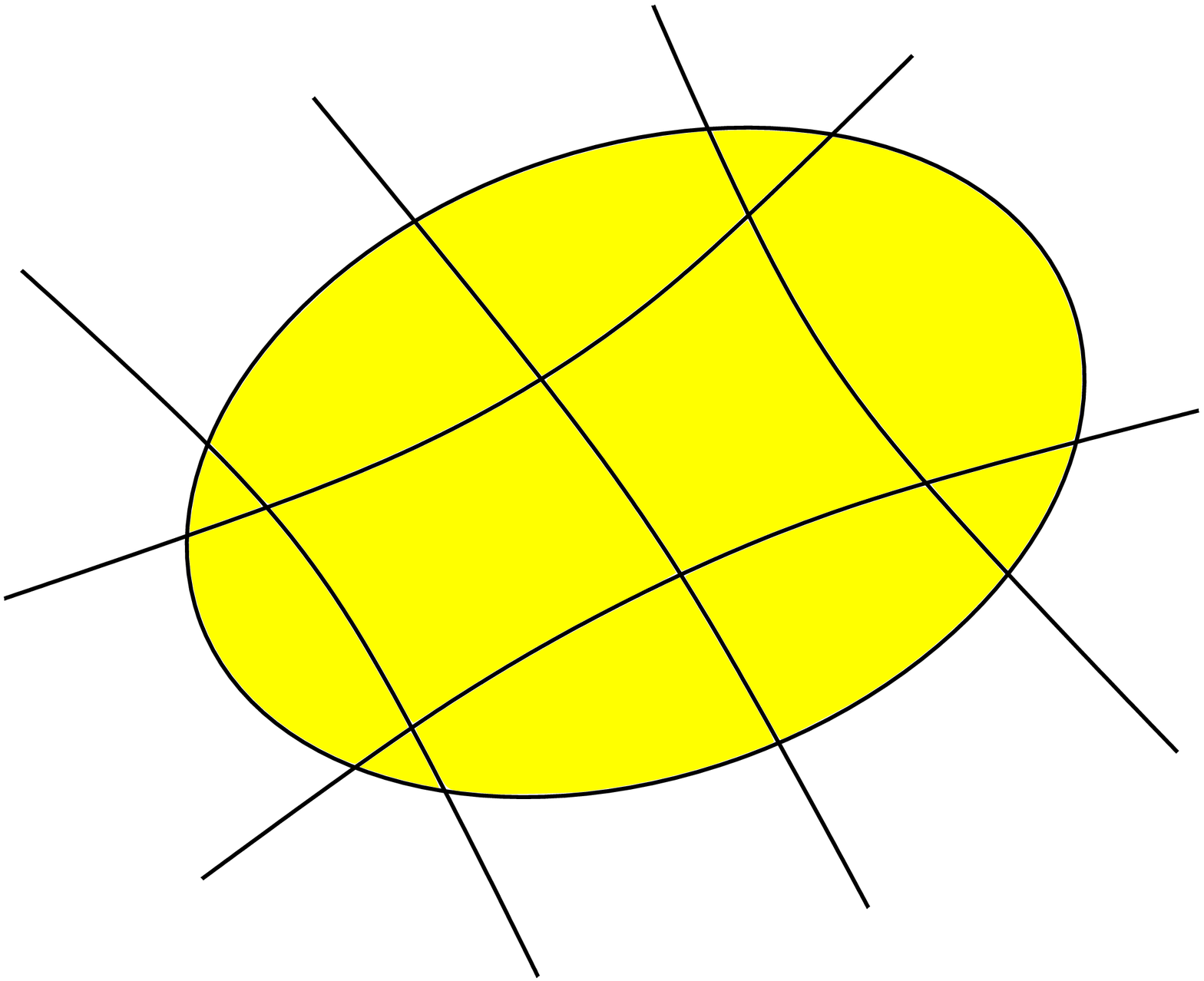}
\end{center}
\caption{The domain on the sphere $\, \mathbb{S}^{2} \, $, 
corresponding to the existence of a certain cylinder of closed 
trajectories on the Fermi surface (schematically).}
\label{fig41}
\end{figure}

 To all of the above, we can add one more remark about the 
complete structure of the set of directions of $ \, {\bf B} \, $, 
corresponding to the reconstructions of the structure of 
system (\ref{MFSyst}). Namely, consider some structure of the 
system (\ref{MFSyst}) on the Fermi surface containing a certain 
number of cylinders of closed trajectories. Then for each of the 
cylinders of closed trajectories it is possible to determine 
the region in the space of directions of $\, {\bf B} \, $ 
(on the sphere $\, \mathbb{S}^{2} $) corresponding to the 
conservation of this cylinder (Fig. \ref{fig41}). The boundary 
of this region represents a (generally piecewise smooth) curve, 
at the intersection of which the selected cylinder disappears. 
The net of directions of $\, {\bf B} \, $ we consider (including 
the boundaries of the Stability Zones) is the union of all the 
boundaries of the regions introduced in this way (for all 
different structures of (\ref{MFSyst}) on the Fermi surface) 
intersecting with each other.

\section{On the connection of the diagram of reconstructions
of the structure of (\ref{MFSyst}) with the geometry of the Fermi 
surface.}
\setcounter{equation}{0}

\begin{figure}[t]
\begin{center}
\includegraphics[width=\linewidth]{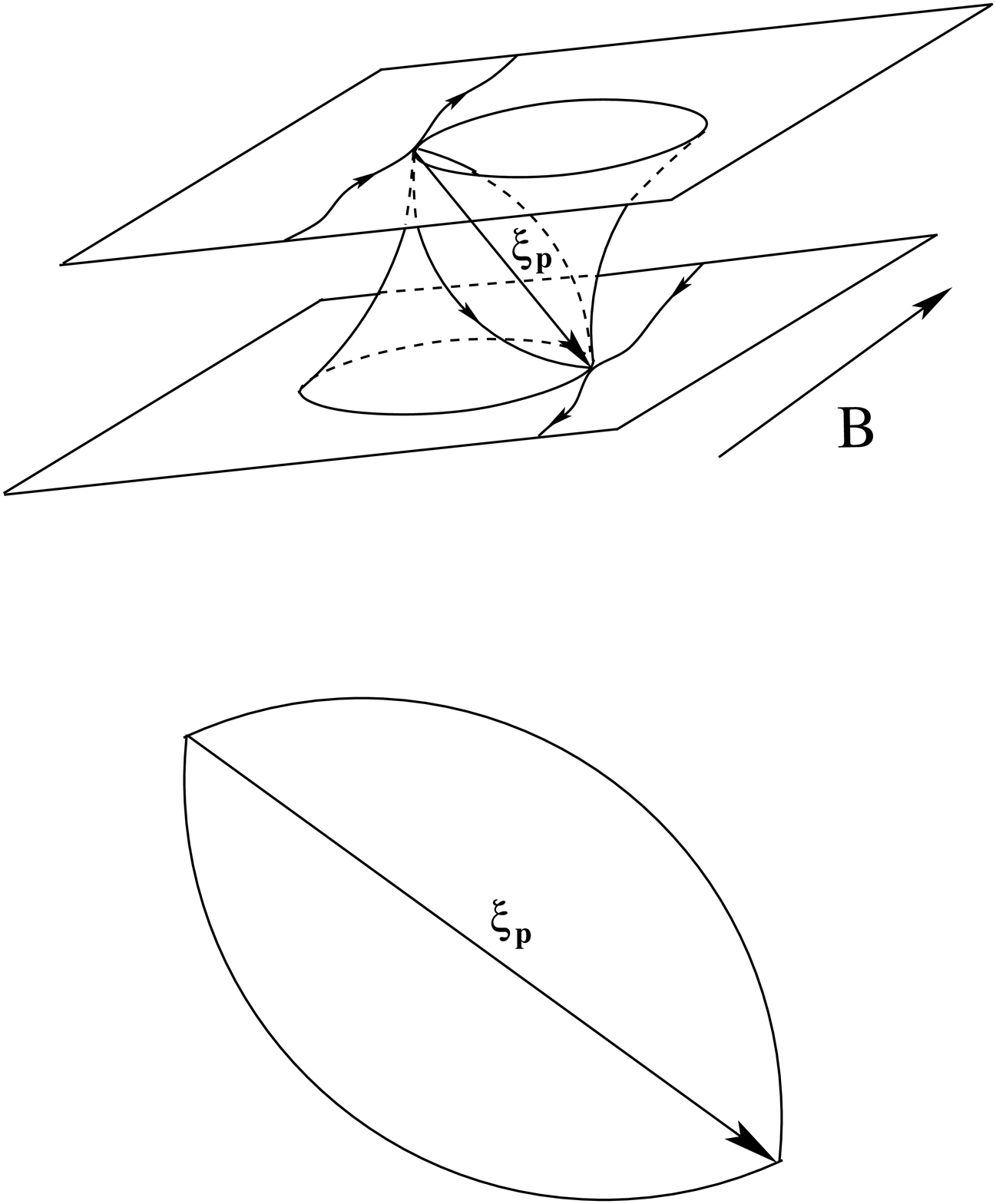}
\end{center}
\caption{The vector $\, \boldmath{\xi}_{{\bf p}} \, $ connecting 
two singular points on a special singular trajectory in the
$\, {\bf p}$ - space.}
\label{fig42}
\end{figure}

 Here we briefly describe how the picture of reconstructions of 
the structure of system (\ref{MFSyst}) on the angular diagram and 
the observation of oscillation phenomena on special extremal 
trajectories can be used to reconstruct the geometry of the Fermi 
surface.  As we have already said, the main feature of the special 
extremal trajectories is their very close approximation to singular 
trajectories for the directions of $\, {\bf B} \, $ close to 
directions of the reconstruction of the structure of (\ref{MFSyst}).
The limit of each of the special extremal trajectories is a part 
(or the whole cylinder) of a ``cylinder of zero height'', 
representing two singular points connected by singular trajectories.
Let us connect the corresponding singular points of (\ref{MFSyst}) 
with a vector (segment) $\, \boldmath{\xi}_{{\bf p}} \, $ lying in 
the plane orthogonal to $\, {\bf B} \, $ 
(see e.g. Fig. \ref{fig42}).

\begin{figure}[t]
\begin{center}
\includegraphics[width=0.9\linewidth]{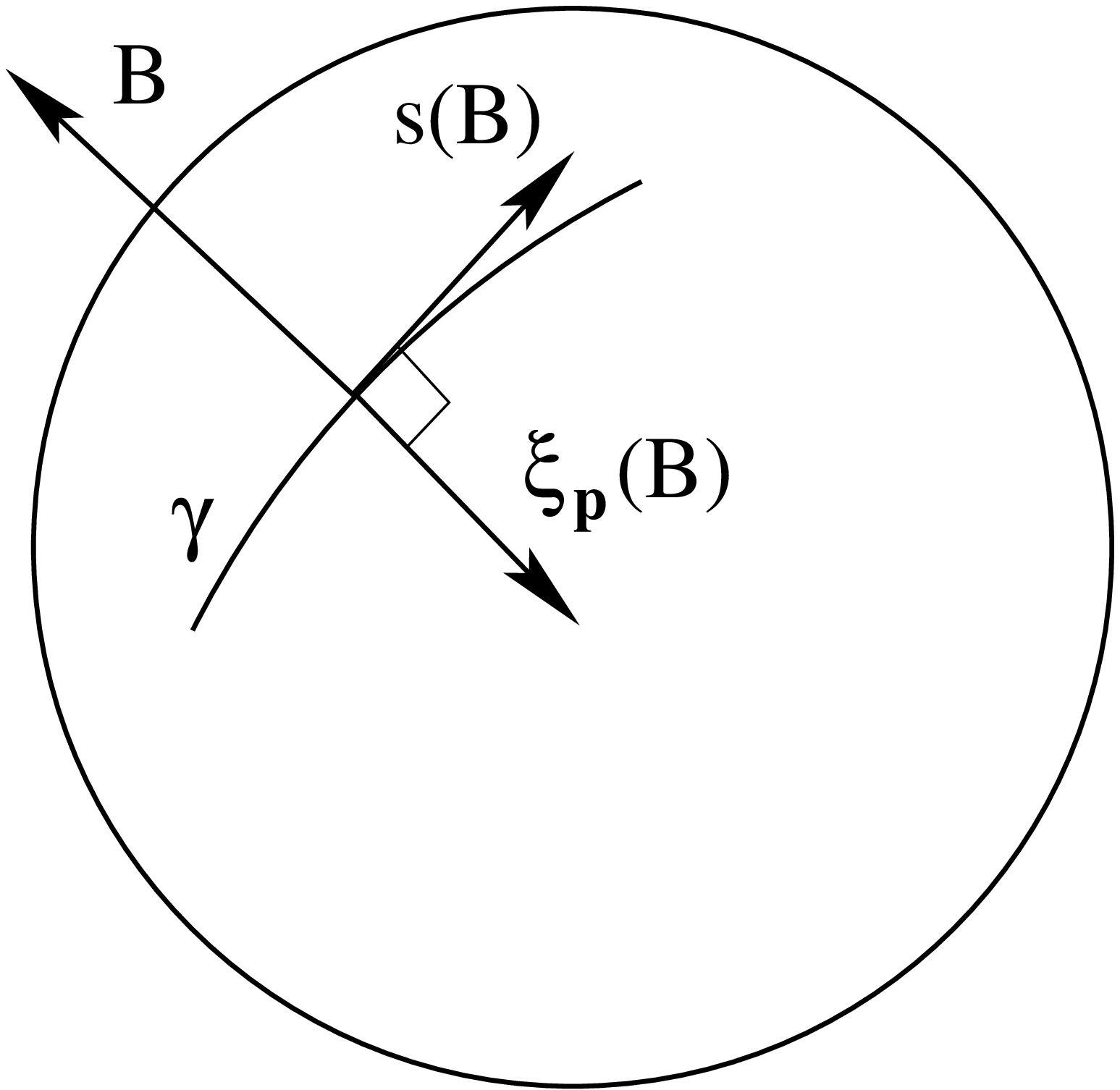}
\end{center}
\caption{The direction of the vector 
$\, \boldmath{\xi}_{{\bf p}} \, $ corresponding to a given 
direction of $\, {\bf B} \, $ on a curve of the change of 
the structure of system (\ref{MFSyst}) on the Fermi surface. }
\label{fig43}
\end{figure}

 As we saw above, the set of directions of $\, {\bf B} \, $, 
corresponding to reconstructions of the structure of system 
(\ref{MFSyst}) on the Fermi surface, is in the general case 
a union (generally speaking, of an infinite number) of 
one-dimensional curves on the unit sphere. It is easy to see 
that small rotations of $\, {\bf B} \, $ along the direction 
of $\, \boldmath{\xi_{{\bf p}}} \, $ correspond to linear 
variations in the height of the corresponding cylinder of 
closed trajectories, while the rotations of $\, {\bf B} \, $ 
in the direction orthogonal to $\, \boldmath{\xi}_{{\bf p}} \, $ 
do not change the height of the cylinder in the linear 
approximation. One can see, therefore, that the tangent 
to the arc on $\, \mathbb{S}^{2} \, $, corresponding to a 
reconstruction of the topological structure of (\ref{MFSyst}), 
is always orthogonal to the corresponding vector 
$\, \boldmath{\xi}_{{\bf p}} \, $ defined above. 
In other words, for a direction of $\, {\bf B} \, $ lying on 
an arc $\, \gamma \, $, corresponding to the disappearance of 
a certain cylinder of closed trajectories on the Fermi surface, 
the corresponding vector 
$\, \boldmath{\xi}_{{\bf p}} ({\bf p}) \, $ in 
$\, {\bf p}$ - space is parallel to the vector 
$\, \, {\bf B} \times {\bf s} ({\bf B}) \, $, where
$\, {\bf s} ({\bf B}) \, $ is the tangent vector to the 
arc $\, \gamma \, $ on $\, \mathbb{S}^{2} \, $ 
(Fig. \ref{fig43}).

 When the direction of $\, {\bf B} \, $ moves along the curve 
of the change of the structure of system (\ref{MFSyst}), the 
beginning and end of the vector $\, \boldmath{\xi}_{{\bf p}} \, $ 
move along one-dimensional curves $\, \hat{\gamma}_{1,2} \, $ in 
the $\, {\bf p} $ - space, and at each point of these curves
the vector $\, \boldmath{\xi}_{{\bf p}} \, $ is tangent to the 
Fermi surface. The direction of the normal to the Fermi surface 
(i.e., $\, {\bf v}_{gr} ({\bf p}) $) coincides at these points 
with the corresponding direction of $\, {\bf B} \, $ and we can 
also say that the Gauss map
$$S_{F} \, \, \, \rightarrow \, \, \, \mathbb{S}^{2} $$
maps the curves $\, \hat{\gamma}_{1} \, $ and 
$\, \hat{\gamma}_{2} \, $ to the curve $\, \gamma \, $ and the 
diametrically opposite curve on $\, \mathbb{S}^{2} \, $.

 Unlike trajectories of the system (\ref{MFSyst}), the curves 
$\, \hat{\gamma}_{1,2} \, $ are not, generally speaking, plane 
curves in the $\, {\bf p} $ - space.

\begin{figure}[t]
\begin{center}
\vspace{5mm}
\includegraphics[width=\linewidth]{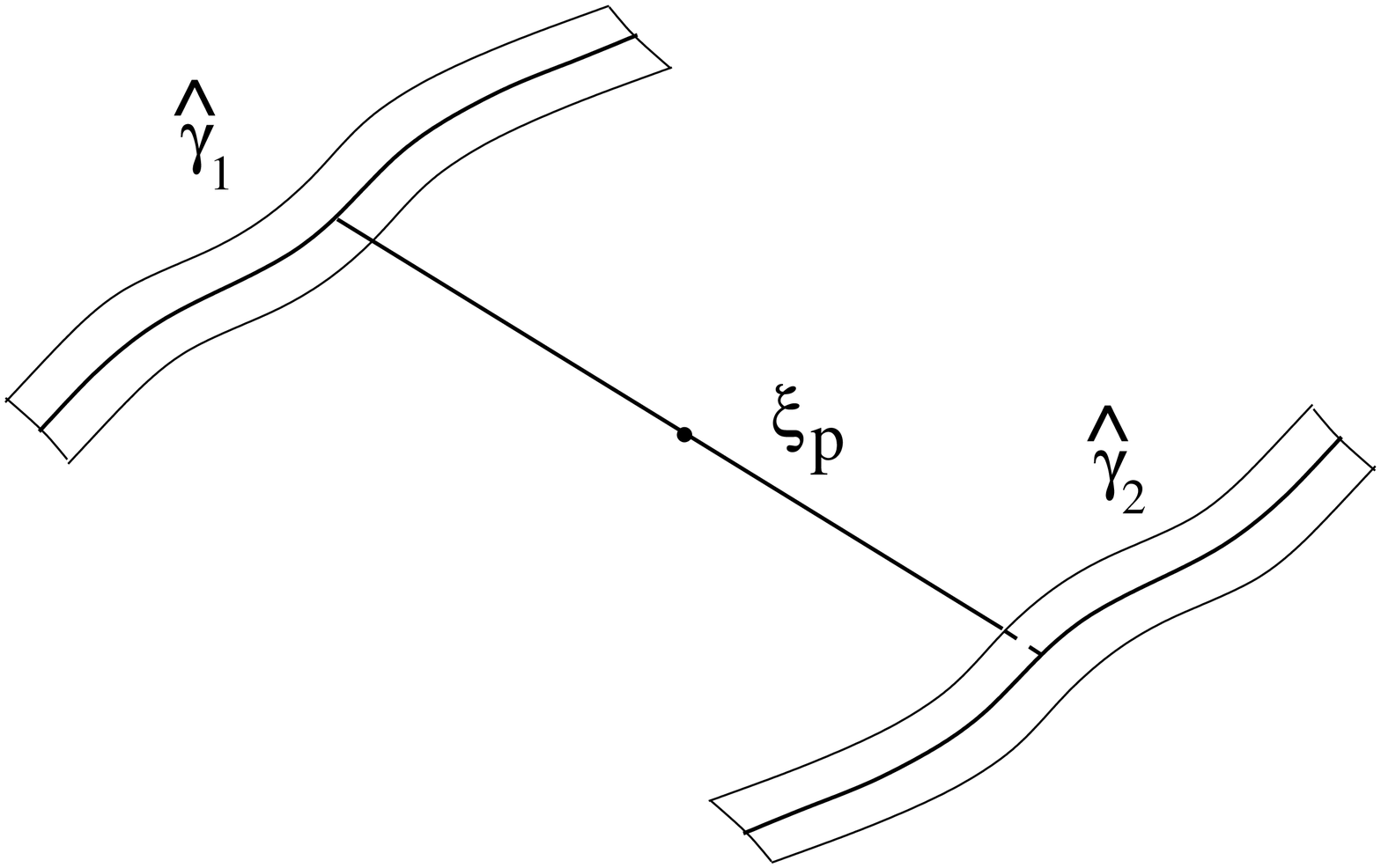}
\end{center}
\caption{Two infinitely narrow bands on the Fermi surface defined 
by the movement of the ends of the segment 
$\, \boldmath{\xi}_{{\bf p}} \, $ when the direction of 
$\, {\bf B} \, $ rotates along a curve of the change of the 
structure of system (\ref{MFSyst}).}
\label{fig44}
\end{figure}

  If the ``disappearing'' cylinder of closed trajectories 
corresponding to the arc $\, \gamma \, $ has central symmetry, 
then the center of the segment $\, \boldmath{\xi}_{{\bf p}} \, $ 
remains motionless when the direction of $\, {\bf B} \, $ moves 
along $\, \gamma \, $ and coincides with one of the centers of 
symmetry of the Fermi surface in $\, {\bf p} $ - space. We can 
see that in this case the exact knowledge of the shape of the 
curve $\, \gamma \, $ (for example, the first or second boundary 
of a Stability Zone), as well as the length of the vector 
$\, \boldmath{\xi}_{{\bf p}} \, $ for each of the points of 
$\, \gamma \, $, allows us to restore two ``infinitely narrow bands'' 
on the Fermi surface corresponding to the movement of the end points 
of the segment $\, \boldmath{\xi}_{{\bf p}} \, $ 
(Fig. \ref{fig44}). (We note here that the interior points of 
$\, \boldmath{\xi}_{{\bf p}} \, $ can intersect the Fermi surface 
in this situation).

 One can see, therefore, that on the arcs $\, \gamma \, $, 
corresponding to the disappearance of the centrally symmetric 
cylinders of closed trajectories, we can indicate an effective 
procedure for reversing the Gauss map and restoring the inverse 
images $\, \hat{\gamma}_{1,2} \, $ on the Fermi surface along 
with the direction of the normal to $\, S_{F} \, $. As we could 
see above, for sufficiently complex Fermi surfaces, the sets of 
directions of $\, {\bf B} \, $, corresponding to reconstructions 
of the structure of system (\ref{MFSyst}), form a fairly dense 
net at the angular diagram, which corresponds to an equally dense 
net of inverse images $\, \hat{\gamma}_{1,2} \, $ on the Fermi 
surface. The curves $\, \hat{\gamma}_{1,2} \, $ are formed by 
saddle singular points of the system (\ref{MFSyst}) and, 
therefore, always lie on a part of the Fermi surface that 
has negative (Gaussian) curvature. It can be noted here that, 
with a more general consideration, the methods for reconstructing 
the Fermi surface, based on the study of the presented special 
extremal trajectories, are connected, first of all, with that 
part of the Fermi surface where it has negative Gaussian 
curvature.

 In a more general situation, when the experimental design does 
not imply a separate measurement of the length of the vector 
$\, \boldmath{\xi}_{{\bf p}} \, $, each curve $\, \gamma \, $, 
corresponding to the disappearance of a centrally symmetric cylinder 
of closed trajectories, gives us a (one-parameter) family of lines 
passing through a given center of symmetry and tangent to the 
Fermi surface.  Each such family provides important information 
about the Fermi surface (its part having negative curvature), which 
is also convenient to use to refine its shape. 

\begin{figure}[t]
\begin{center}
\includegraphics[width=0.95\linewidth]{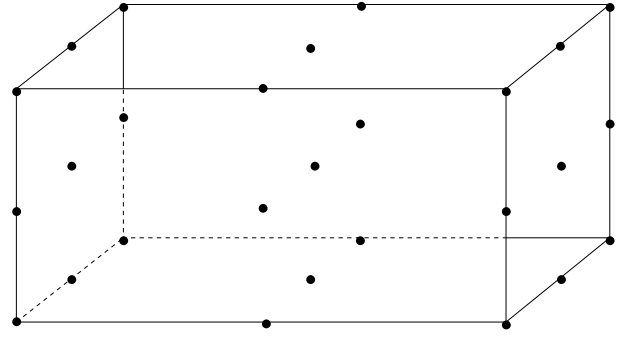}
\end{center}
\caption{Brillouin zone and symmetry centers of a dispersion 
relation $\, \epsilon ({\bf p}) \, $ in $\, {\bf p} $ - space.}
\label{fig45}
\end{figure}

 In considering the formation and disappearance of centrally 
symmetric cylinders of closed trajectories, we must, of course, 
discuss also the following question. Considering centers of symmetry 
of a periodic dispersion relation $\, \epsilon ({\bf p}) \, $, 
we can immediately say that there are always actually several such 
symmetry centers (nonequivalent to each other) in the
$\, {\bf p} $ - space. It is easy to see that in the most general 
case it is possible to choose the Brillouin zone (parallelepiped in 
$\, {\bf p} $ - space) so that such centers are: the center of the
parallelepiped, the centers of its faces, the centers of its edges 
and its vertices (Fig. \ref{fig45}).

 The centers of the opposite faces of the parallelepiped are 
pairwise equivalent to each other, the centers of the edges form 
3 fours of equivalent centers, and all the vertices of the 
parallelepiped represent one center after factorization by the 
reciprocal lattice vectors. Thus, in the general case, we always 
have 8 nonequivalent centers of symmetry in the 
$\, {\bf p} $ - space. In other words, we have 8 nonequivalent 
periodic families of centers of symmetry that transform into each 
other when shifted by half-integer reciprocal lattice vectors.

 The presence of the central symmetry in a cylinder of closed 
trajectories means that when reflected with respect to any of 
the centers of symmetry, such a cylinder transforms into an 
equivalent cylinder (shifted by a reciprocal lattice vector) in 
$\, {\bf p} $ - space. For each such cylinder there is a unique 
point in $\, {\bf p} $ - space, the reflection with respect to 
which takes the cylinder into itself. It is easy to see that for 
the construction of the curves $\, \hat{\gamma}_{1,2} \, $ 
described above, we need, among other things, to know which 
of the periodic families of centers of symmetry this point 
belongs to.

 It can also be seen that the correspondence between the 
centrally symmetric cylinders of closed trajectories and their 
``own'' centers of symmetry in the $\, {\bf p} $ - space 
remains unchanged until any of these cylinders disappears and, 
therefore, also applies to the definition of the topological 
structure of the system (\ref{MFSyst}) on the Fermi surface.  
In addition, for any elementary reconstruction of the structure 
of (\ref{MFSyst}), the disappearance of one centrally symmetric
cylinder of closed trajectories also leads to the appearance 
of a centrally symmetric cylinder of closed trajectories with 
the same ``own'' center of symmetry (also coinciding with the 
``own'' center of symmetry of the corresponding 
``cylinder of zero height''). In particular, we can see that if 
for two directions of $\, {\bf B} \, $:

1) it is possible to move from one direction to another with 
a finite number of elementary reconstructions of the structure 
of (\ref{MFSyst});

2) each of the appearing ``cylinders of zero height'' during the 
motion corresponds to the appearance and disappearance of only one 
cylinder of closed trajectories; 

then the structure of the system (\ref{MFSyst}) for such directions 
contains the same number of centrally symmetric cylinders of closed 
trajectories with the same ``own'' centers of symmetry. 

 Thus, under the above conditions, connecting two directions 
$\, {\bf B}_{1} \, $ and $\, {\bf B}_{2} \, $ by a suitable curve 
at the angular diagram, we can establish a correspondence between 
the centrally symmetric cylinders of closed trajectories that occur 
for both directions, matching the disappearing cylinder the one that 
appears during each reconstruction. In particular, in such a 
correspondence, we can indicate, for example, the ``own'' symmetry 
centers of such cylinders for the direction $\, {\bf B}_{2} \, $, 
if they were known for the direction $\, {\bf B}_{1} \, $.

 This circumstance may be convenient in the experimental study 
of reconstructions of the structure of system (\ref{MFSyst}), 
since it allows one to obtain additional data on cylinders of 
closed trajectories for the directions of $\, {\bf B} \, $, 
for which an independent study of the structure of (\ref{MFSyst}) 
can be difficult. In particular, if approximate information about 
the Fermi surface geometry is sufficient to determine the topological 
structure of system (\ref{MFSyst}) for the direction 
$\, {\bf B}_{1}\,$, then by observing the reconstructions along 
the path connecting $\, {\bf B}_{1} \, $ and $\, {\bf B}_{2} \, $, 
we can also obtain the necessary data for the direction 
$\, {\bf B}_{2} \, $, where such information is not sufficient.
This situation is actually quite general, thus, the first case 
usually occurs in those areas of the angle diagram where the net of 
directions of $\, {\bf B}\, $ corresponding to the reconstructions 
of the structure of (\ref{MFSyst}) is not dense, and the second - 
in areas where such a net is very dense. In this case, each of the 
cylinders of closed trajectories can be associated with the 
corresponding contribution in the overall picture of oscillations, 
and we can continue this matching 
from the direction $\, {\bf B}_{1} \, $ to $\, {\bf B}_{2} \, $, 
associating in reconstructions the arising cylinder with a 
new oscillation term. This matching allows, in particular, 
to indicate the ``own'' center of symmetry of a centrally 
symmetric cylinder corresponding to any oscillation term 
observed at the direction $\, {\bf B}_{2} \, $.

 We note again that we represent the general picture of 
oscillations when observing oscillations of various types as 
the sum of a finite number of oscillating terms corresponding 
to extremal trajectories on the Fermi surface. As we noted above, 
the main feature in the behavior of such oscillations for us is 
a sharp change in the overall picture with a change in the 
topological structure of (\ref{MFSyst}), which allows us to 
observe the net of directions of $\, {\bf B} \, $ on 
$\, \mathbb{S}^{2} \, $, corresponding to changes in this 
structure. More precisely, as we have already said, a change 
in the oscillation picture during each ``elementary'' 
reconstruction of the system (\ref{MFSyst}) consists in the 
disappearance of one of the oscillation terms in the total sum 
of oscillations and its replacement by another one (or simply the 
disappearance of one of the oscillation terms at the boundary of 
a Stability Zone).

 It can be noted here again that, as we have already said, 
reconstructions containing ``cylinders of zero height'' 
corresponding to the disappearance and appearance of pairs of 
cylinders of closed trajectories can appear only in special 
cases on rather complex Fermi surfaces. For a large class of 
real Fermi surfaces, the above conditions will actually be 
satisfied for any two directions of $\, {\bf B} \, $, which 
can be connected by a path that does not intersect the boundaries 
of the Stability Zones or segments of the appearance of 
periodic trajectories.

\vspace{1mm}

 Coming back to the reconstructions corresponding to the 
disappearance and appearance of cylinders of closed trajectories 
that do not have central symmetry, we can immediately note that 
the number of such reconstructions at the angular diagram is 
rather small even for fairly complex Fermi surfaces. The 
determination of the shape of the corresponding curve 
$\, \gamma \, $ at the angular diagram also determines here 
the direction of the vector 
$\, \boldmath{\xi}_{{\bf p}} ({\bf B}) \, $ connecting two 
saddle singular points of the system (\ref{MFSyst}). In the 
$\, {\bf p} $ - space we have now two non-equivalent segments 
with the common direction 
$\, \boldmath{\xi}_{{\bf p}} ({\bf B}) \, $, each of which 
is tangent to the Fermi surface at two points. Thus, in this 
case, we can say that we have a segment of a given direction
which is tangent to the Fermi surface at two points that do not 
pass into each other upon reflection (and also a segment 
symmetrical to it). Such data on the Fermi surface look 
somewhat more complicated in comparison with the centrally 
symmetric case, however, we can verify that they have the same 
information content from the functional point of view.

 We can still make a small remark regarding the reconstructions, 
in which the disappearance (and appearance) of a pair of 
cylinders of closed trajectories occurs (Fig. \ref{fig39}).  
If the Fermi surface is not ultra-complex, it is natural to 
assume that in such reconstruction we observe a disappearance 
of a pair of cylinders with central symmetry, and the appearance 
of a pair of cylinders that pass into each other under the 
central reflection. From the experimental point of view, this 
reconstruction differs from the others by the disappearance of 
two oscillation terms in the total picture of oscillations and 
their replacement by one term (in fact, two coinciding ones 
coming from a pair of extremal trajectories that are symmetrical 
to each other). In this situation, as in the case of the 
disappearance of a single centrally symmetric cylinder, the 
segment $\, \boldmath{\xi}_{{\bf}} ({\bf B})\, $ is unique and 
passes through one of the symmetry centers in the 
$\, {\bf p}$ - space. The corresponding center of symmetry 
coincides with the ``own'' center of symmetry of the disappearing 
cylinders and the refinement of the geometry of the Fermi surface 
can be carried out in the same way as in the case of the 
disappearance of one centrally symmetric cylinder.

\vspace{1mm}

 Let us consider at the end of this chapter some aspects of 
observing the cyclotron resonance phenomenon on the special 
extremal trajectories, related to the peculiarities of their 
geometry.

\vspace{1mm}

 As is well known, electron trajectories in the 
$\, {\bf x} $ - space have geometric properties that are somewhat 
similar to the geometric properties of the trajectories of the 
system (\ref{MFSyst}) in the quasimomenta space. In particular, 
the projections of the trajectories in the $\, {\bf x} $ - space 
onto the plane orthogonal to $\, {\bf B} \, $ are similar to the 
trajectories in the $\, {\bf p} $ - space rotated by $90^{\circ}$.
The similarity coefficient between the trajectories in the momentum 
space and their projections onto the plane orthogonal to 
$\, {\bf B} \, $ in the coordinate space is equal to 
$\, c / eB \, $. In general, all the trajectories in the 
$\, {\bf x} $ - space decrease in size with the growth of 
$\, B \, $, like $\, B^{- 1} \, $, without changing their 
geometric shape.

\begin{figure}[t]
\begin{center}
\vspace{5mm}
\includegraphics[width=0.95\linewidth]{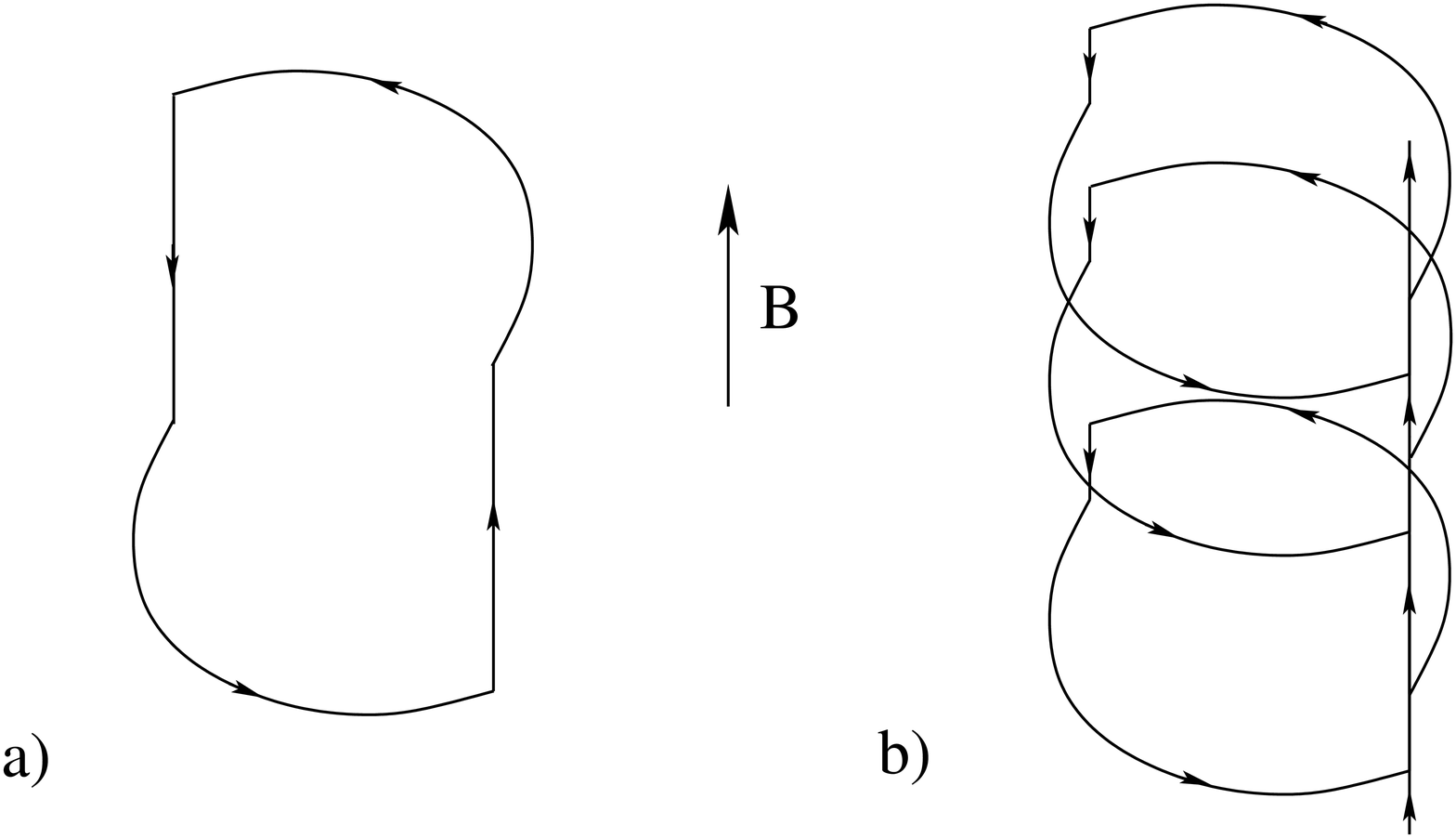}
\end{center}
\caption{Trajectories in $\, {\bf x} $ - space corresponding to 
special extremal trajectories of the system (\ref{MFSyst}) 
in the presence (a) and absence (b) of central symmetry. }
\label{fig46}
\end{figure}

 It is easy to see that the special trajectories we consider 
correspond in the $\, {\bf x} $ - space to either closed 
trajectories (in the presence of central symmetry) or spiral 
trajectories of a special shape (in the absence of central 
symmetry) that describe a constant drift along the magnetic 
field (see e.g. Fig. \ref{fig46}). In both cases, one 
can note the presence of almost vertical (parallel to 
$\, {\bf B} $) sections due to the presence of sections 
in $\, {\bf p} $ - space, close to singular points of the system 
(\ref{MFSyst}), at which the group velocity is almost parallel 
to $\, {\bf B} \, $ and the speed of motion along the trajectory 
is very small. Here we will consider the observation of cyclotron 
resonance at $\, {\bf B} \, $, parallel to the surface of the sample, 
so we can see that when such areas get into the skin layer 
(Fig. \ref{fig55}), the electrons can spend quite a lot 
of time in the skin layer, while moving mainly along the direction 
of $\, {\bf B} \, $. One can immediately note, therefore, that 
the amplitude of the absorption oscillations in the cyclotron 
resonance should in this situation be maximal when the electric 
field in the incident wave $\, {\bf E}_{w} \, $ is parallel to 
$\, {\bf B} \, $. This, in particular, can serve as one of the 
signs that distinguishes the situation under consideration from 
the general one (getting into the skin layer of an arbitrary part 
of the trajectory), when the maximal absorption is observed when 
$\, {\bf E}_{w} \, \parallel \, {\bf v}_{gr} \, $, where the 
direction of $\, {\bf v}_{gr} \, $ at the section falling into the 
skin layer is different from the direction of $\, {\bf B} \, $.

\begin{figure}[t]
\begin{center}
\vspace{5mm}
\includegraphics[width=\linewidth]{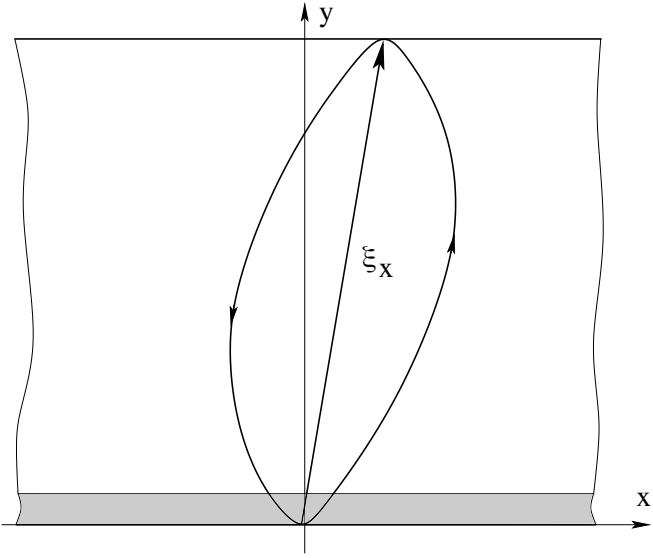}
\end{center}
\caption{Measuring of $\, | \boldmath{\xi}_{\bf x} | \, $ 
by cutting off cyclotron orbits for a special extremal 
trajectory of the most common geometry. }
\label{fig47}
\end{figure}

 The situation where the ``deceleration region'' gets into the 
skin layer can be typical, for example, when measuring 
$\, | \boldmath{\xi}_{\bf x} | \, $ (and with it 
$ \, | \boldmath{\xi}_{\bf p} | $) by cutting off cyclotron 
orbits (see \cite{Kaner}) or by a non-resonant size effect 
(\cite{Gantmaher}) for the most common special extremal 
trajectories (Fig. \ref{fig47}). We note here that 
such a method of measuring of $\, | \boldmath{\xi}_{\bf p} | \, $ 
is perhaps the most convenient, since, as we have already said, 
the direction of $\, \boldmath{\xi}_{\bf x} \, $ relative to the 
crystal lattice is known for any point on the curve of the 
reconstruction of the structure of (\ref{MFSyst}), so that the 
thickness of the film on which the extremal orbits are cut off 
can easily be correlated with the value of 
$\, | \boldmath{\xi}_{\bf x} | \, $.

\begin{figure}[t]
\begin{center}
\vspace{5mm}
\includegraphics[width=0.95\linewidth]{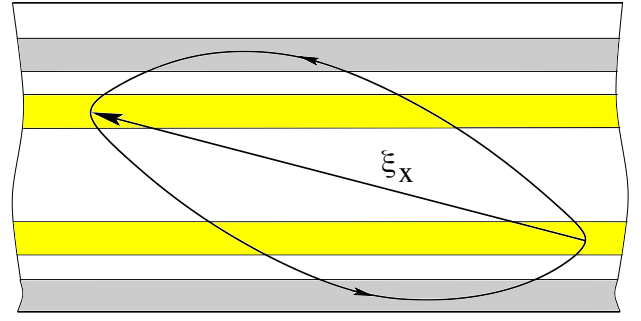}
\end{center}
\caption{Current layers of the first and second type generated 
by special extremal trajectories inside the sample.}
\label{fig48}
\end{figure}

 It can be noted here that the ``deceleration regions'' 
on the trajectories in $\, {\bf p} $ - space can also manifest 
themselves in the case when an ``ordinary'' trajectory segment 
gets into the skin layer in a cyclotron resonance situation.  
Namely, such sections can create narrow current layers inside 
the sample with the direction of the current along the magnetic 
field. Such current layers should arise together with the layers 
corresponding to the points of the trajectory at which the group 
velocity is parallel to the sample boundary, differing 
from them in the direction and amplitude of the arising 
currents. Each special extremal trajectory gives 
rise to two different current layers of the first type, 
corresponding to two saddle singular points present on the 
corresponding ``cylinder of zero height''. Like the layers of 
the second type, the layers corresponding to the 
``deceleration regions'' are parallel to the boundary of the 
sample (Fig. \ref{fig48}). The distance between two 
layers generated by two different ``deceleration regions'', as is 
easily seen, is equal to the length of projection of the vector 
$\, \boldmath{\xi}_{\bf x} \, $ on the normal to the sample 
boundary.

 It should also be noted here, in addition, that among the types 
of special extremal trajectories given in the previous chapter, only 
a part allows the deceleration section to enter the skin layer at 
the sample boundary, which is due to the local geometry of such 
trajectories near special points on the corresponding 
``zero-height cylinder''. Let us list these types here.

\begin{figure}[t]
\begin{center}
\vspace{5mm}
\includegraphics[width=0.7\linewidth]{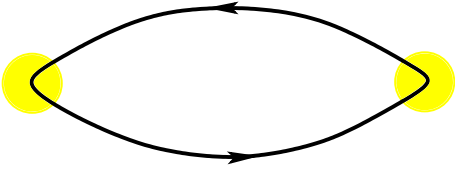}
\end{center}
\caption{A special extremal trajectory of the most common type.}
\label{fig49}
\end{figure}

 We begin here with the most basic type, corresponding to the 
trajectory arising at the reconstruction shown at
Fig. \ref{fig25} - \ref{fig26} 
(Fig. \ref{fig49}). Trajectories of the same type also arise 
during the reconstruction shown at Fig. \ref{fig12}, 
the reconstruction shown at Fig. \ref{fig35} (trajectory (c)), 
the reconstruction shown at Fig. \ref{fig39} (one of the 
trajectories (f)), as well as near the boundaries of the Stability 
Zones $\, \Omega_{\alpha} \, $. As for the reconstruction shown 
at Fig. \ref{fig25} - \ref{fig26}, and for the 
reconstructions shown at Fig. \ref{fig12}, \ref{fig35}, 
and Fig. \ref{fig39}, such trajectories occur only on one side 
from the line of reconstruction of the structure of (\ref{MFSyst}) 
(as well as only on the inner side of a Stability Zone boundary).
It is easy to see that any of the ``deceleration sections'' on the 
presented trajectory can enter the skin layer at a suitable 
orientation of the crystal lattice of the sample. Group velocities 
in two ``deceleration regions'' on such trajectories are directed 
opposite to each other and, as we have already said, in the 
overwhelming majority of cases, trajectories of this type possess 
central symmetry. In rare cases, however, such trajectories can also 
occur in pairs, turning into each other under the transformation of 
central symmetry. In the latter case, group velocities at different 
``deceleration sections'' can differ from each other in absolute 
value.

\begin{figure}[t]
\begin{center}
\vspace{5mm}
\includegraphics[width=0.9\linewidth]{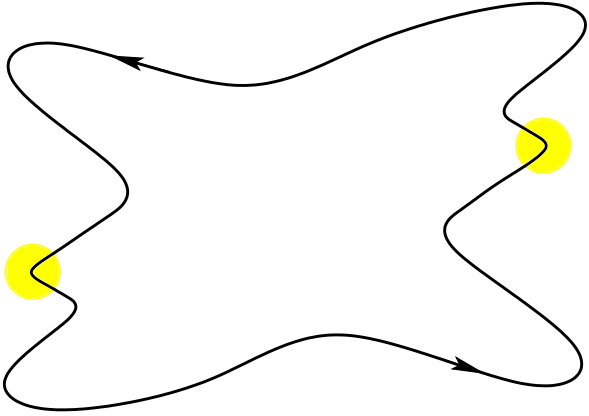}
\end{center}
\caption{A special extremal trajectory of complex shape that 
does not allow the ``deceleration sections'' to enter the skin 
layer.}
\label{fig50}
\end{figure}

 Let us make one more remark here. Namely, the trajectory shown 
at Fig. \ref{fig49} has, in fact, to some extent an ideal shape 
and can be significantly more complex geometrically in some cases.  
In particular, despite the consistency of the required situation 
(the ``deceleration section'' in the skin layer) with the local 
trajectory geometry, this situation can still be prohibited by 
the global properties of the trajectory (see for example, 
Fig. \ref{fig50}). It should be said that in such cases 
the shape of such a trajectory changes quite quickly when the 
direction of $\, {\bf B}\, $ moves along the corresponding curve 
of the reconstruction of the structure of system (\ref{MFSyst}) 
at the angular diagram. The situation we require then, as a rule, 
becomes possible not on the entire curve $\, \gamma \, $, but 
only on certain sections of it. We also note that the remark 
made will equally apply to the other types of trajectories 
considered below.

\begin{figure}[t]
\begin{center}
\vspace{5mm}
\includegraphics[width=\linewidth]{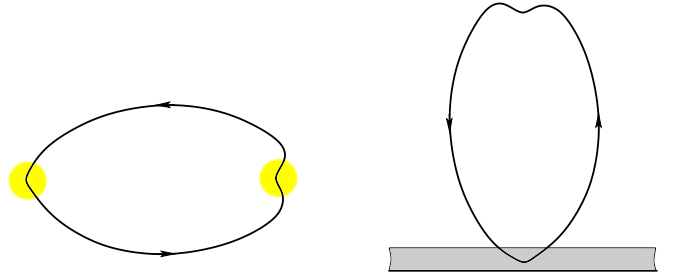}
\end{center}
\caption{A special extremal trajectory of the second type, 
allowing the ``deceleration section'' to enter the skin layer.}
\label{fig51}
\end{figure}

 The second type of special extremal trajectories that allow 
the ``deceleration section'' to enter the skin layer can include 
trajectories of similar shape, shown at Fig. \ref{fig32}(c) 
and \ref{fig37}(c) (Fig. \ref{fig51}). In both cases, 
trajectories of this shape arise only on one side of the 
corresponding reconstruction curve $ \, \gamma \, $ at the 
angular diagram. Both cases, presented at Fig. \ref{fig32} 
and \ref{fig37}, do not have central symmetry, therefore, 
the corresponding reconstructions can arise only in pairs and, 
in particular, each of the trajectories shown at 
Fig. \ref{fig32} (c) and \ref{fig37} (c) always arise 
in pair with a trajectory symmetric to it (located on a part of 
the Fermi surface symmetric to the one under consideration). 
However, there is also a significant difference between the 
trajectories shown at Fig. \ref{fig32} (c) and 
\ref{fig37} (c). Namely, for the trajectory shown at 
Fig. \ref{fig32} (c) the group velocities 
$\, {\bf v}_{gr} ({\bf p}) \, $ at the two 
``deceleration sections'' are co-directional, while for the 
trajectory, shown at Fig. \ref{fig37} (c), they are directed 
opposite to each other. The latter circumstance, as is easily 
seen, should play a significant role in considering the 
penetration of the ``current layers'' (parallel to the boundary 
of the sample) into the sample in our situation.

\begin{figure}[t]
\begin{center}
\vspace{5mm}
\includegraphics[width=\linewidth]{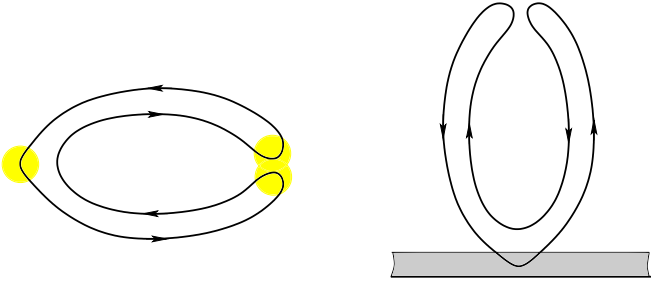}
\end{center}
\caption{A special extremal trajectory of the third type, 
allowing the ``deceleration section'' to enter the skin layer.}
\label{fig52}
\end{figure}

 The third type of special extremal trajectories that allow 
the ``deceleration section'' to enter the skin layer can be 
trajectories of similar shape, shown at Fig. \ref{fig31} (f) 
and \ref{fig38} (c) (Fig. \ref{fig52}). As in the previous 
case, in both cases trajectories of this shape arise only on one 
side of the corresponding reconstruction curve $\, \gamma \, $ 
at the angular diagram. Both cases, presented at 
Fig. \ref{fig31} and \ref{fig38}, do not have central 
symmetry, so the corresponding reconstructions can occur only 
in pairs and, in particular, each of the trajectories shown at 
Fig. \ref{fig31} (f) and \ref{fig38} (c) always arise 
in pair with a trajectory symmetrical to it. Between the 
trajectories shown at Fig. \ref{fig31} (f) and 
\ref{fig38} (c), as in the previous case, there is 
a difference in the direction of the group velocity at 
the ``deceleration sections''. Thus, for the trajectory shown 
at Fig. \ref{fig31} (f), the group velocity at the 
``deceleration section'' falling into the skin layer is opposite 
in direction to the group velocity at the other two 
``deceleration sections'', while for the trajectory, shown at 
Fig. \ref{fig38} (c), group velocities at all 
``deceleration sections'' are co-directed.

\begin{figure}[t]
\begin{center}
\vspace{5mm}
\includegraphics[width=\linewidth]{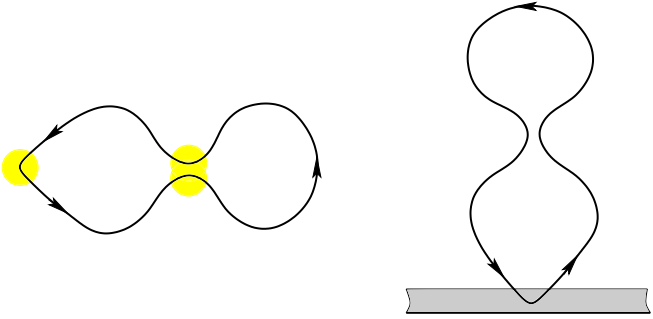}
\end{center}
\caption{A special extremal trajectory of the fourth type, 
allowing the ``deceleration section'' to enter the skin layer.}
\label{fig53}
\end{figure}

 The next type of special extremal trajectories that allow 
the ``deceleration section'' to fall into the skin layer can be 
the trajectories shown at Fig. \ref{fig28} (c,f) and 
\ref{fig36} (c) (Fig. \ref{fig53}). In the case of the 
reconstruction shown at Fig. \ref{fig27}, trajectories 
of this form arise on both sides of the curve of reconstruction of 
the structure of (\ref{MFSyst}). In this case, however, the 
trajectories on one side of the reconstruction curve, generally 
speaking, are not identical to the trajectories on the other side 
of the reconstruction curve, therefore, the parameters of the 
corresponding oscillating terms change during the reconstruction 
of the structure of (\ref{MFSyst}). For the reconstruction, shown 
at Fig. \ref{fig36} (c), trajectories of this shape arise 
only on one side of the corresponding reconstruction curve 
$\, \gamma \, $ on the angular diagram. Both cases, presented at 
Fig. \ref{fig27} and \ref{fig36}, do not have 
central symmetry, so the corresponding reconstructions can occur 
only in pairs and, in particular, each of the trajectories shown 
at Fig. \ref{fig28} (c,f) and \ref{fig36} (c) 
always appears in pair with a symmetrical trajectory to it. 
For all the presented trajectories, the group velocities are 
co-directional to each other at all three ``deceleration sections'' 
on the trajectory.

\begin{figure}[t]
\begin{center}
\vspace{5mm}
\includegraphics[width=\linewidth]{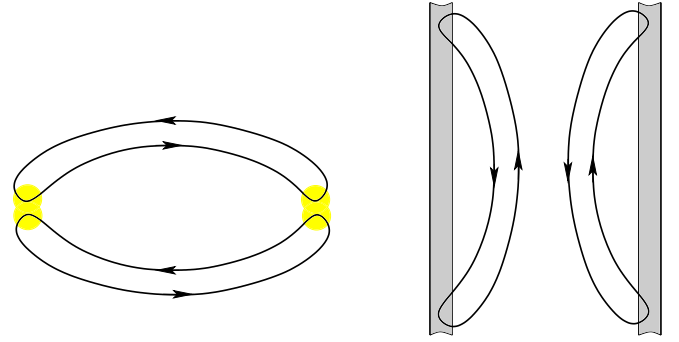}
\end{center}
\caption{A special extremal trajectory of the fifth type, allowing 
two ``deceleration sections'' to enter the skin layer at the
boundary of the sample. }
\label{fig54}
\end{figure}

 Finally, the last type of special extremal trajectories that 
allow the ``deceleration sections'' to enter the skin layer 
can be attributed to the trajectories shown at 
Fig. \ref{fig39} (c) (Fig. \ref{fig54}). In the case of 
the reconstruction shown at Fig. \ref{fig39}, trajectories 
of this shape arise only on one side of the corresponding 
reconstruction curve $\, \gamma \, $ on the angular diagram. 
The case presented at Fig. \ref{fig39} has central symmetry 
and can occur on a single part of the Fermi surface. The two 
trajectories shown at Fig. \ref{fig39} (c), are not 
connected to each other in the coordinate space and appear 
independently of each other. The main feature of the trajectories 
shown at Fig. \ref{fig39}(c) is that it is possible for them 
to place two ``deceleration sections'' at once into the skin layer 
at the sample boundary. Observation of cyclotron resonance in 
such a situation is, of course, more complicated due to 
interference from the contributions originating from two 
boundary points. On each of the trajectories presented at 
Fig. \ref{fig39} (c), the group velocities at the two 
``deceleration sections'' are opposite to each other.

\vspace{1mm}

 As it is easy to see, except the trajectories shown above, 
the other special extremal trajectories presented at 
Fig. \ref{fig26}, \ref{fig28}, 
\ref{fig31} - \ref{fig40} 
do not allow ``deceleration sections'' to fall into the skin 
layer due to the features of their local geometry near these
sections.

\section{Some features of the observation of oscillation 
phenomena on special extremal trajectories. }
\setcounter{equation}{0}

 This chapter is, to some extent, technical and is devoted to 
the peculiarities of observing of oscillation phenomena on the 
extremal trajectories described above. Most of it will actually 
be devoted to considering the features of observing cyclotron 
resonance when a ``deceleration section'' enters the skin layer 
near the surface of the sample. Besides that, we will consider 
some general features and limitations of observing the 
oscillation phenomena at very small angles of deviation of 
the direction of $\, {\bf B} \, $ from the boundaries of the 
reconstruction of the structure of system (\ref{MFSyst}).

 We note here that different effects of the 
``deceleration sections'' on semiclassical trajectories 
on the oscillation phenomena and, in particular, on the 
cyclotron resonance phenomenon, were considered, of course, 
earlier from different points of view. Let us also note that the 
deceleration sections on the trajectories can arise not only 
due to the presence of singular points of system (\ref{MFSyst}) 
on the Fermi surface, but also due to the presence of 
a reconstruction of the constant-energy surface (Van Hoff 
singularity) near the Fermi level
(see e.g. \cite{OrNeugFuagBarPotPelBas}). It can be noted 
here that the main difference between the two described 
situations consists, apparently, in preserving the finite 
value of the component of $\, {\bf v}_{gr} \, $ along the 
direction of $\, {\bf B} \, $ in the first case and the 
proximity to zero of all components of $\, {\bf v}_{gr} \, $ 
in the second.

 From the most general conditions, we note, first of all, 
that all the phenomena considered by us can occur only at 
rather big mean free path of an electron in a crystal. 
In the purest single-crystal samples at low temperatures, 
the mean free time of electrons reaches 
$\, 10^{-9} - 10^{-8} \, c \, $, and the corresponding mean 
free path $\, l \simeq 10^{-1} \, cm \, $. The value of the skin 
layer depth $\, \delta \, $ for the incident wave frequency 
$\, \nu \simeq 10^{10} \, Hz \, $, which is typical in a 
situation of cyclotron resonance observation, can be 
estimated in order of magnitude as 
$\, 10^{-5} - 10^{-4} \, cm \, $. 
In our situation, the resonant frequencies can, in fact, be 
lower than the value of $\, \nu \, $ given here, since the 
period of circulation along special extremal trajectories 
can be longer than the period of circulation along 
``ordinary'' closed trajectories. The value of 
$\, \delta \, $, however, is rather weakly dependent on the 
frequency in the situation of anomalous skin effect 
($\delta \sim \nu^{-1/3} $), therefore we will also use here 
the estimate $\, \delta \simeq 10^{-5} - 10^{-4} \, cm \, $. 
The size of a special extremal trajectory in the 
$\, {\bf x} $ - space (its projection onto the plane 
orthogonal to $\, {\bf B} $) depends on the value of $\, B $. 
For the values $\, B \simeq 1 \, Tl \, $ for electrons in 
different metals we can take the estimate 
$\, r_{B} \simeq \, 10^{-3} - 10^{-2} \, cm \, $ and 
$\, r_{B} \simeq \, 10^{-4} - 10^{-3} \, cm \, $ for 
$\, B \simeq 10 \, Tl \, $.

 As we have already said, we will be particularly interested 
in the case where the ``deceleration section'' on a special 
extremal trajectory falls into the skin layer at the boundary 
of the sample. For simplicity, we will assume here that the 
extremal trajectories have the most common centrally symmetric 
shape, shown at Fig. \ref{fig49}. 

 As is well known (\cite{AzKan1,AzKan2}), the main role in the 
phenomenon of classical cyclotron resonance is played by
the accumulation of energy by electrons in the skin layer and the 
synchronization of the incident radiation frequency with an 
integer multiple of the frequency of the electron circulation 
along extremal closed trajectories on the Fermi surface.  
As a rule, it is assumed that electron spends a rather small 
part of the time in the skin layer and manages to return to 
the skin layer many times between two scattering acts. 
In a standard situation, the fraction of time spent by an 
electron in the skin layer can be estimated as 
$\, \sqrt{\delta / r_{B}} \, $, where $\, \delta \, $ is the 
depth of the skin layer, and the number of electron returns to 
the skin layer varies from several tens to several hundred. 
It is usually assumed that the phase of the incident wave 
remains almost unchanged during the time spent by electron in 
the skin layer for relatively small $\, n \, $ in the relation 
$ \, \Omega = n \omega_{B} \, $, where $\, \Omega \, $ is 
the frequency of the incident radiation. In the purest samples, 
cyclotron resonance can be observed up to rather large values 
of $\, n \, $, reaching several tens.

 It can be seen that the conditions of observing the cyclotron 
resonance when a ``deceleration section'' enters the skin 
layer can, in fact, differ from the above. We should therefore 
consider the corresponding situation in more detail here.

 For our purposes, we will need to reproduce the (standard) 
consideration of the shape of special extremal trajectories 
and the motion along them at the ``deceleration sections''. 

\begin{figure}[t]
\begin{center}
\vspace{5mm}
\includegraphics[width=\linewidth]{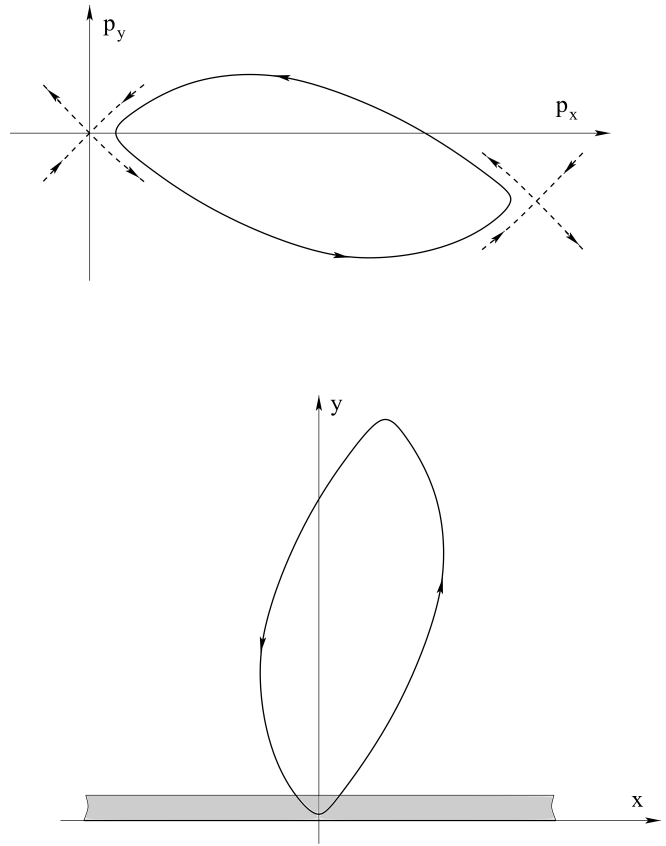}
\end{center}
\caption{Extremal trajectory in $\, {\bf p} $ - space and 
the corresponding trajectory in $\, {\bf x} $ - space passing 
through the skin layer at the surface of the sample 
(projection onto the plane orthogonal to $\, {\bf B} $). }
\label{fig55}
\end{figure}

 In our considerations, we will always assume that the axis 
$\, z \, $ is directed along the magnetic field. In addition, 
we choose the axes $\, x \, $ and $\, y \, $ parallel to the 
directions of the principle curvatures at the currently considered 
saddle singular point close to the special trajectory on the 
``disappearing'' cylinder of closed trajectories. For simplicity, 
we will also assume here that the boundary of the sample coincides 
with the plane $\, x z \, $ (Fig. \ref{fig55}).

 In general, the shape of a trajectory near a critical point 
in the $\, {\bf p} $ - space is given by the equation 
\begin{equation}
\label{TrImpPr}
{1 \over 2} \,\, \Big( K_{1} \, (\Delta p_{x})^{2} \,\, - \,\,
K_{2} \, (\Delta p_{y})^{2} \Big) \,\,\, = \,\,\, \Delta p_{z} \,\,\, ,
\end{equation}
where $\, K_{1} \, $ and $\, K_{2} \, $ are the values of the 
principal curvatures of the Fermi surface at the critical point.
The value of $\, \Delta p_{z} \, $ is directly related to the angle 
of deviation of $\, {\bf B} \, $ from the boundary of the 
reconstruction of the structure of system (\ref{MFSyst}), since 
$$\Delta p_{z} \,\,\, = \,\,\, \alpha \, |\xi_{\bf p}| / 2 \quad \quad
(0 < \alpha \ll 1) $$
 
 Thus, we can write
\begin{equation}
\label{TrImpPr2}
K_{1} \, (\Delta p_{x})^{2} \,\, - \,\, K_{2} \, (\Delta p_{y})^{2} 
\,\,\, = \,\,\, \alpha \, |\xi_{\bf p}|
\end{equation}

 In coordinate space directly from the equation (\ref{TrImpPr2}) 
follows the equation
$$K_{1} \, (\Delta y)^{2} \,\, - \,\, K_{2} \, (\Delta x)^{2}
\,\,\, = \,\,\, {c^{2} \over e^{2} B^{2}} \, \alpha |\xi_{\bf p}|
\,\,\, = \,\,\, {c \over e B} \, \alpha |\xi_{\bf x}| $$
for the projection of a special extremal trajectory onto the plane 
orthogonal to $ \, {\bf B} $. In particular, for 
$\, \Delta x = 0 \, $ we get
$$\Delta y \,\,\, = \,\,\, {c \over e B} \,
\sqrt{ {\alpha |\xi_{\bf p}| \over K_{1}} } $$
for the distance from the trajectory to the intersection point 
of the asymptotes of the hyperbola.

 By entering the notation $\, r_{B} \,= \, c p_{F} / e B \, $, 
the conditions $\, \Delta y \leq \delta \, $ and 
$\, \Delta y \geq \delta \, $ can be written as
$${1 \over p_{F}} \sqrt{{\alpha |\xi_{\bf p}| \over K_{1}} }
\,\,\, \leq \,\,\, {\delta \over r_{B}}  \quad \quad \text{and}
\quad \quad 
{1 \over p_{F}} \sqrt{{\alpha |\xi_{\bf p}| \over K_{1}} }
\,\,\, \geq \,\,\, {\delta \over r_{B}} $$

 Assuming approximately $\, |\xi_{\bf p}| \simeq p_{F} \, $ 
and $\, K_{1} \simeq p_{F}^{-1} \,$, the above conditions can also 
be written as
$$\sqrt{\alpha} \,\, \leq \,\, \delta / r_{B} 
\quad \quad \text{and} \quad \quad
\sqrt{\alpha} \,\, \geq \,\, \delta / r_{B}$$

 The projection of the group velocity onto the plane orthogonal 
to $\, {\bf B} $ can be approximately written near the 
singular point as 
$$v_{\perp} \,\,\, \simeq \,\,\, v^{0}_{gr} \,
\sqrt{ K_{1}^{2} \, (\Delta p_{x})^{2} \, + \,
K_{2}^{2} \, (\Delta p_{y})^{2} } \,\,\, , $$
where $\, v^{0}_{gr} \, $ is the group velocity at the singular 
point. Thus, on the trajectory we can, using (\ref{TrImpPr2}), 
write near the singular point 
$$v_{\perp} \,\,\, \simeq \,\,\, v^{0}_{gr} \,
\sqrt{ (K_{1} K_{2} + K_{2}^{2}) \, (\Delta p_{y})^{2}  \,\, + \,\,
\alpha K_{1} |\xi_{\bf p}| } \,\,\,\,\, , $$
and the velocity of passing the trajectory in 
$\, {\bf p} $ - space is respectively 
$${e B \over c} \, v^{0}_{gr} \,
\sqrt{ (K_{1} K_{2} + K_{2}^{2}) \, (\Delta p_{y})^{2}  \,\, + \,\,
\alpha K_{1} |\xi_{\bf p}| } $$

 The trajectory length element in the $\, {\bf p} $ - space can be 
written in the form 
\begin{multline*}
d l_{\bf p} \,\,\, = \,\,\, 
\sqrt{(d p_{x})^{2} \, + \, (d p_{y})^{2}} \,\,\, =  \\
= \,\,\, d p_{y} \, \sqrt{ \left( 
{K_{2} \Delta p_{y} \over K_{1} \Delta p_{x}} \right)^{2} 
\, + \, 1 } \,\,\, =  \\
= \,\,\, d p_{y} \, { 
\sqrt{(K_{1} K_{2} + K_{2}^{2}) \, (\Delta p_{y})^{2}  \,\, + \,\,
\alpha K_{1} |\xi_{\bf p}| } \over
\sqrt{ K_{1} K_{2} \, (\Delta p_{y})^{2}  \,\, + \,\,
\alpha K_{1} |\xi_{\bf p}| } } 
\end{multline*}

 Thus, the time of passing the section $\, d l_{\bf p} \, $ is 
equal to 
$$d t \,\,\, = \,\,\, {c \over e B v^{0}_{gr}} \,\,
{d p_{y} \over 
\sqrt{ K_{1} K_{2} \, (\Delta p_{y})^{2}  \,\, + \,\,
\alpha K_{1} |\xi_{\bf p}| } } $$

 The total time of increasing of the circulation period along 
a special extremal trajectory due to the presence of the 
``deceleration section'' can be (somewhat formally) estimated as 
$$T_{1} \,\, \simeq \,\, {c \over e B v^{0}_{gr}}
\int_{\Delta p_{y} = - p_{F}}^{\Delta p_{y} = p_{F}} \,\,
{d p_{y} \over \sqrt{K_{1} K_{2} (\Delta p_{y})^{2} \, + \,
\alpha K_{1} |\xi_{\bf p}|} }  \,\,\, = $$
$$= \,\,\, {c
 \over e B v^{0}_{gr}} {1 \over \sqrt{K_{1} K_{2}}}
\,\, \ln \left( {\sqrt{\alpha \, |\xi_{\bf p}| + K_{2} \, p_{F}^{2}} 
\,\, + \,\, \sqrt{K_{2}} \, p_{F}  \over 
\sqrt{\alpha \, |\xi_{\bf p}| + K_{2} \, p_{F}^{2}} \,\, - \,\, 
\sqrt{K_{2}} \, p_{F} } \right)  $$

 In general, we have 
\begin{equation}
\label{Integral1}
{c \over e B v^{0}_{gr}}
\int_{\Delta p_{y} = - P}^{\Delta p_{y} = P} \,\,
{d p_{y} \over \sqrt{K_{1} K_{2} (\Delta p_{y})^{2} \, + \,
\alpha K_{1} |\xi_{\bf p}|} }  \,\,\, = 
\end{equation}
$$ = \,\,\, {c \over e B v^{0}_{gr}} {1 \over \sqrt{K_{1} K_{2}}}
\,\, \ln \left( {\sqrt{\alpha \, |\xi_{\bf p}| + K_{2} \, P^{2}}
\,\, + \,\, \sqrt{K_{2}} \, P  \over
\sqrt{\alpha \, |\xi_{\bf p}| + K_{2} \, P^{2}} \,\, - \,\,
\sqrt{K_{2}} \, P } \right) $$

 In the limit
$${K_{2} \, p_{F}^{2} \over \alpha \, |\xi_{\bf p}| } \,\, \gg \,\, 1 $$
or just
$$\alpha \ll 1 $$
we can write
$$T_{1} \,\,\,\,\, \simeq \,\,\,\,\, {c \over e B v^{0}_{gr}} \,
{1 \over \sqrt{K_{1} K_{2}}} \,\, \ln 
{4 K_{2} \, p_{F}^{2} \over \alpha \, |\xi_{\bf p}|} $$

 It is easy to see that in the centrally symmetric case we are 
considering, the time $\, T_{1} \, $ should actually be doubled 
due to the presence of two identical ``deceleration sections'' 
on the extremal trajectory. Here we can also see that for any 
definition of $\, T_{1} \, $, the total period of circulation 
along the trajectory under consideration grows linearly in 
$\, \ln 1 / \alpha \, $ with decreasing $\, \alpha \, $ with the 
coefficient $\, 2 c / e B v^{0}_{gr} \sqrt{K_{1} K_{2}} \, $.
This circumstance can be used, in particular, to determine the 
value of $\, v_{gr} \sqrt{K_{1} K_{2}} \, $ at the corresponding 
limit singular points of (\ref{MFSyst}), i.e. on the curves 
$\, \hat{\gamma}_{1,2}\,$.

 For the principle curvatures here, as above, we can use the 
estimates $\, K_{1} \simeq K_{2} \simeq p_{F}^{-1} \, $ and write
the estimate
$$T_{1} \,\,\, \simeq \,\,\, {c \, p_{F} \over e B \, v^{0}_{gr}} \,\,
\ln {4 p_{F} \over \alpha \, |\xi_{\bf p}|} $$

 As for the value of $\, | \xi_{\bf p} | \, $, it can be of the 
order of $\, p_{F} \, $ (for the simplest trajectories) or several 
times larger (or an order of magnitude). For a rough estimate of 
the time $\, T_{1} \, $, we can use the relation 
$$T_{1} \,\,\, \simeq \,\,\, {r_{B} \over v_{F}} \,\,
\ln {1 \over \alpha} $$

 For very small values of $\, \alpha \, $ the doubled time 
$\, T_{1} \, $ can be considered approximately equal to the 
period of circulation along a special extremal trajectory 
(if it exceeds the time 
$\, T_{0} \, \simeq \, 2\pi r_{B} / v_{F} \, $ or is comparable 
with it).

 To estimate the time spent by electrons in the skin layer, note 
that the corresponding section of the trajectory in 
$\, {\bf p}$ - space corresponds to the values of 
$\, \Delta p_{x} \,$, satisfying the relation
$$\sqrt{ {\alpha \, |\xi_{\bf p}| \over K_{1}} } 
\,\,\,\,\, \leq \,\,\,\,\, \Delta p_{x} \,\,\, \leq \,\,\, 
\sqrt{ {\alpha \, |\xi_{\bf p}| \over K_{1}} } \,\, + \,\,
{\delta \over r_{B}} \, p_{F} $$

 Substituting the second value in the equation (\ref{TrImpPr2}), 
we get the relation for the boundary values of $\, \Delta p_{y} \, $
$$K_{2} \, (\Delta p_{y})^{2}
 \,\,\,\,\, = \,\,\,\,\,
2 \, \sqrt{K_{1} \, \alpha \, |\xi_{\bf p}|} \,\, 
{\delta \over r_{B}} \, p_{F} \,\,\, + \,\,\,
K_{1} \, {\delta^{2} \over r_{B}^{2}} \, p_{F}^{2} $$

 Substituting the found values into the integral 
(\ref{Integral1}), we will thus find the time $\, T_{2} \, $ 
spent by an electron in the skin layer. Here, however, we will 
not present these calculations in a general form, but consider 
only two limiting cases: 
$$\sqrt{ {\alpha \, |\xi_{\bf p}| \over K_{1}} } \,\,\, \ll \,\,\,
{\delta \over r_{B}} \, p_{F} \quad \quad {\text {\rm and}} \quad \quad
\sqrt{ {\alpha \, |\xi_{\bf p}| \over K_{1}} } \,\,\, \gg \,\,\,
{\delta \over r_{B}} \, p_{F} $$
or, in a rough approximation: 
$$\sqrt{\alpha} \,\, \ll \,\, \delta / r_{B} 
\quad \quad \text{and} \quad \quad
\sqrt{\alpha} \,\, \gg \,\, \delta / r_{B}$$

 In the first case, we can write 
$$\Delta p_{y} \,\,\, = \,\,\, \pm \, \sqrt{{K_{1} \over K_{2}}} \,\,
{\delta \over r_{B}} \, p_{F} $$
and write the corresponding value for $\, T_{2} \, $ in the form 
\begin{multline*}
T_{2} \,\,\, \simeq \,\,\, 
{c \over e B v^{0}_{gr}} \, {1 \over \sqrt{K_{1} K_{2}}} \,\, \times  \\
\times \,\, \ln \left( 
{\sqrt{\alpha \, |\xi_{\bf p}| + K_{1} \, \delta^{2} \, 
p_{F}^{2} / r_{B}^{2} }
\,\, + \,\, \sqrt{K_{1}} \, \, \delta \, p_{F} / r_{B}  \over
\sqrt{\alpha \, |\xi_{\bf p}| + K_{1} \, \delta^{2} \, 
p_{F}^{2} / r_{B}^{2} }
\,\, - \,\, \sqrt{K_{1}} \, \delta \, p_{F} / r_{B} } \right) 
\end{multline*}

 Using the same assumption, we can now write also 
$$T_{2} \,\,\, \simeq \,\,\, {c \over e B v^{0}_{gr}} \,
{1 \over \sqrt{K_{1} K_{2}}} \,\, \ln 
{4 K_{1} \, p_{F}^{2} \, \delta^{2} \over \alpha \, |\xi_{\bf p}| \,
r_{B}^{2} } $$

 We can see that the ratio of times  $\, T_{2} / T_{1} \, $ 
in this case is equal to 
$$\left(  \ln 
{4 K_{1} \, p_{F}^{2} \over \alpha \, |\xi_{\bf p}| } \,\, - \,\,
\ln {r_{B}^{2} \over \delta^{2}} \right) \Big/
\ln {4 K_{1} \, p_{F}^{2} \over \alpha \, |\xi_{\bf p}| } $$
and under our assumption is close to unity. It can be noted that 
the same ratio in the case of getting of a ``normal part'' of 
a trajectory into the skin layer is of the order of 
$\, \sqrt{\delta/r_{B}} \, $ and represents a small value.  
As a consequence of this, the amplitude of oscillations in the 
absorption of the incident radiation can be noticeably higher 
here in comparison with the ``ordinary'' case. 
It should also be noted that the condition 
$${\alpha \, |\xi_{\bf p}| \over 4 K_{1} \, p_{F}^{2} } 
\,\,\, \ll \,\,\, {\delta^{2} \over r_{B}^{2}} $$
is actually quite strong and can be observed only in sufficiently 
strong magnetic fields and at very small angles 
$\, \alpha $. Thus, using again rough estimates for 
$\, K_{1} \, $ and $\, |\xi_{\bf p}| \,$, this condition 
can be written as
$\,\, \alpha \, \ll \, \delta^{2} / r_{B}^{2} \, $.
We can see that this condition in reality implies the relations 
$\,\, \delta / r_{B} \, \simeq \, 10^{-1} \, $
(strong magnetic fields $\, B \, \simeq \, 10 \, Tl $) and 
$\, \alpha \leq 10^{-3} \, $.

\vspace{2mm}

 As for the opposite limit
$$\sqrt{ {\alpha \, |\xi_{\bf p}| \over K_{1}} } \,\,\, \gg \,\,\,
{\delta \over r_{B}} \, p_{F} $$
(but, as before, $\, \alpha \ll 1$), it is, on the contrary, observed 
at not very strong magnetic fields. Here, as above, we can write the 
relation for rough estimation
$$\sqrt{\alpha} \,\,\, \gg \,\,\, \delta / r_{B} $$

 For example, for the values of $\, B \, $ of order of $ 1 \,  Tl $ 
we can take $\, \delta / r_{B} \simeq 10^{-2} \, $ and the required 
condition is satisfied for $\, \alpha \geq 10^{-3} $. For 
$\, B \simeq 10 \,  Tl \, $ we have the relation
$\, \delta / r_{B} \simeq 10^{-1} \, $ and then we should assume 
$\, \alpha \gg 10^{-2} $.

 Let us also use a rough estimate for the values $\, \Delta p_{y} \, $, 
bounding the region corresponding to the stay of the electron in the 
skin layer 
$$\Delta p_{y} \,\,\, = \,\,\, p_{F} \, 
\sqrt{\sqrt{\alpha} \, \delta / r_{B} } $$

 After substituting these values into the integration limits in 
(\ref{Integral1}), we get 
\begin{multline*}
T_{2} \,\,\, \simeq \,\,\,
{c \over e B v^{0}_{gr}} \, {1 \over \sqrt{K_{1} K_{2}}} \,\, \times  \\
\times \,\, \ln \left(
{ \sqrt{1  \, + \,  \delta / ( r_{B} \sqrt{\alpha}) }
\,\, + \,\, \sqrt{\delta / ( r_{B} \sqrt{\alpha}) }   \over
\sqrt{1  \, + \,  \delta / ( r_{B} \sqrt{\alpha}) }
\,\, - \,\, \sqrt{\delta / ( r_{B} \sqrt{\alpha}) } } \right) 
\,\,\, \simeq 
\end{multline*}
$$\simeq \,\,\, 
{2c \over e B v^{0}_{gr}} \, {1 \over \sqrt{K_{1} K_{2}}} \,\,
\sqrt{\delta / ( r_{B} \sqrt{\alpha})} \,\,\, \simeq   \,\,\, 
{2 r_{B} \over v_{F}} \,\, \sqrt{\delta / ( r_{B} \sqrt{\alpha})} $$

 Note that the time ratio  $\, T_{2} / T \, $ here is formally larger 
than $\, \sqrt{\delta / r_{B}} \, $ for $\, \alpha \ll 1 \, $. 
However, substituting the above real values of $\, \alpha \, $
for this limit, we can see in fact that the value of
$\, 1 / \sqrt[4]{\alpha} \, $ is not very large. It can be seen, 
therefore, that in the case under consideration, this ratio is of the
order of the same ratio for the case when an ``ordinary'' section of 
an extremal trajectory enters the skin layer. In fact, this is not 
a contradiction, because, although the section under consideration 
is a ``deceleration section'', the curvature of the trajectory on it is 
much larger than the curvature on ``ordinary'' sections. Thus, in the 
limit we are considering, we should expect approximately the same
amplitude of oscillations in the absorption of the incident radiation 
compared with the cyclotron resonance in the ``standard'' situation.

 In general, we can see that when the ``deceleration section'' hits 
the skin layer, the behavior of the amplitude of oscillations of 
absorbtion of the incident radiation differs noticeably from its 
behavior in the ``usual'' situation. In particular, the amplitude 
of the absorption oscillations can be of the order of the same 
amplitude in the ``standard'' case when the direction of 
$\, {\bf B} \, $  is not too close to the boundary of the 
reconstruction of the structure of system (\ref{MFSyst}). 
With more precise approximation of the direction of 
$\, {\bf B} \, $ to the boundary of reconstruction of the 
structure of (\ref{MFSyst}), the amplitude of the oscillations 
increases and at very small $\, \alpha \, $ 
and in sufficiently strong magnetic fields can noticeably exceed 
the amplitude of such oscillations in the ``standard'' case.

 It can also be seen that, for the same reasons, the intensity 
of the current layers, corresponding to the ``deceleration sections'', 
inside the sample (when the ``normal'' section of the trajectory 
enters the skin layer (Fig. \ref{fig48})) should not be very large
for not very precise approach of the direction of $\, {\bf B} \, $ 
to the boundary of the reconstruction of the structure of system 
(\ref{MFSyst}) and should increase markedly with a more precise 
approach of the direction of $\, {\bf B} \, $ to this boundary. 
In particular, their intensity should be comparable with  the 
intensity of the current layers corresponding to the sections 
at which the group velocity is parallel to the sample boundary, 
in the limit $\, \sqrt{\alpha} \gg \delta/r_{B} \, $, and can  
noticeably exceed it in the limit 
$\, \sqrt{\alpha} \ll \delta/r_{B} \, $.

\vspace{1mm}

 Here, however, we need to make a few more remarks about the 
amplitude of oscillations at a very close approximation of the 
direction of $\, {\bf B} \, $ to a boundary of the reconstruction 
of the structure of system (\ref{MFSyst}).

 First, as we said above, in this approximation the height of the 
corresponding cylinder of closed trajectories becomes very small. 
Due to this circumstance, the measure of trajectories near the 
special extremal trajectory, contributing to the corresponding 
oscillation term, can also decrease. This effect, as we see, is 
opposite to the effect described above, so that, possibly, 
as a result, the amplitude of the corresponding oscillation term 
may not significantly exceed its amplitude in the ``standard''
situation. The exact picture 
of the behavior of the corresponding oscillation amplitude depends, 
in fact, on the features of the geometry of the Fermi surface near 
the special extremal trajectory and on the experimental conditions.

 Another distinctive feature in observing the cyclotron resonance 
when a ``deceleration section'' enters the skin layer is that, at 
very small $\, \alpha \, $, the decreasing of the amplitude of the 
observed oscillations with increasing number 
$\, n \, = \, T \Omega / 2 \pi \, = \, T \nu \, $ here should be 
much faster compared to the case of ``ordinary'' extremal closed 
trajectories. The reason for this is that, due to the long time 
spent by electron in the skin layer, $\, n \, $ cannot be too large, 
otherwise the field $\, {\bf E}_{w} \, $ can become oscillating 
already at the time of the electron's presence in the skin layer, 
which significantly reduces the efficiency of energy accumulation 
during this time. As we saw above, in the limit of the extreme 
proximity of the singular points of system (\ref{MFSyst}) to the 
trajectories under consideration, the time spent by the electron 
in the skin layer can become comparable with the period 
$\, T \, $ of circulation along the entire trajectory. In this 
situation, only a few peaks of oscillations of rapidly decreasing 
amplitude are likely to have a relatively large value. (For 
comparison, it can be noted here that for cyclotron resonance 
on ordinary trajectories, the number of observed peaks in similar 
oscillating terms can reach several tens). A sharp decrease in the 
number of observed peaks in the absorption amplitude can also serve 
as an indication of the approach of the direction of 
$\, {\bf B} \, $ to the boundary of the reconstruction of the 
topological structure
 of system (\ref{MFSyst}).

\vspace{2mm}

 Finally, let us make one more remark here. According to the 
general idea of this paper, it is obvious that we would like to 
study oscillation phenomena as close as possible to the ``net'' 
of directions of $\, {\bf B} \, $, corresponding to the 
reconstructions of the structure of (\ref{MFSyst}). As for the
restrictions on the approximation of the direction of 
$\, {\bf B} \, $ to the curve of a reconstruction of the 
structure of (\ref{MFSyst}) while maintaining the observation 
conditions for oscillation phenomena on special extremal 
trajectories, we can immediately distinguish two such 
restrictions. The first can be attributed to an increase in 
the period $\, T \, $ of circulation along extremal trajectories, 
while the observation of the oscillation phenomena requires 
the condition $\, \tau / T \gg 1 \, $. As we noted above, this 
condition is not really too restrictive, since the growth of 
$\, T \, $ is rather slow (logarithmic) when approaching the 
boundary of a reconstruction.

\begin{figure}[t]
\begin{center}
\includegraphics[width=0.9\linewidth]{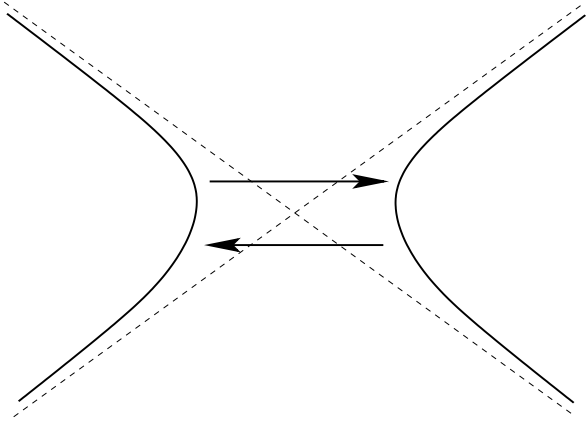}
\end{center}
\caption{The phenomenon of magnetic breakdown between 
two ``deceleration sections'' on trajectories approaching 
singular ones.}
\label{fig56}
\end{figure}

 The second limitation can be attributed to the phenomenon 
of magnetic breakdown between two close sections of trajectories 
close to a singular one, as the direction of $\, {\bf B} \, $ 
approaches the boundary of the reconstruction of system 
(\ref{MFSyst}) (Fig. \ref{fig56}).

 Here we will also not consider in detail the theory of 
magnetic breakdown, which leads to many different effects for 
various trajectories of the system (\ref{MFSyst}) 
(see e.g. \cite{Zilberman1,Zilberman2,Zilberman3,Azbel,
etm,Kittel,Abrikosov,AlexsandradinGlazman}). In our case, it is 
obvious that the ``simple'' picture of oscillation phenomena 
that we assume on special extremal trajectories should arise 
in a situation where the probability of magnetic breakdown is 
close to zero. In our situation we are dealing with the 
intraband magnetic breakdown, the probability of which becomes 
significant when the height of the classical potential barrier 
between close sections of the trajectory approaches the value 
$\, \mu_{B} B \, $. The height of the classical potential 
barrier is directly related to the angle $\, \alpha \, $ and, 
as it is not difficult to see, can be estimated as 
$$\Delta \epsilon \,\,\, \simeq \,\,\, v^{0}_{gr} 
{\alpha |\xi_{\bf p}| \over 2} \,\,\, \simeq \,\,\, 
v_{F} {\alpha |\xi_{\bf p}| \over 2} \,\,\, \simeq \,\,\,
{\alpha \epsilon_{F} \over 2} {|\xi_{\bf p}| \over p_{F}} 
\,\,\, , $$
such that
$${\Delta \epsilon \over \mu_{B} B} \,\,\, \simeq \,\,\,
{\epsilon_{F} \over \mu_{B} B} \, {\alpha \over 2} \,
{|\xi_{\bf p}| \over p_{F}} $$

 In the roughest approximation, we can actually write
$$ {\Delta \epsilon \over \mu_{B} B} \,\,\, \simeq \,\,\,
{\epsilon_{F} \over \mu_{B} B} \, \alpha $$

 Taking the values
$\, \mu_{B} \, \simeq \, 5.8 \cdot 10^{-5} \, eV/Tl \, $ and
$\, \epsilon_{F} \, \simeq \, 5 \, eV \, $, we can see that
$${\Delta \epsilon \over \mu_{B} B} \,\,\, \simeq \,\,\,
10^{5} \, \alpha \quad \quad \text{at} \quad \quad
B \, \simeq \, 1 \, Tl $$
and
$${\Delta \epsilon \over \mu_{B} B} \,\,\, \simeq \,\,\,
10^{4} \, \alpha \quad \quad \text{at} \quad \quad 
B \, \simeq \, 10 \, Tl $$

\vspace{1mm}

 Thus, we can see that the ``simple'' picture of oscillation 
phenomena (classical or quantum) is limited by the values  
$\, \alpha \, \geq \, 10^{-5} \, $ at
$\,  B \, \simeq \, 1 \, Tl \,\, $ and
$\,\, \alpha \, \geq \, 10^{-4} \,\, $ at 
$\,  B \, \simeq \, 10 \, Tl \, $. In particular, one may ask 
about the possibility of observing such a picture in the above 
limit 
$$\sqrt{{\alpha |\xi_{\bf p}| \over K_{1}}} \,\,\, \ll \,\,\,
{\delta \over r_{B}} \, p_{F} \quad \quad \text{or} \quad \quad
\sqrt{\alpha} \,\,\, \ll \,\,\, \delta / r_{B} $$
in the case of the cyclotron resonance. 

 Going back to the previous estimates
$\, \delta / r_{B} \, \simeq \, 10^{-1} \,\, $
($B \simeq 10 \, Tl$) and $\, \alpha \leq 10^{-3} \, $ 
for such a situation, it can be seen that this limit, in fact, 
may have certain restrictions due to the magnetic breakdown.

\vspace{1cm}

\section{Conclusions}
\setcounter{equation}{0}

 The paper considers issues related to reconstructions of 
the topological structure of a dynamical system that describes 
the semiclassical motion of electrons on complex Fermi surfaces 
in the presence of an external magnetic field. In particular, 
the paper lists all types of elementary reconstructions of the 
structure of this system and describes special closed extremal 
trajectories, the appearance of which always accompanies such 
reconstructions. It is shown that the oscillation phenomena 
corresponding to the appearance of such trajectories may have 
certain special features. The study of oscillation phenomena 
associated with special extremal trajectories can be a useful 
tool for the study of electronic spectra in metals.

\vspace{1cm}

The study was carried out at the expense of a grant from the 
Russian Science Foundation (project \textnumero $\, $ 18-11-00316).

\end{document}